\documentclass[traditabstract]{aa}

\usepackage{graphicx}
\usepackage{txfonts}
\usepackage{color}
\usepackage{natbib} 
\usepackage{hyperref}

\def\ts{$T^*$}
\def\n3{$n_u$}
\def\n1{$n_l$}
\def\gu{$g_u$}
\def\gl {$g_l$}
\def \be{\begin{equation}}
\def \ee{   \end{equation}}

\newcommand{\Msun}{M_{\odot}}
\newcommand{\Zsun}{Z_{\odot}}

\begin{document}

   \title{The [CII] 158 $\mathrm{\mu}$m line emission in high-redshift galaxies}

   \author{G. Lagache
          \inst{1}
          \and
          M. Cousin
          \inst{1}
        \and
          M. Chatzikos
         \inst{2}
          }

   \institute{Aix Marseille Univ, CNRS, LAM, Laboratoire d'Astrophysique de Marseille, Marseille, France
   \email{guilaine.lagache@lam.fr}
  \and Department of Physics and Astronomy, University of Kentucky, Lexington, KY 40506, United States}
   \date{Received September 30, Accepted November 1, 2017}
 
  \abstract{Gas is a crucial component of galaxies, providing the fuel to form stars, and it is impossible to understand the evolution of galaxies without knowing their gas properties. The [CII] fine structure transition at 158\,$\mu$m is the dominant cooling line of cool interstellar gas, and is the brightest of emission lines from star forming galaxies from FIR through meter wavelengths, almost unaffected by attenuation. With the advent of ALMA and NOEMA, capable of detecting [CII]-line emission in high-redshift galaxies, there has been a growing interest in using the [CII] line as a probe of the physical conditions of the gas in galaxies, and as a star formation rate (SFR) indicator at $z\ge4$. In this paper, we use a semi-analytical model of galaxy evolution ({\tt G.A.S.}) combined with the photoionisation code {\tt CLOUDY} to predict the [CII] luminosity of a large number of galaxies (25,000 at $z\simeq$5) at $4\le z \le 8$. We assume that the [CII]-line emission originates from photo-dominated regions.  At such high redshift, the CMB represents a strong background and we discuss its effects on the luminosity of the [CII] line. We study the L$_{\mathrm{[CII]}}$--SFR  and L$_{\mathrm{[CII]}}$--$Z_g$ relations and show that they do not strongly evolve with redshift from z=4 and to z=8.  Galaxies with higher [CII] luminosities tend to have higher metallicities and higher star formation rates but the correlations are very broad, with a scatter of about 0.5 and 0.8\,dex for L$_{\mathrm{[CII]}}$--SFR  and L$_{\mathrm{[CII]}}$--$Z_g$, respectively. Our model reproduces the L$_{\mathrm{[CII]}}$--SFR relations observed in high-redshift star-forming galaxies, with [CII] luminosities lower than expected from local L$_{\mathrm{[CII]}}$--SFR relations. Accordingly, the local observed L$_{[CII]}$--SFR relation does not apply at high-z (z$\gtrsim$5), even when CMB effects are ignored. Our model naturally produces the [CII] deficit (i.e. the decrease of L$_{\mathrm [CII]}$/L$_{IR}$ with L$_{IR}$), which appears to be strongly correlated with the intensity of the radiation field in our simulated galaxies.
 We then predict the [CII] luminosity function, and show that it has a power law form in the range of L$_{\mathrm{[CII]}}$ probed by the model (1$\times$10$^7$ - 2$\times$10$^9$\,L$_{\odot}$ at z=6) with a slope $\alpha$=-1. The slope is not evolving from z=4 to z=8 but the number density of [CII]-emitters decreases by a factor of 20$\times$. We discuss our predictions in the context of current observational estimates on both the differential and cumulative luminosity functions. The outputs from the model are distributed as FITS-formatted files at the CDS.}


   \maketitle

\section{Introduction}

One of the final frontiers in piecing together a coherent picture of cosmic history relates to the period 300-900 million years after the Big Bang (redshifts $6<z<15$). During this time, the Universe underwent two major changes. Firstly, the earliest stars and galaxies began to shine, bathing the Universe in starlight. Secondly, the intergalactic medium transitioned from a neutral to a fully ionized gas, a timespan known as the epoch of reionization (EoR). Connecting these two changes is highly desirable and after years of effort, recent breakthroughs showed that reionization occured at $6<z<10$ \citep{planck16} and that UV-selected star-forming galaxies likely dominated the reionization process \citep[e.g.,][]{robertson15}. Active galactic nuclei can also potentially contribute to reionization \citep{giallongo15}; the exact role of the two populations is still unclear.

Another remarkable result of cosmology in the last decade is the realization that the star formation rate (SFR) density at redshifts z$>$1 is higher than at present by about an order of magnitude and that half of the energy produced since the surface of last scattering has been absorbed and reemitted by dust \citep{dole06}, in dusty star-forming galaxies (DSFG). Most of the light produced at high redshift thus reaches us in the wavelength range
100$\mu$m-1mm \citep{lagache05}. Contribution of DSFG to the global star formation history is roughly known up to $z=3$ \citep{madau14}. But at higher redshifts and in the EoR, it is an uncharted territory. At such early epochs (z$>$5) dust is surely present even if in small amounts \citep{riechers13, watson15}  and can strongly affect SFR measurements based on UV-luminosity. 

With the advent of the Atacama Large Millimeter Array (ALMA) and NOEMA, it is now possible to measure the dust content of very high redshift galaxies, but also to use far-infrared fine-structure lines (as [OIII] or [CII]) to study the physical conditions of their interstellar medium (ISM). The [OIII] line, originating from diffuse and highly ionized regions near young O stars, is a promising line \citep{inoue16} that might gain in importance in low-metallicity environments where photo-dominated regions (PDRs) may occupy only a limited volume of the ISM.  The [CII] line, predominantly originating from PDRs at high redshift \citep{stacey10, gullberg15}, can provide SFR estimates that are not biased by dust extinction, although it has been found to depend strongly on the metallicity \citep{vallini15, olsen17}. This line can also be used to measure the systemic redshift of the galaxies \citep[e.g.,][]{pentericci16}.  In addition, the [CII]-line ALMA surveys will derive the line luminosity functions, thus measuring the abundance and intensity distributions of [CII] emitters \citep{aravena16}.

Due to its relatively low ionization potential, [CII] is the dominant form of the element under a large variety of conditions.  The C$^{+}$ ion has only two fine structure levels in the ground electronic state. The lower J = 1/2 level has statistical weight \gl\ = 2.  The upper J = 3/2 level has statistical weight \gu\ = 4, and lies at  equivalent temperature \ts\ = $\Delta E/k$ = 91.25 K above the ground state. The measured transition frequency is 1900.537 GHz \citep{cooksy86} corresponding to a transition wavelength of 157.74 $\mu$m, making the [CII] line easily accessible from the ground for $4.5\lesssim z \lesssim 8.5$. These redshifts marks an important epoch when the ISM in typical galaxies matures
from a nearly primordial, dust-free state at $z\sim8$, during the EoR, to the dust- and metallicity-enriched state observed at $z\sim4$.

Consequently, we investigate in this paper the correlation between SFR, [CII] luminosity and metallicity, and predict the luminosity function of [CII] line emitters at $z\ge4$. We use the Semi-Analytical Model (SAM) described in \cite{cousin15a}, that we combine with the {\tt CLOUDY} photoionisation code \citep{ferland13, C17}.  For each galaxy in the SAM (that has its own mass, SFR, metallicity, size, etc) we define an equivalent photo-dominated region characterised by its own properties (i.e. interstellar radiation field, gas metallicity, mean hydrogen density) and run {\tt CLOUDY} to derive its [CII] emission, taking into account the CMB (heating and attenuation). We are well aware that using global galaxy characteristics to predict the [CII] line emission ignores the complex properties of galaxies at very high redshift in which differential dust extinction, excitation and metal enrichment levels may be associated with different subsystems assembling the galaxies \citep[e.g,][]{carniani17, katz17,pallottini17}. Complex hydrodynamical simulations are being undertaken \citep[e.g.,][]{olsen17} but future developments and more statistics are needed to make detailed comparisons with observations \citep[see the discussion in][]{katz17}. In the meantime, the low computational cost of SAMs makes them a powerful tool to model large volumes of the sky and to sample a large diversity of galaxy properties. 

The paper is organized as followed: in Sect\,\ref{model}, we present briefly our SAM and validate its use for predicting the [CII] emission at very high redshift. Then, we describe our model for [CII]-line emission, and we quantify the effects of CMB and galaxy properties (as gas metallicity) on [CII] luminosity  (Sect.\,\ref{model_CII}). We discuss in Sect.\,\ref{sect_SFR_LCII} the L$_{\mathrm{[CII]}}$ -- SFR relation, and compare it with recent observations. Section\,\ref{sect_CII_def} is dedicated to the [CII] deficit. 
In Sect.\,\ref{sect_LF} we present the [CII] luminosity function from z=4 to 8 and discuss its evolution. Finally, we conclude in Sect.\,\ref{sect_cl}. Throughout the paper, we use \cite{chabrier2003} initial mass function.

\section{\label{model} Galaxy formation in the early Universe}

\subsection{Brief description of the semi-analytical model}

We use the SAM presented in \cite{cousin15a, cousin15b, cousin16}.  In addition to the original prescriptions detailed in \cite{cousin15a}, the extension of the model described in \cite{cousin16} tracks the metal enrichment in both the gas phase and stellar populations, which is essential to predict the [CII] emission. The chemodynamical model is applicable from metal-free primordial accretion to very enriched interstellar gas contents. 

The SAM is combined with dark-matter merger trees extracted from a pure N-body simulation. The simulation is based on a WMAP-5yr cosmology ($\Omega_m = 0.28$, $\Omega_{\Lambda} = 0.72$, $f_b = 0.16$, $h = 0.70$) and covers a volume of $[100/h]^3 Mpc$ with $1024^3$ particles. Each particle has a mass $m_p = 1.025~10^8~\Msun$. Haloes and sub-structures (satellites) are identified using {\tt HaloMaker} \citep{Tweed_2009}.

Dark matter haloes grow following a smooth accretion, with a dark-matter accretion rate $\dot{M}_{dm}$ derived from particles that are newly detected in the halo and that have never been identified in an other halo. Baryons are then progressively accreted following
\begin{equation}
\dot{M}_b = f_b^{ph-ion}(M_h,z)\dot{M}_{dm} \,,
\end{equation}
where $f_b^{ph-ion}(M_h,z)$ is the effective baryonic fraction depending on the virial halo mass and redshift. This fraction is computed following \cite{gnedin_2000} and \cite{kravtsov_2004} photoionization models but with an effective filtering mass as defined in \cite{okamoto_2008}.

Our SAM assumes a bimodal accretion \citep{khochfar2009, benson2011}, based on a cold and a hot reservoirs that are both fed by the metal-free cosmological accretion. In addition, the hot reservoir receives the galactic metal-rich ejecta. As the metallicity of the wind phase depends on the galaxy metal enrichment process, the metallicity of the hot reservoir evolves with time. Metals are initially formed by stars in the galaxies. The enriched gas is then ejected by supernova and active galactic nuclei (AGN) feedback (see \cite{cousin15a} for the detailed implementation of the supernovae and AGN feedback). 

The chemodynamical model  \citep{cousin16} can follow the $^1H$, $^4He$, $^{12}C$, $^{14}N$, $^{16}O$ and $^{56}Fe$ elements in the gas phase. Their production in stars and re-injection in the ISM are taken into account for stars with initial mass between 0.1\,M$_\odot$ and 100\,M$_\odot$ and for metal-free to super-solar metal fraction. It is assumed that stars are formed following a \cite{chabrier2003} initial mass function.

One of the particularity of the \cite{cousin15a, cousin15b} model is that the freshly accreted gas is assumed to be no-star-forming. It is progressively converted into star-forming gas and then into stars. The main idea behind the existence of the no-star-forming gas reservoir is that only a fraction of the total gas mass in a galaxy is available to form stars. The reservoir generates a delay between the accretion of the gas and the star formation. In the present paper, we use the conversion between the no-star-forming and star-forming gas as described in \cite{cousin17a}, thus assuming an inertial turbulent cascade in the gas. This updated version of \cite{cousin15a} SAM is called {\tt G.A.S.}, for Galaxy Assembly from dark-matter Simulations. 

Cousin et al. 2017b (in prep) present the model for dust extinction and emission. This modelling is not used for the [CII]-line emission prediction but is mandatory to compute the UV and IR luminosities of {\tt G.A.S.} galaxies.  Stellar spectra are based on \cite{Bruzual_2003} library. Extinction curves and dust spectral energy distribution are computed using {\tt DustEM} \citep{compiegne_2011} and are self consistently applied to the disc and the bulge of the galaxy. A standard slab geometry for old stars in the disc is used \citep{guiderdoni87}. Additional extinction from burst clouds is applied for young stars in disc using a screen geometry \citep{charlot00}. For the bulge, a standard Dwek geometry is used \citep[][and references therein]{devriendt99}. Effective extinctions predicted by this model are in excellent agreement with \cite{calzetti_2000} extinction law. 

\subsection{High-redshift stellar-mass and UV luminosity functions}

The {\tt G.A.S.} model is quite successful in predicting a vast number of observations from z=4 to z=0 (see \cite{cousin15b}), including the stellar mass function, stellar-to-dark-matter halo mass relation, star-formation rate density, stellar mass density, and specific star formation rate.  Also, it reproduces well the stellar mass to gas-phase metallicity relation observed in the local universe and the shape of the average stellar mass to stellar metallicity relations \citep{cousin16}.

\begin{figure}[h]
\centering
\includegraphics[scale=0.4]{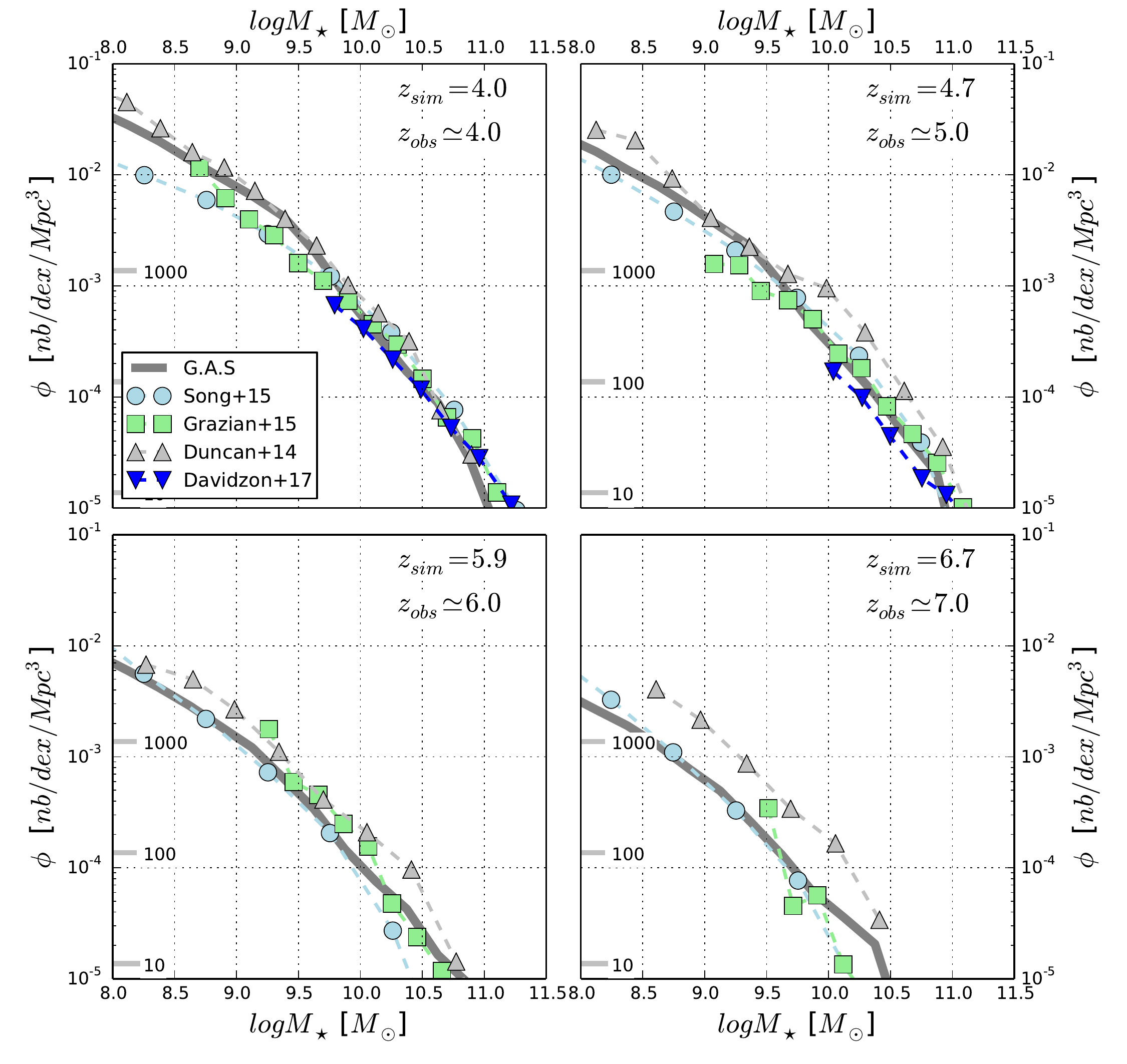}
\caption{\label{fig_SMF} Observed and predicted stellar mass functions from z=4 to z=7. Observations are from \cite{song16, duncan14, grazian15, caputi15}. Our SAM predictions are shown in grey.}         
\end{figure}

In this section, we extend the comparison between the model and observations to $z>4$ to check the model validity at very high redshift $z\sim$4-9. At such high $z$, stellar mass functions and UV-luminosity functions are the only observables that can be used.\\

We first compare the model prediction with the stellar mass functions (SMF). Stellar mass assembly is one of the most fundamental property of galaxy evolution, that does not depend in a SAM on e.g., complex metal-dependent extinction curve. SMF has been measured up to z$\simeq$7, although with a quite large dispersion in the data points at z$\simeq$7. The comparison between model predictions and observations is shown in Fig.\,\ref{fig_SMF}. We have an overall excellent agreement between the two.

\begin{figure}[t]
\centering
\includegraphics[scale=0.5]{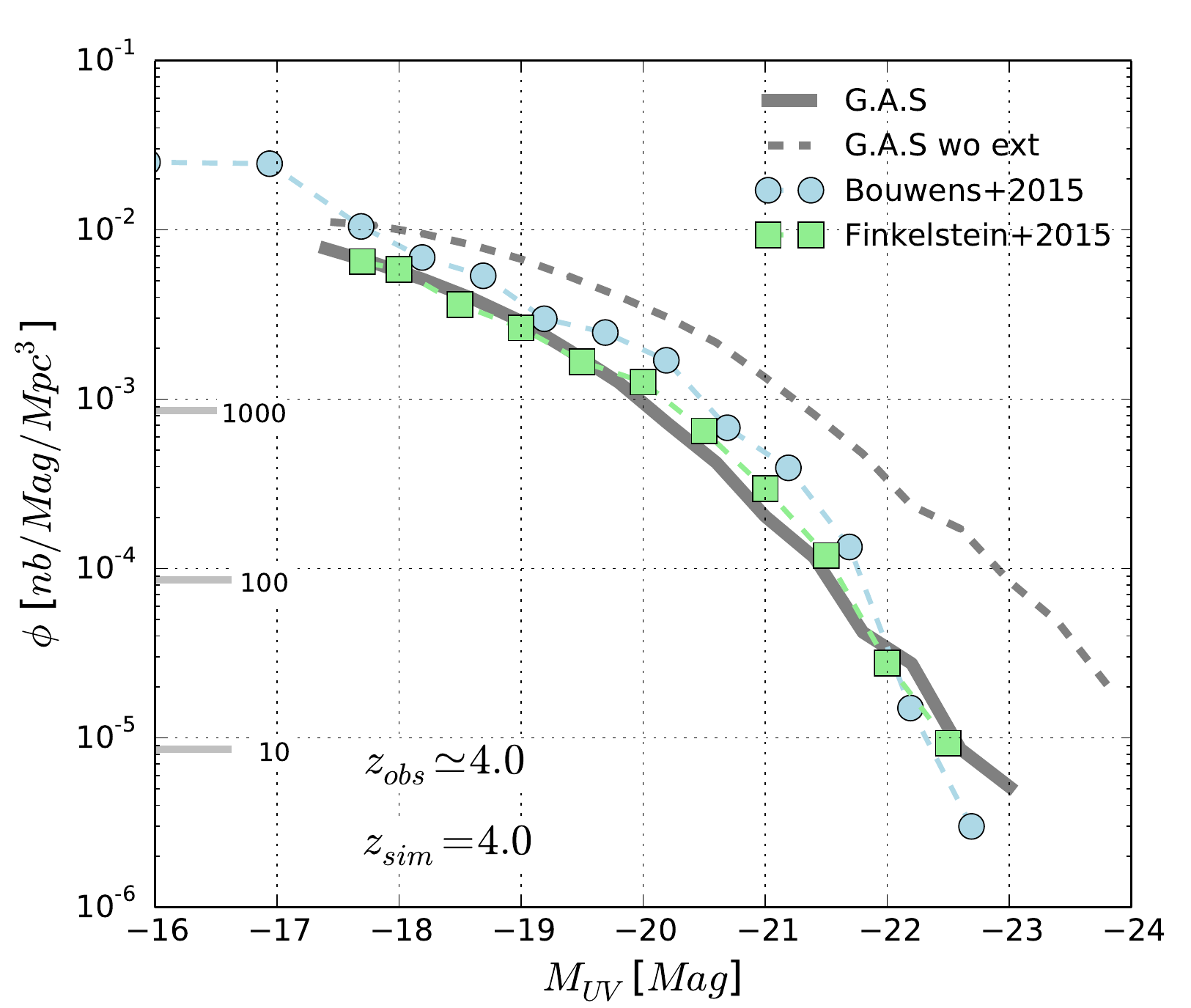}
\caption{ \label{fig_UVLF} Observed and predicted UV luminosity functions at z$\sim$4. 
Observations are from \cite{bouwens15, finkelstein15}. Our SAM predictions are shown in grey (with and without extinction correction, solid and dotted lines, respectively).}
\end{figure}

We show on Fig.\,\ref{fig_UVLF} the model prediction for the UV luminosity function together with the most recent observations at z=4 (the comparison at higher $z$ is similar). The model has no constraint at the faint end due to our mass resolution (our model contains the contribution of all galaxies only for M$_{\star} \ge 10^7$ M$_{\odot}$). At the bright end, we limit the comparison when the number of galaxies in the simulation is $>$5 in the given luminosity bin. Observations are not corrected for extinction so we show the model prediction with and without extinction corrections. We can see that such corrections are important only for $M_{UV}\le$-19.  We have a very good agreement between the model and observations up to $z\simeq6$. At z$\simeq7-8$, our model slightly overestimates the number of $M_{UV}\le$-21 objects. This may be caused by an underestimate of extinction corrections, which are very large for bright-UV galaxies.

These comparisons between model predictions and observations at $z\gtrsim4$ give us confidence in using our SAM as a reference model to predict the [CII]-line emission.

\section{Modeling [CII] emission \label{model_CII}}
[CII] line from high-z galaxies has been computed both through numerical simulations \cite[e.g.,][]{nagamine06, vallini13, vallini15,olsen17}
and semi-analytical models \citep[e.g.,][]{gong12, munoz14, popping16}. Here we take advantage of the excellent agreement of our SAM predictions at z$>$4 with current constraints to revisit the expected [CII] signal from high-$z$ galaxies.

\subsection{Origin of [CII] emission in distant galaxies \label{orig_CII}}
The single [CII] fine structure transition is a very important coolant of the atomic ISM and of PDRs in which carbon is partially or completely in ionized form. Carbon has an ionization potential of 11.3\,eV (compared to 13.6\,eV for hydrogen), implying that line emission can originate from a variety of phases of the ISM:  cold atomic clouds (CNM), diffuse warm neutral and ionized medium (WNM and WIM) and HII regions. Excitation of the [CII] fine structure transition can be via collisions with hydrogen molecules, atoms, and electrons. For example, for the WNM and WIM conditions (Tk = 8000\,K; \citet{wolfire03}) the critical density for excitation of [CII] by H atoms is $\sim$1300\,cm$^{-3}$, and for electrons $\sim$45\,cm$^{-3}$. \\

\begin{figure}
\centering
\includegraphics[scale=0.52]{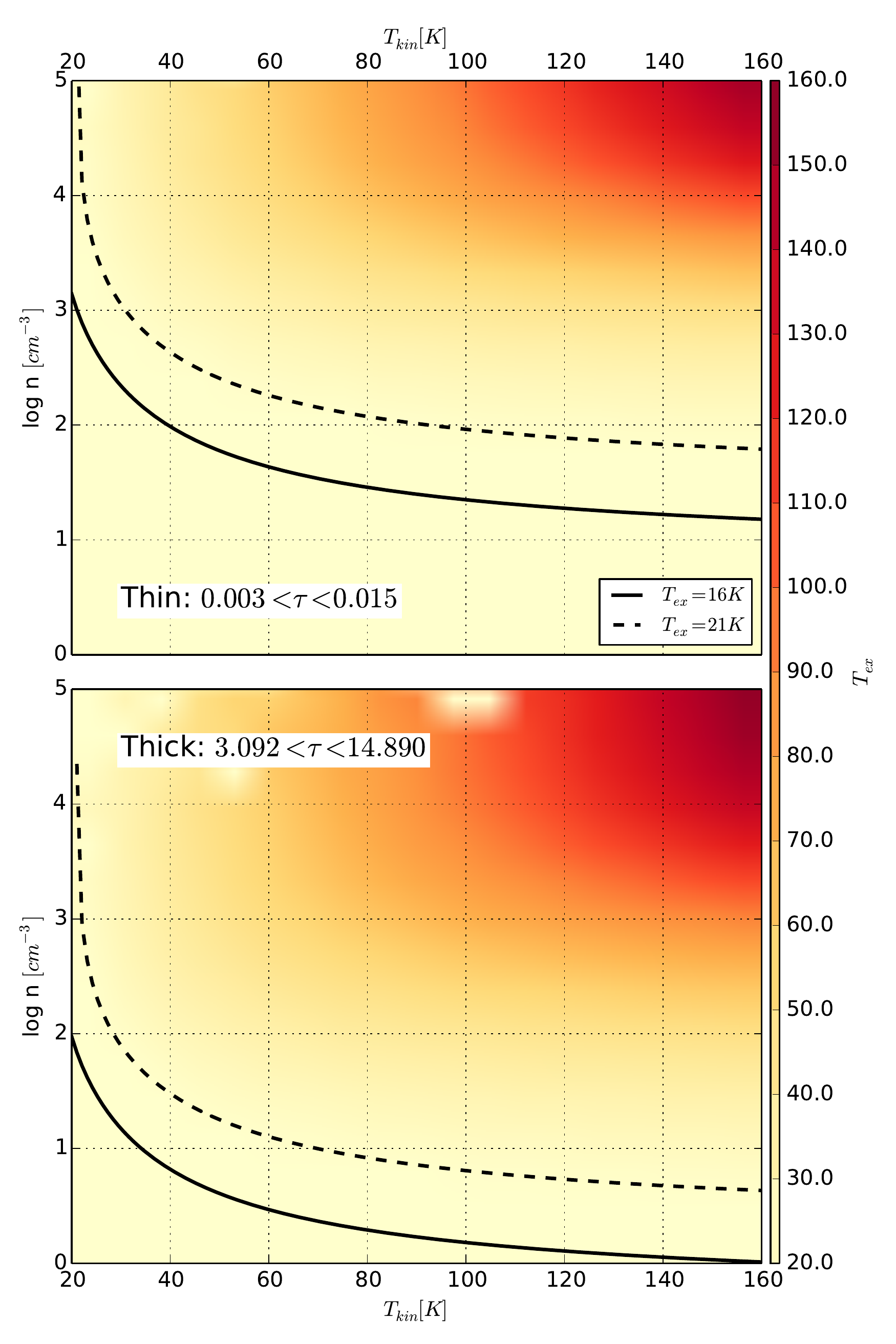}
\caption{\label{fig_Tkin_Tex} [CII] excitation (or spin) temperature, T$^{ex}$, as a function of total gas density $n$ and kinetic temperature on interstellar gas, T${kin}$ (computed for T$^{bg}$=2.726\,K). The upper (lower) panel is dedicated to optically thin (thick) medium. The black solid and dot-dash lines correspond to T$^{ex}$=16\,K and T$^{ex}$=21\,K, i.e. the CMB temperature at z=5.0 and z=6.7, respectively. At these redshifts, [CII] emission is suppressed for kinetic temperature below these lines, due to the CMB. For optically thin emission, this suppression affects mostly the cold neutral medium (T$^{kin} \sim$50-120\,K, $n \sim$20-200\,cm$^{-3}$).}
\end{figure}

Observationally, it is tremendously difficult to separate the contribution of [CII] emission from all different components. Analysis of [CII] observations is also complicated by the fact that it is difficult to determine the optical depth of the line \citep[e.g.,][]{neri14}. 
In the ISM of our Galaxy, because of the density contrast between the CNM and WNM, the [CII] emission associated with the WNM is expected to be a factor of $\sim$20 weaker than that associated with the CNM for a given HI column density \citep{pineda13}.
In the Galactic plane, \cite{pineda14} estimate that 80\% of the [CII] comes from atomic and molecular regions, and 20\% from ionized gas. In local star-forming galaxies, \cite{croxall17} show that 60--80\% of [CII] emission originates from neutral gas. This fraction has a weak dependence on the dust temperature and surface density of star formation, and a stronger dependence on the gas-phase metallicity. For metallicities corresponding to the bulk of our galaxies at high redshift (see Fig.\,\ref{LCII_grid_Cloudy}), the fraction of [CII] emission originating in the neutral phase approaches 90\%.
At higher redshift, in the interacting system BR1202-0725 at z = 4.7, while [CII] emission arises primarily in the neutral gas for the sub-millimeter galaxy and the quasar, [CII] emission seems to be associated with the ionized medium (H II regions) for one Lyman-$\alpha$ emitter of the system \citep{decarli14}. Studying 20 dusty star-forming galaxies from the SPT sample at 2.1$<$z$<$5.7, \cite{gullberg15} found that [CII] emission is consistent with PDRs. Similarly, \cite{stacey10} found that the bulk of the [CII] emission line (70\%) is originating from PDRs in twelve z$\sim$1-2 galaxies.  \\
Theoretically, \cite{olsen15} compute the [CII] emission from cosmological smoothed particle hydrodynamics simulations in seven main sequence galaxies at $z=2$ and found the ionized gas to have a negligible contribution ($<$3\%). Most of [CII] emission ($\gtrsim$70\%) originates from the molecular gas phase in the central $\le$1kpc of their galaxies, whereas the atomic/PDR gas dominates ($>$90\%) further out ($>$ 2 kpc).
In two  zoom-in high-resolution (30 pc) simulations of prototypical M$_{\star} \sim$10$^{10}$\,M$_{\odot}$ galaxies at
z= 6, representative of typical lyman break galaxies at this redshift,  95\% of [C II] emission comes from dense gas located in the H$_{2}$ disk \citep{pallottini17, pallottini17b}. In their simulations of galaxy formation during the epoch of reionisation, \cite{katz17} found that the majority of [CII] mass is associated with cold neutral clumps and that the [CII] emission (although not computed) is likely to originate in cold, neutral gas, or in PDRs close to young stars. \\
Thus, it is reasonable to assume that [CII] at high redshift originates mainly from the CNM and PDRs. \\

However, at the redshifts of interest, one has to consider that the CMB represents a strong background against which the line flux is detected. Indeed, the fraction of the intrinsic line flux observed against the CMB radiation approaches to zero when the excitation temperature ($T^{ex}$) is close to the CMB temperature. One has thus to check in which physical conditions the [CII] line is being attenuated. To get a first hint, we follow \cite{goldsmith12} to compute the [CII] excitation temperature and transpose part of their results to the case of distant unresolved galaxies. The computation is detailed in Appendix\,\ref{CII_ex}. The deexcitation collision rate coefficients (valid for the range of temperature probed here) are extracted from \cite{barinovs05, goldsmith12, wiesenfeld14}:
\be
R_{ul} (e^-) = 8.7\times10^{-8} (T^{kin}/2000)^{-0.37} ~~\mathrm{cm^3s^{-1}} \,.
\ee
\be 
R_{ul}(H^0) = 7.6\times10^{-10} (T^{kin}/100)^{0.14} ~~\mathrm{cm^3s^{-1}} \,.
\ee
\be
R_{ul}(H_2) = (4.9 + 0.22 \times T^{kin}/100)\times 10^{-10}   ~~\mathrm{cm^3s^{-1}} \,,
\ee
\be
R_{ul}(He) = 0.38 \times R_{ul}(H^0) \,.
\ee

Unlike the situation for molecular tracers, for [CII] emission we do not have the possibility of multiple transitions and many isotopologues to allow determination of the volume density, temperature, and optical depth, and thus obtain a reasonably accurate determination of the column density.  
While \cite{gullberg15} found low to moderate [CII] optical depth, $\tau_{\mathrm{[CII]}} \lesssim 1$, in a sample of lensed dusty star forming galaxies covering the redshift range z = 2.1-5.7,  optically thick [CII] emission was proposed by \cite{neri14} for the high-$z$ sub-millimetre source HDF850.1. 
As the situation is not clear, we use a range of $N(C^+)$ and $\delta v$ such as to cover the optically thin and thick case (Eqs.\,\ref{eq_tau} and \ref{tau_0}). We consider $T^{bg}$ =$T_{CMB}(z=0)$ and calculate the excitation (or spin) temperature, T$^{ex}$, of the [C II] transition. We show in Fig.~\ref{fig_Tkin_Tex} the relation between the total gas density $n$, the kinetic temperature T$^{kin}$ and T$^{ex}$ for the optically thin and thick cases. We also show the curve corresponding to the CMB temperature at z=5 and z=6.7. In any case, [CII] emission from the warm ($\simeq$10$^4$\,K), low density ($\lesssim$ 0.1\,cm$^{-3}$) component of the ISM is suppressed at high redshift by the CMB.
For the optically thin case, [CII] emission from gas density $n \lesssim 100$ cm$^{-3}$ (the CNM) will be mostly completely attenuated for z$>$6.5 (as also noticed by \cite{vallini15}), the CMB effect becoming negligible only for galaxies at z$\le$4.5. In that case, only [CII] from PDRs will reach the instrument. In the optically thick case, only the very cold, low density neutral medium will be affected by CMB attenuation.\\
Based on these arguments and observational and theoretical constraints detailed above, we can assume that [CII] emission in high-$z$ galaxies arises predominantly from PDRs.

We caution the readers that the above simple calculations do not account for the temperature structure of the clouds. Indeed, as shown in the next section (Sect.\,\ref{CMB_eff}), large CMB attenuation is seen at higher cloud depths, where the temperatures are below $\sim$100\,K.  At the deepest point of the cloud, where the temperature is only $\sim$50\,K, radiative and collisional excitation rates are comparable, but deexcitations are primarily spontaneous.  Thus the A$_{ul}$ term dominates the other terms by a few dex in Eq.\,\ref{rate}. \\

PDRs are well-studied structures with intense characteristic emissions. Theoretical models addressing the structure of PDRs have been available for approximately 40 years \citep{hollenbach71,jura74,glassgold75,black77}. The PDR gas mass fraction in star forming galaxies ranges from a few percent for quiescent systems like the Milky Way up to more than 50\% for starburst galaxies like M82 making PDRs important on galactic scales \citep[e.g.,][]{stacey91, malhotra01}. 

Consequently, we assume that the [CII]-line emission originates from PDRs and use {\tt CLOUDY} to compute its luminosity. Similarly, \cite{popping16} only account for the contribution by PDRs to the [CII] luminosity of galaxies when coupling their semi-analytic model of galaxy formation with a radiative transfer code.

\subsection{CMB effects on [CII] emission \label{CMB_eff}}
\begin{figure}
   \centering
   \includegraphics[scale=0.65]{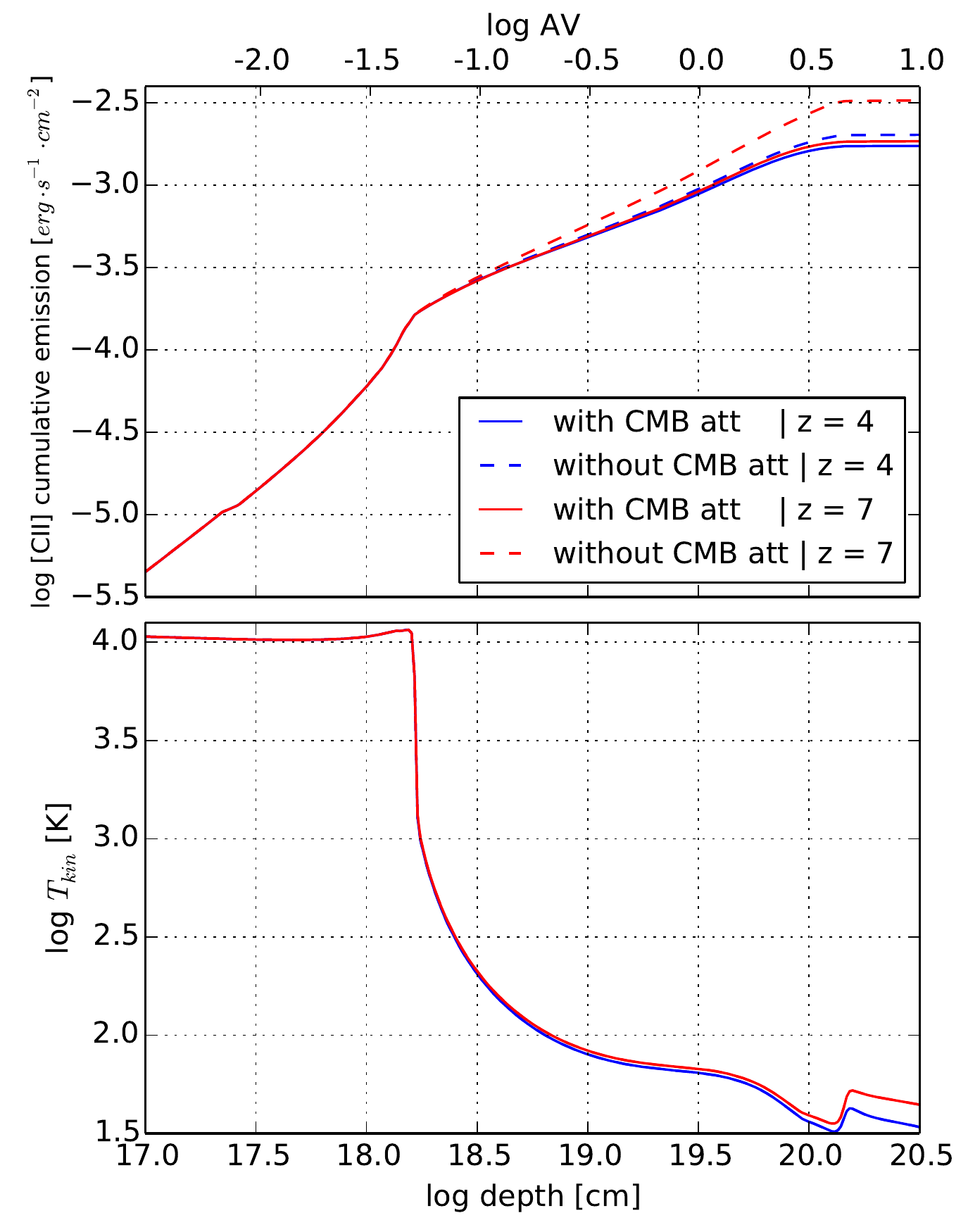}
  \caption{Example [CII] emission (top) and kinetic temperature (bottom) computed by {\tt CLOUDY}. Two redshifts are shown (blue and red lines). On the top figure one can see the effect of CMB attenuation on the [CII] line luminosity. In this example, the PDR is homogeneous with $\log{n_H} = 2.4$, and exposed to a radiation field with $\log{\mathrm{ISRF}} = 3.5$, in addition to the CMB radiation. \label{Profile_pdr}}
 \end{figure}

The structure of a PDR is well established \citep{HollenbachTielens1999}. The outermost layer, which is exposed to the ambient radiation is ionized, and its thickness determined by the ionization parameter. This is followed by a neutral layer, and yet deeper lies a molecular layer.
[CII] is present at varying degrees across the PDR. In Fig.~\ref{Profile_pdr} we show the structure of a PDR of modest density ($\log{n_H} = 2.4$), exposed to an intense interstellar radiation field, ISRF (ISRF=3.2$\times$10$^3\,G_0$, where $G_0$ is the Habing Field, see Sect. \ref{sect:eq_PDR_mod}).
These conditions are close to those predicted in high-$z$ galaxies (see Fig.\,\ref{LCII_grid_Cloudy}). In this example, the ionized ``skin'' of the cloud has a thickness of $\sim$0.1\% of the total, and is at a temperature of 10,000\,K. The temperature drops sharply in the neutral layer of the cloud, to below 100\,K, and the [CII] 157\,$\mu$m emissivity is reduced. This is illustrated by the shallower slope between 10$^{18.5}$ and 10$^{20}$~cm in the top panel of the figure, which presents the emergent intensity in the line (integrating inward). The ionization fraction of [CII] (which is the singly ionized carbon to total carbon mass fraction) drops to below 1\% in the molecular core of the cloud, and the intensity flattens at these depths.\\

Apart from the ISRF, the [CII] emission is influenced by the CMB at high redshift, which peaks at $\sim$130--210~$\mu$m at redshifts 4--7. The photon occupation number (Eq.~\ref{eq:photon-occ}) of the CMB dominates the corresponding number in the ISRF for intensities less
than $\sim$10$^3$--10$^4$~$G_0$. In our {\tt CLOUDY} models, both fields are isotropic and subject to removal when corrected line intensities are computed.

On account of the high temperature, the level populations of the transition in the ionized layer are set primarily by collisions,
leading to a very small correction for isotropic radiation. The insignificance of radiation in this layer is illustrated by the coincidence of the line intensity at $z=4$ and 7 as a function of depth, shown in the top panel of Fig.~\ref{Profile_pdr}.
As the temperature drops in the neutral zone, radiative pumping becomes more important, as does the correction for isotropic radiation. The correction is more important at $z=7$, because of the higher photon occupation number. The net effect is that the emergent line intensity corrected for isotropic radiation does not vary significantly between $z=4$ and 7. 

\subsection{[CII] emission from photon dominated regions}
\label{PDR_eq_model_SECT}
\subsubsection{The equivalent PDR model \label{sect:eq_PDR_mod}}
For each galaxy in our SAM, we need to define an equivalent PDR characterised by three parameters: the mean hydrogen density ($n_{H}$), gas metallicity ($Z_g$) and interstellar radiation field (ISRF).\\
    
The mean Hydrogen density ($n_{H}$) is computed from the mean hydrogen surface density ($\Sigma_{H}$) and disc scale height ($h_d$) following
$n_{H} = \frac{\Sigma_H}{h_{pdr}} \mbox{with } h_{pdr} = h_d/5$. 
The average disc scale height ($h_d$) is derived at half-mass radius (which is 1.68$r_d$ where $r_d$ is the exponential disc radius) by assuming a vertical equilibrium in the disc between the gas and stars, and an homogeneous gas velocity dispersion depending on physical properties of galaxies, which is $\sim$15-25~km/sec \citep{cousin17a}. In our simulated galaxies, $h_d\sim$200\,pc at $z=5$. Hydrogen densities for the PDRs are computed using a height 5 times smaller, to take into account the fact that the gas is more concentrated to the equatorial plane. Taking h$_d/10$ or  h$_d/2.5$, rather than h$_d/5$, modifies the [CII] luminosities by less than 0.1\,dex. The distribution of the PDR scale height $h_{pdr}$ is shown in Fig.\,\ref{rpdr_hpdr}.

$\Sigma_{H}$ is computed inside a critical radius $r_c$ which sets the limit outside of which the gas is not dense enough to form stars (we consider $\Sigma_{H}(r>r_c)<$50\,$M_{\odot}$ pc$^2$). By assuming an exponential gaseous disc, the average hydrogen surface density is given by:
\begin{equation}
\Sigma_{H} = \frac{M_H}{\pi r_c^2}\left[1-exp(-x_c)(1+x_c)\right]~\mbox{with}~x_c = \frac{r_c}{r_d}
\label{Sigma_H}
\end{equation}
$r_d$ and $M_H$ are the exponential disc radius and mass of hydrogen in the disc, respectively. \\

Our equivalent PDR model depends also on gas metallicity. Oxygen is the most abundant element formed in stars. It is therefore commonly used as a tracer of the gas-phase metallicity. We define the gas metallicity $Z_g$ as the number of oxygen atoms to hydrogen atoms with a logarithmic scale: $Z_g$ = 12 + log$(O/H)$. We adopt $\Zsun = 8.94$ \citep{karakas2010}. Correspondance between $Z_g$ and metals mass fraction is given in Table\,\ref{Met_abundance}. We assume that the gas is homogeneously distributed
in a given galaxy and consequently in a given equivalent PDR. We thus consider average metallicities.\\

Finally we need to compute the ISRF produced by young stars. It is is defined as the flux of stellar radiation integrated between 6 and 13.6\,eV. We assume a mean distance between gas and stars of $D = 50$\,pc. The exact choice of this mean distance has a very small impact on the predicted L$_{\mathrm{[CII]}}$. As commonly used in the literature, the ISRF is normalised in units of the Habing Field \citep{habing68}, $G_0$ = 1, which corresponds to $f_0$ = 1.6$\times$10$^{-3}$ erg  s$^{-1}$ cm$^{-2}$.

\subsubsection{{\tt CLOUDY}: model grids, parameters and outputs}
 
 \begin{table*}[ht!]
  \begin{center}
    \footnotesize{
      \begin{tabular}{|c|c|c|c|c|c|c|c|}
        \hline
        12 + log$(O/H)$ & Metallicity & $m_Z/m_H $ & Helium & Carbon & Nitrogen & Oxygen & Iron\\
        & in unit of Z$_{\odot}$ & & & & & & \\
        \hline
           $6.2$ & $3.79\times10^{-3}$ & $9.22\times10^{-5}$ & $5.767\times10^{-2}$ & $9.192\times10^{-7}$ & $3.038\times10^{-6}$ & $1.107\times10^{-6}$ & $1.383\times10^{-7}$\\
           $6.6$ & $6.70\times10^{-3}$ & $1.63\times10^{-4}$ &  $5.770\times10^{-2}$ & $1.932\times10^{-6}$ & $3.493\times10^{-6}$ & $3.301\times10^{-6}$ & $2.361\times10^{-7}$\\
           $7.0$ & $1.23\times10^{-2}$ & $2.99\times10^{-4}$ & $5.773\times10^{-2}$ & $3.920\times10^{-6}$ & $3.477\times10^{-6}$ & $8.295\times10^{-6}$ & $3.796\times10^{-7}$\\
           $7.4$ & $2.65\times10^{-2}$ & $6.43\times10^{-4}$ & $5.786\times10^{-2}$ & $9.181\times10^{-6}$ & $4.304\times10^{-6}$ & $1.992\times10^{-6}$ & $7.577\times10^{-7}$\\
           $7.8$ & $6.38\times10^{-2}$ & $1.55\times10^{-3}$ & $5.823\times10^{-2}$ & $2.262\times10^{-6}$ & $7.561\times10^{-6}$ & $4.880\times10^{-6}$ & $1.765\times10^{-6}$\\
           $8.2$ & $1.60\times10^{-1}$ & $3.89\times10^{-3}$ & $5.912\times10^{-2}$ & $5.772\times10^{-6}$ & $1.749\times10^{-6}$ & $1.119\times10^{-4}$ & $4.898\times10^{-6}$\\
           $8.6$ & $3.80\times10^{-1}$ & $9.23\times10^{-3}$ & $6.101\times10^{-2}$ & $1.499\times10^{-4}$ & $5.246\times10^{-6}$ & $2.505\times10^{-4}$ & $1.037\times10^{-6}$\\
           $9.0$ & $1$ & $2.43\times10^{-2}$ & $7.006\times10^{-2}$ & $4.800\times10^{-4}$ & $2.318\times10^{-4}$ & $5.977\times10^{-4}$ & $2.755\times10^{-6}$\\
           $9.4$ & $2.35$& $5.71\times10^{-2}$ & $8.910\times10^{-2}$ & $1.245\times10^{-3}$ & $6.048\times10^{-4}$ & $1.397\times10^{-4}$ & $5.373\times10^{-6}$\\
           $9.8$ & 4.44& $1.08\times10^{-1}$ & $1.057\times10^{-2}$ & $2.585\times10^{-4}$ & $1.098\times10^{-4}$ & $2.741\times10^{-4}$ & $7.618\times10^{-6}$\\
        \hline
      \end{tabular}}
  \end{center}
  \caption{\label{Met_abundance} The ten metallicities considered in our grids of models and the abundances of the five main elements of the ISM we are including in {\tt CLOUDY} (first column: average metallicity values, second column: equivalent solar metallicity fraction, third column: relative metal mass abundances to hydrogen mass and last five columns: abundances (number of element atoms to hydrogen atoms).}
\end{table*} 

\begin{figure}[h]
  \centering
\includegraphics[scale=0.6]{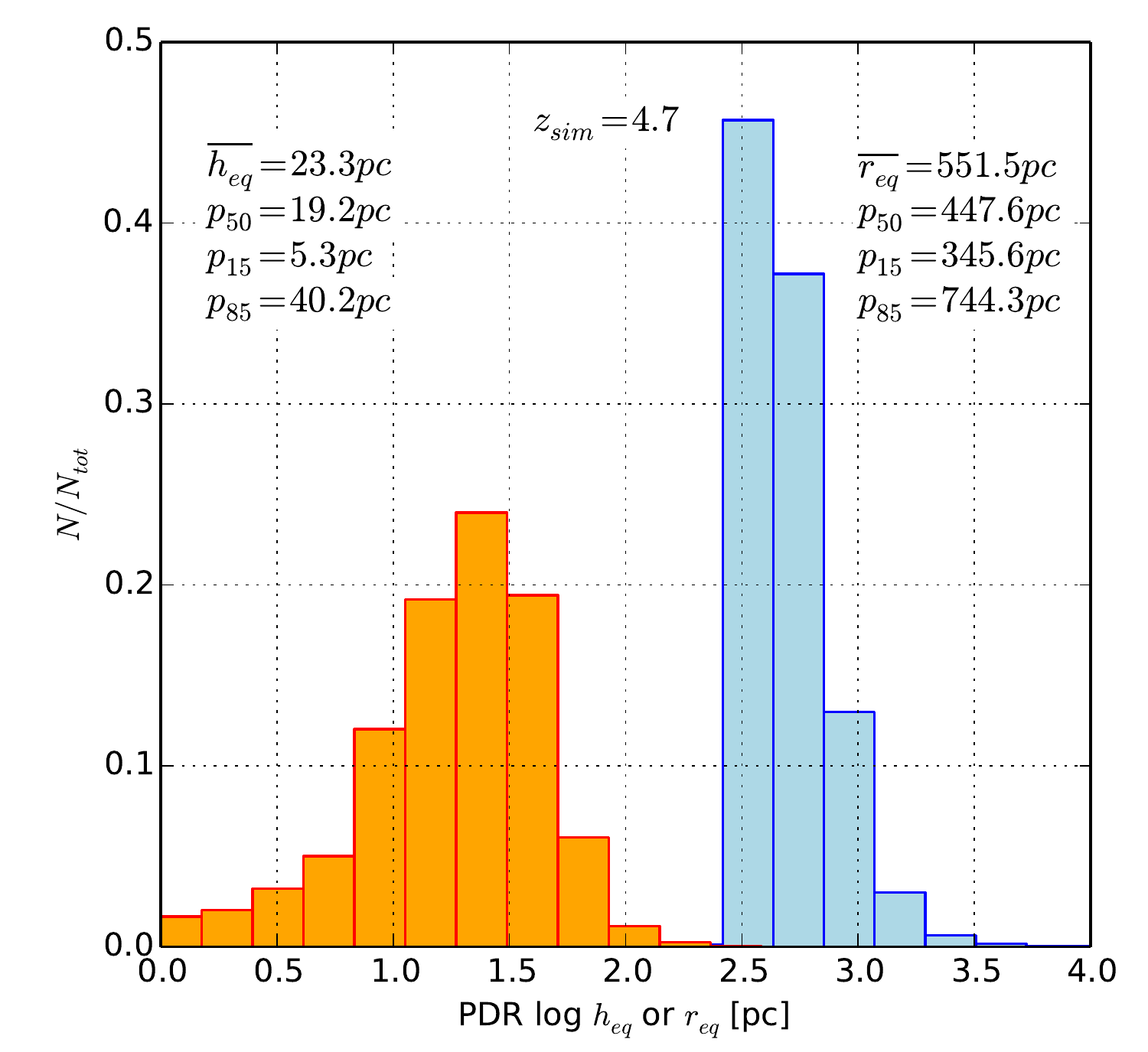}
 \caption{Physical sizes of effective PDRs in our simulated galaxies at z=4.7. Orange and blue histograms are associated to equivalent PDR scale-height ($h_{eq}$) and equivalent PDR radius ($r_{eq}$), respectively.}
 \label{rpdr_hpdr}
\end{figure}

\begin{figure*}[ht!]
  \flushleft
  \includegraphics[scale=0.46]{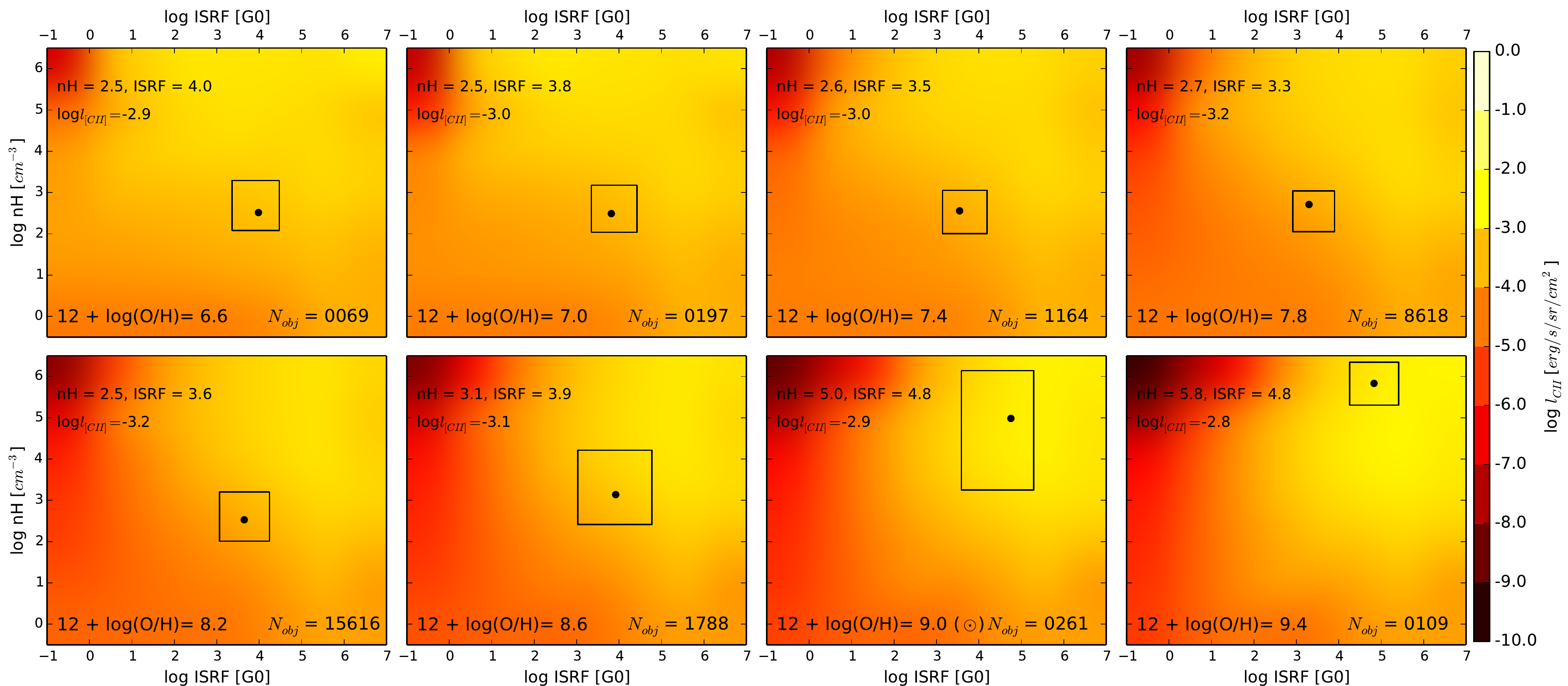}
  \caption{[CII] luminosities predicted by {\tt CLOUDY} (for a CMB temperature at z=5). Each panel is dedicated to a given metallicity bin and shows the [CII] luminosity per unit area as a function of both ISRF and hydrogen density. In each panel we show the location of {\tt G.A.S.} galaxies (extracted from the snapshot at z=4.7). Rectangles represent the 0.15 and 0.85 quantiles of the galaxy distribution. The black point marks the median, with its coordinates given in the top left corners. We also give the number of galaxies $N_{obj}$ present in the considered metallicity bin.}
  \label{LCII_grid_Cloudy}
\end{figure*}

Predictions of [CII] emission are computed with the plasma simulation code {\tt CLOUDY} \citep{ferland13}. We use the C17 version of the code \citep{C17} as it incorporates a diminution factor due to an external isotropic radiation field (both the CMB and the ISRF, in our case) in its line intensity estimates. This factor was derived as an extended radiative transfer theorem \citep{chatzikos13}, and applied to predictions for hyperfine structure line intensities in \cite{chatzikos14}.\\

We have built grids of models based on 560 distinct model parameters. A given grid is divided in ten bins of metallicity $Z_g$. In each metallicity bins, ISRF and hydrogen density are sampled following log$(ISRF)$=[-0.5,0.5,1.5,2.5,3.5,4.5,5.5,6.5] and log($n_H$)=[0,1,2,3,4,5,6]. 

In a given metallicity bin, abundances of the five main ISM elements are fixed (helium, carbon, nitrogen, oxygen and iron, see Table \ref{Met_abundance}). They are fixed to the median abundances of all simulated galaxies that have gas metallicity in the given bin. Abundances of all other elements are set to zero (we checked that this had no impact on the [CII] luminosity). A galaxy, in a given metallicity bin, will have the median metallicity of the bin, but individual values for hydrogen density and ISRF. The {\tt CLOUDY} grids are interpolated to find the corresponding [CII] luminosity and the carbon fraction f$_{\mathrm{[CII]}}$.

PDRs are modelled assuming a plan-parallel geometry. The shape of the ISRF is that of \cite{black87}, as given through the {\tt CLOUDY} option \textquotedblleft Table ISM". Cosmic rays background is fixed to the fiducial value of $2\times 10^{-16}$ s$^{-1}$. 
For a given PDR model the computation is stopped at $A_{V}$ = 10 and gas temperature can decrease until  T=10\,K. 

We have generated five grids of models using five background temperatures (T$_{bg}$) corresponding to CMB temperatures at z=4, 5, 6, and 7. We have also computed the grids for both intrinsic and emergent [CII] luminosities. No significant differences have been observed between these two luminosities. We can therefore assumed that [CII] emission is weakly affected by extinction in our galaxies. 

For each model associated to a given set of parameters ($ISRF-n_{H}-Z_g$) we extract from {\tt CLOUDY} the [CII] luminosity per unit area, $l_{\mathrm{[CII]}}$ (in L$_{\odot}$\,cm$^{-2}$\,sr$^{-1}$), and the [CII] column density, $N_{\mathrm{[CII]}}$.  We then compute the carbon fraction (in number) in [CII], $f_{\mathrm{[CII]}}$, by computing the ratio between the [CII] column density and the sum of the column density of all species containing carbon atom(s).
From the [CII] luminosity and column density ($l_{\mathrm{[CII]}}$, $N_{\mathrm{[CII]}}$) we then define the equivalent surface of the PDR as:
\begin{equation}
S_{PDR} = \frac{M_c}{m_c} \times \frac{f_{\mathrm{[CII]}}}{N_{\mathrm{[CII]}}}
\label{eq_S_PDR}
\end{equation}
where $M_c$ is the carbon mass in the galaxy and $m_c$ is the mass of individual carbon atom. This formulation implies that we have an uniform [CII] column density in the PDR.\\ 
Combined with the [CII] luminosity per unit area ($l_{\mathrm{[CII]}}$), this PDR equivalent surface leads to the following [CII] luminosity,
\begin{equation}
L_{\mathrm{[CII]}}= 4\pi \times S_{PDR} \times l_{\mathrm{[CII]}}\,.
\end{equation}

\begin{figure*}
  \centering
  \includegraphics[scale=0.4]{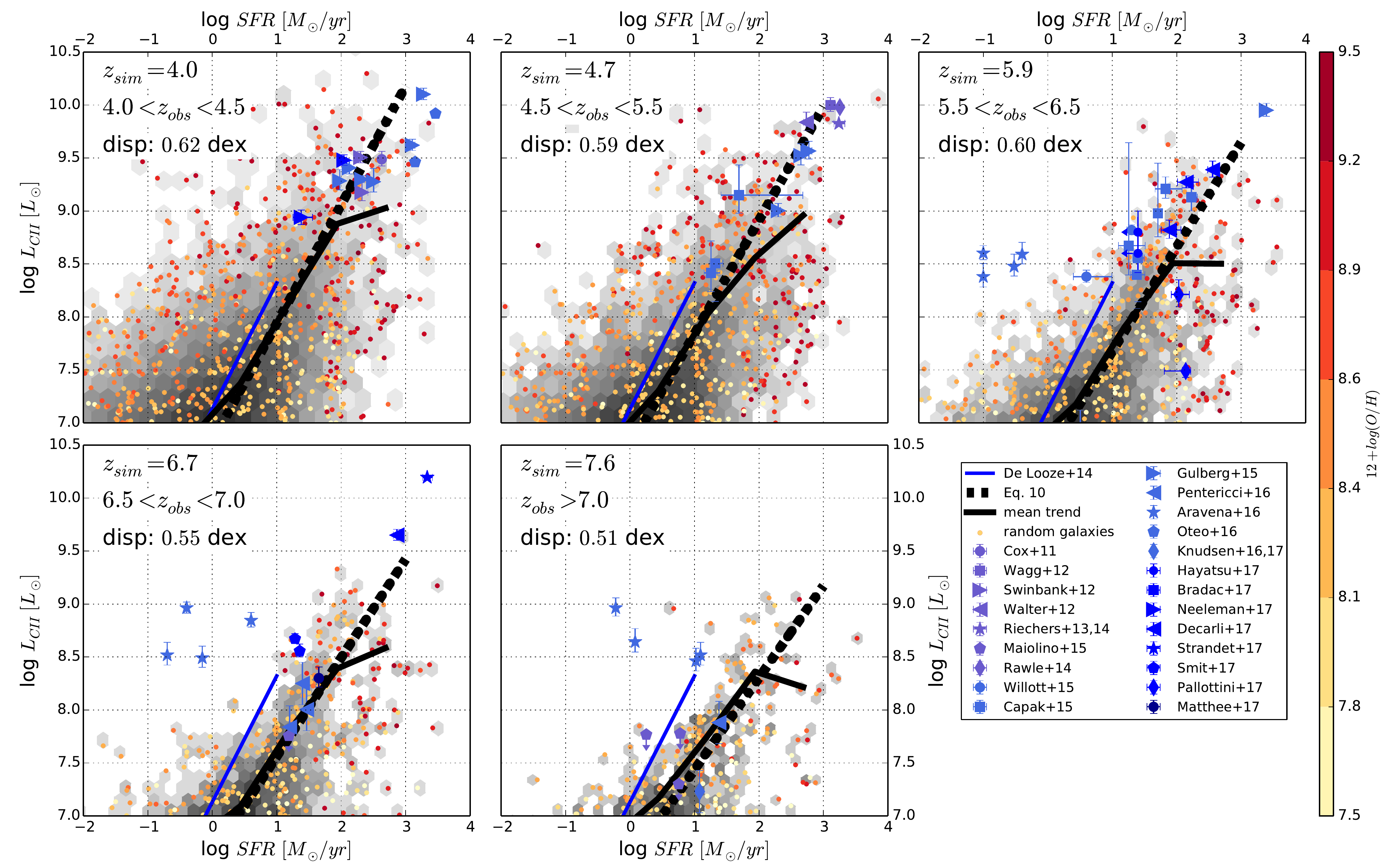}
  \caption{L$_{\mathrm{[CII]}}$--SFR relation. Predictions from our model are shown for a set of redshifts from z = 4 to z = 7.6.  In each panel the whole sample of {\tt G.A.S.} galaxies is shown in grey scale. The average relation is plotted with a solid black line. The black dashed line shows the relation given  in Eq.\,\ref{SFR_CII_eq}. Yellow to red coloured points mark the gas metallicity of a randomly selected sample of simulated galaxies (note that the observed tendency of high-metallicity galaxies to fall either above or below the mean trend, i.e. making an \textquotedblleft envelop", is only a trick of the eye caused by the plotting; galaxies with high metallicities ($Z_g>8.8$) are spread over the whole area, with a higher density of objects at high SFR). Our predictions are compared to a large sample of observational data that are detailed in Table\,\ref{cii_lfir}. Amplification corrections on luminosity and SFR, when available, are applied.  For dusty star forming galaxies, SFR are converted directly from L$_{IR}$ using the \cite{kennicutt98} conversion factor assuming a \cite{chabrier2003} IMF 
where SFR (M$_{\odot}$\,yr$^{-1}$)= $1.0 \times 10^{-10}\,\mathrm{L_{IR}(L_{\odot})}$.
 The blue solid line shows the \cite{delooze14} relation for the local dwarf galaxy sample. \label{SFR_CII}}
\end{figure*}

We show on Fig\,\ref{rpdr_hpdr} the distribution of PDR sizes in our simulated galaxies, which is computed using $S_{PDR} = \pi \times r_{eq}^2$. 
The median size is around 450\,pc. Sizes range from 345 to 745\,pc for 70\% of the objects. These sizes are in line with the PDR region of M82 (ranging from 300\,pc for \citet{joy87} to 600\,pc \citet{carlstrom91}). They are in general smaller by a factor $\sim$2 than the estimated sizes of the lensed DSFGs found by SPT and covering the redshift range z=2.1-5.7 \cite[see Table\,2 of][]{gullberg15}. This is not surprising as those SPT DSFGs have de-magnified far-IR luminosities (42$<\lambda<$500\,$\mu$m) of L$_{FIR}\sim5 \times10^{12} L_{\odot}$, thus mean SFR much larger than the average population (see Fig\,\ref{SFR_CII}). 

\begin{figure*}
  \centering
  \includegraphics[scale=0.45]{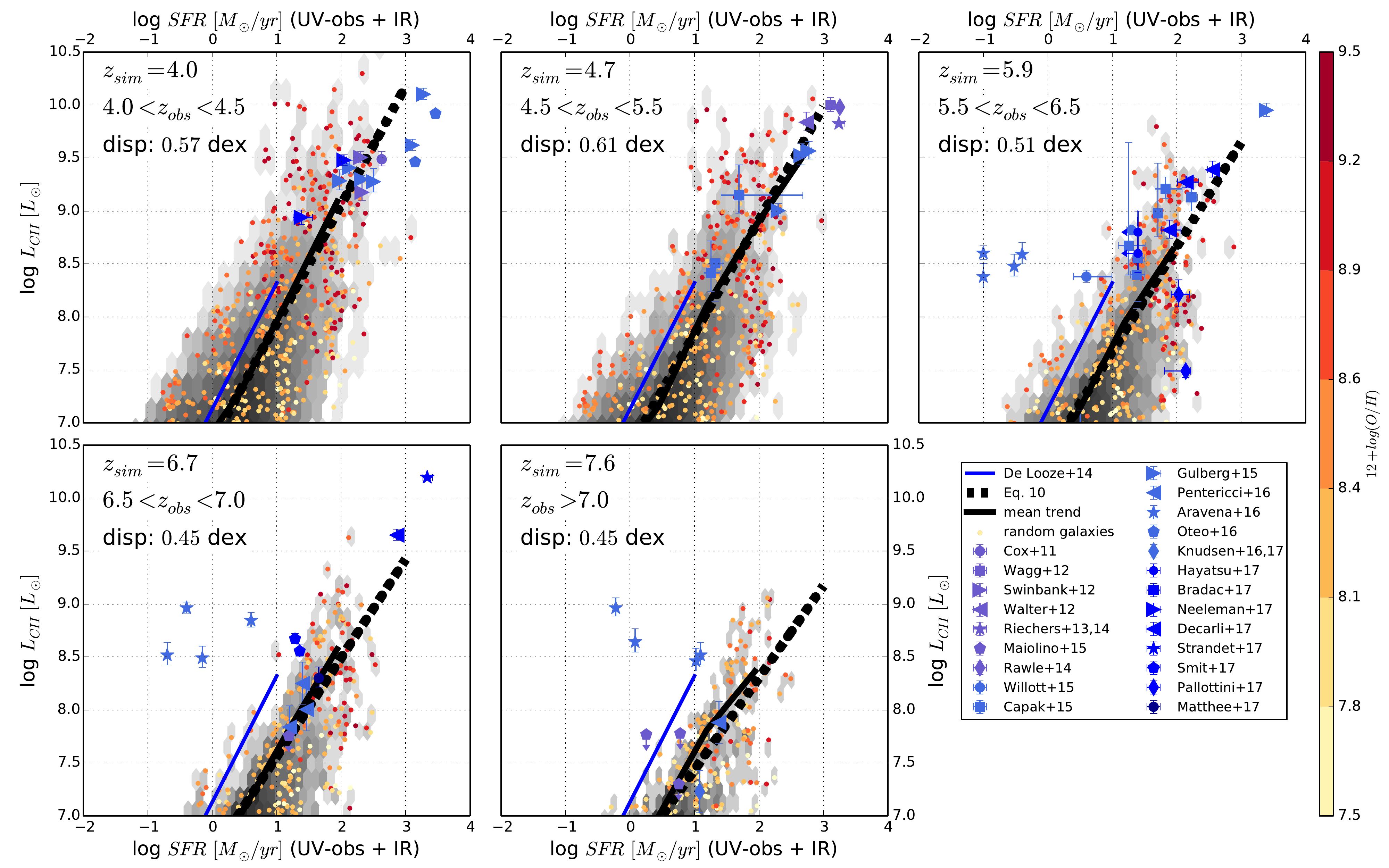}
  \caption{Same as Fig.\,\ref{SFR_CII} but with the SFR determined from the UV (observed, so attenuated) and IR emission, following \cite{kennicutt98} standard conversions between L$_{UV}$ and L$_{IR}$ and SFR. This mimics what is being doing when computing the SFR from the UV and IR emission of galaxies. Taking the \textquotedblleft observed" SFR (this figure) rather than the \textquotedblleft true" SFR from the model (Fig.\,\ref{SFR_CII}) decreases the dispersion and removes a large fraction of the outliers.
  \label{ObsSFR_CII}}
\end{figure*}

\subsubsection{Effect of metallicity, ISRF and density on [CII] emission}

We show on Fig.\,\ref{LCII_grid_Cloudy} the [CII] luminosity variation as a function of ISRF, hydrogen density and gas metallicity. The grids of models are shown for a background temperature corresponding to the CMB at z=5.
We also show the location of our simulated galaxies extracted from the SAM at z=4.7. The majority is found in regions where ISRF and hydrogen densities are in the ranges (log\,$ISRF$, log\,$n_H$) = ([3.0, 4.5], [2.0, 3.5]). These ranges of ISRF and hydrogen densities are mainly associated with low gas metallicities ($>$98\% of galaxies with $Z_g \le Z_{\odot}$). In the small fraction of galaxies ($\sim$2\%) with higher gas metallicities ($Z_g > Z_{\odot}$), the physical conditions are different: both the radiation field and hydrogen density are higher, with (log\,$ISRF$, log\,$n_H$) = ([3.5, 5.5], [3.0, 6.0]). 
These galaxies have discs that are smaller (r$_\mathrm{d}$=0.1$\pm$0.2\,kpc versus r$_\mathrm{d}=$0.4$\pm$0.2\,kpc) and flatter (h$_\mathrm{d}$=0.01$\pm$0.04\,kpc versus h$_\mathrm{d}$=0.12$\pm$0.10\,kpc).
Smaller sizes and scale heights lead to a higher gas density. The star formation is thus higher than in the other discs of the sample (SFR=156$\pm$441\,M$_{\odot}$\,yr$^{-1}$ versus SFR=5$\pm$56\,M$_{\odot}$\,yr$^{-1}$). The strong ISRF associated to the high-metallicity galaxies is therefore explained by the high star formation activity. We finally note that the high gas metallicity of these galaxies is also associated to a high stellar metallicity (Z$_{\star}$=0.79$\pm$0.48\,Z$_{\odot}$ versus Z$_{\star}$=0.05$\pm$0.12\,Z$_{\odot}$).

Comparing the SPT data with PDR model tracks from \cite{kaufman99}, \cite{gullberg15} obtained a rough estimate of the radiation field strength and gas density of $10^{3}<G_0<10^{4}$ and $10^2<$n$<10^5$\,cm$^{-3}$ for the $z>4$ objects, which is in line with our model.

\section{L$_{\mathrm{[CII]}}$ -- SFR relation \label{sect_SFR_LCII}}

The [CII] transition has great potential as a star formation rate tracer at high redshift. In this section we examine the correlation between L$_{\mathrm{[CII]}}$ and SFR at  $4<z<8$ obtained from our model.

\cite{delooze14} analyze the applicability of the [CII] line to reliably trace the SFR in a sample of low-metallicity dwarf galaxies from the Herschel Dwarf Galaxy Survey and, furthermore, extend the analysis to a broad sample of galaxies of various types and metallicities in the literature (see also \citet{herrera15, Sargasyan14}). They found that the L$_{\mathrm{[CII]}}$--SFR relation has a quite high dispersion, with 1$\sigma$=0.38\,dex. Including all the samples from the literature (ex: AGNs, ULIRGS, high-$z$ galaxies) the dispersion increases to 0.42\,dex. The scatter in the L$_{\mathrm{[CII]}}$--SFR relation increases toward low metallicities, warm dust temperatures, and large filling factors of diffuse, highly ionized gas.
At high redshift ($z\simeq7$) and using numerical simulations, \cite{vallini15} find that the L$_{\mathrm{[CII]}}$--SFR relation holds (and is consistent
with observations of local dwarf galaxies), with eventual displacements due to extremely low metallicities or a modified Kennicutt-Schmidt relation. The results from their models (obtained assuming a constant metallicity and $\Sigma_{\rm SFR}\propto \Sigma_{\rm H_2}$) are well described by the following best-fitting formula:
\begin{eqnarray}
{\rm log}{L_{\mathrm{[CII]}}}=7.0+1.2~{\log} (\mathrm{SFR}) +0.021~{\log} (Z_g) +\nonumber\\
0.012~{\log (\mathrm{SFR})\log (Z_g)}-0.74~{\log ^2(Z_g)},
\label{eq_vallini}
\end{eqnarray} 
where $\rm L_{\mathrm{[CII]}}$ is expressed in solar units, and the SFR in $\rm M_{\odot}\,yr^{-1}$.\\

We show on Fig.\,\ref{SFR_CII} the L$_{\mathrm{[CII]}}$--SFR relation derived from the coupling between  {\tt G.A.S.} and {\tt CLOUDY}.
The predictions are compared to a large set of observational measurements mostly based on UV- or submillimeter-selected galaxies where SFRs are either derived from UV flux, deduced from SED-fitting analysis, or computed from L$_{\mathrm{IR}}$. All observational data are compiled in Table\,\ref{cii_lfir}\footnote{We did not add the upper limits coming from the [CII] search in bright Lyman-alpha emitters (LAEs), but the locations of such galaxies in the L$_{\mathrm{[CII]}}$--SFR plot are not unexpected. For example, at the SFR of the three LAEs of \cite{gonzalez14}, most galaxies in our simulation have much fainter L$_{\mathrm{[CII]}}$ than their reported upper limits. And the galaxies that lie above the upper limits are those with high metallicities (thus they are not dust-poor, bright LAEs).}. In Fig. \ref{SFR_CII}, we also compare our predictions with the local L$_{\mathrm{[CII]}}$--SFR relation measured by \cite{delooze14} for local dwarf galaxies. Most of these galaxies have metallicities comparable with the bulk of our simulated galaxies at high redshift. \\

As expected, at each redshift, there is a relation between SFR and the [CII] luminosity, albeit with a very large scatter (0.62\,dex at z =4.0 and 0.51\,dex at z=7.6). To investigate the origin of the scatter, and which one of the three parameters (ISRF, n$_H$ or Z$_g$) contributes the most, we compute for each galaxy its [CII] luminosity, fixing two parameters to their median value while keeping the third one to its original value. We find that ISRF, n$_H$, Z$_g$ contribute roughly equally to the scatter.

As shown on Fig.\,\ref{SFR_CII}, predictions are in remarkably good agreement with the majority of observational data points. A source of dispersion in the observed L$_{\mathrm{[CII]}}$--SFR relation may come from the fact that the [CII] emission may not overlap with the bulk of UV emission that is used to determine the SFR \citep{maiolino15, carniani17}. Nevertheless, the current observations seem to be less scattered than the model. This may be explained by the different timescales that are implicitly assumed when measuring the SFR in galaxies. Our SAM is based on the progressive structuring of the diffuse accreted gas, following an inertial cascade which depends on the fraction of gas already \textquotedblleft structured" in the galaxies (i.e., the gas mass fraction in giant star-forming clouds). This leads to brief and intense star-forming episodes, separated by phases in which the gas starts again its cascade structure \citep{cousin17a}. Most of these star-forming episodes occurs on timescales shorter than those assumed when converting UV and far-IR luminosities in SFR using \cite{kennicutt98} relations. To investigate the impact on timescales used to determine the SFR, we compute the SFR of galaxies in our model using the UV and IR luminosities, following what is being done from the observations (SFR=SFR$_{UV-obs}$ + SFR$_{IR}$ using \cite{kennicutt98} relations to convert luminosities into SFR). We show on Fig.\,\ref{ObsSFR_CII} the L$_{\mathrm{[CII]}}$--SFR relation using this \textquotedblleft observed\textquotedblright \,\, SFR. We see that the scatter in the relation decreases ; we also see that using the \textquotedblleft observed\textquotedblright \,\, SFR removes a large fraction of outliers. This clearly illustrates the importance of timescales when using instantaneous quantities as SFR, and shows that using an average conversion that is the same for all galaxies smooth the variations.

Two kinds of sample are not falling into our L$_{\mathrm{[CII]}}$--SFR relation: (i) galaxies with very high SFR and bright [CII] emission \citep[e.g.,][]{gullberg15, oteo16, strandet17}. They correspond to SFR and L$_{\mathrm{[CII]}}$ that are not probed in our simulation. Indeed, such objects are rare and are not present in our simulated volume; (ii) galaxies with a strong  [CII]-excess emission as compared to their SFR, as tentatively detected in the blind ASPECS survey \citep{aravena16}. These galaxies cannot be produced using our assumption that [CII] emission only arises from PDR. The galaxy named \textquotedblleft CR7" at z=6.6 \citep{matthee17} is also a particular case, as it lies close to the average relation but in a region where our simulated galaxies have IR luminosities higher than the upper limit L$_{IR}<3.14\times$10$^{10}$\,L$_{\odot}$ reported by \cite{matthee17}. The minimum L$_{IR}$ of our simulated galaxies around the location of CR7 in the L$_{\mathrm{[CII]}}$--SFR diagram is 4.2$\times$10$^{10}$\,L$_{\odot}$.\\

\begin{figure}
  \centering
  \includegraphics[scale=0.53]{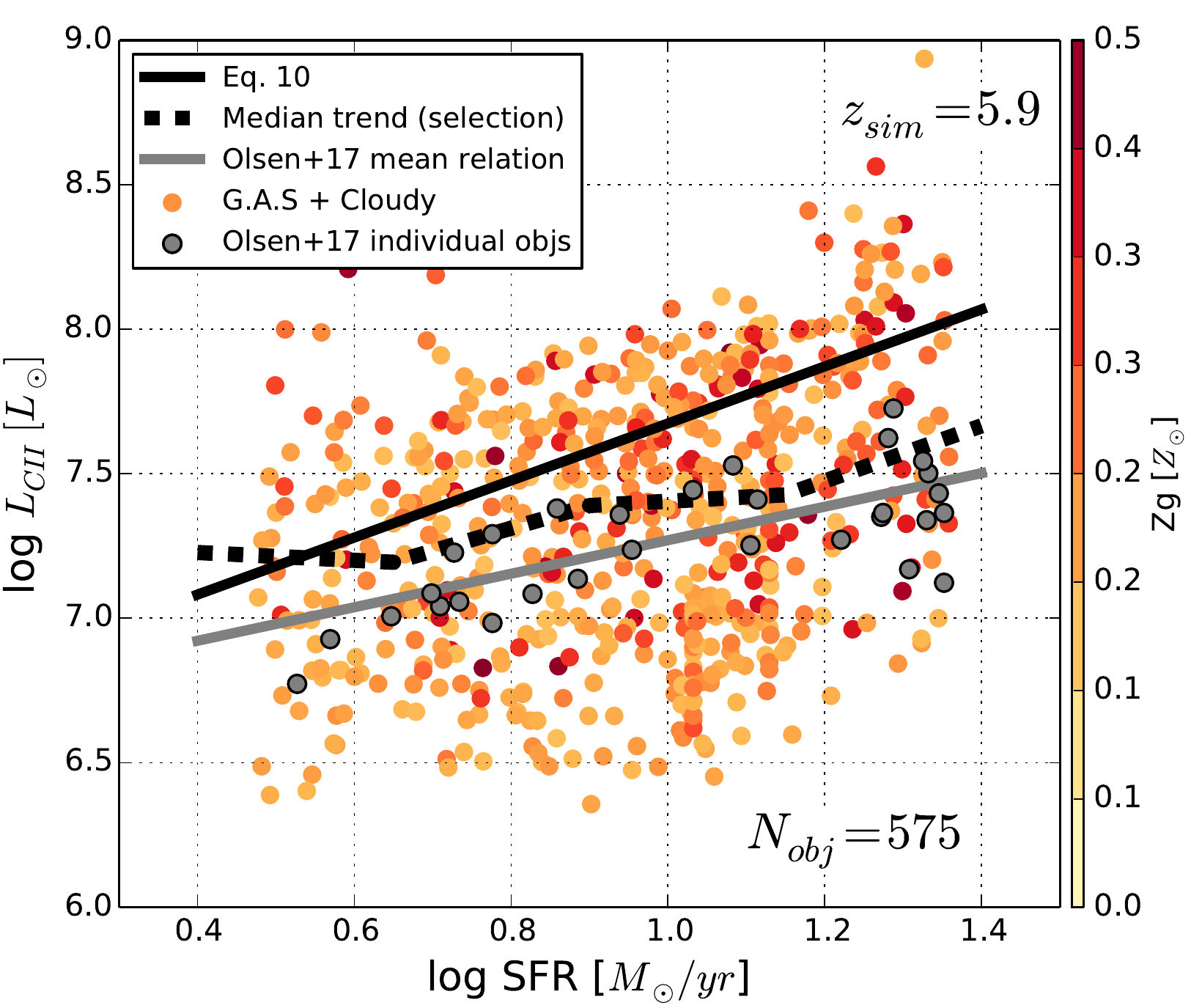}
  \caption{The L$_{\mathrm{[CII]}}$--SFR relation for galaxies that follow the selection of the simulated galaxy sample by \cite{olsen17} at z$\sim$6. Our 575 selected galaxies are shown with coloured points (with colour scale reflecting the gas metallicity). \cite{olsen17} sample is shown with grey points. The black and the grey solid lines show the mean trend of our whole galaxy sample (Eq. \ref{SFR_CII_eq}) and of \cite{olsen17} sample (their Eq.\,6), respectively. The black dashed line marks the mean trend of our galaxy selection. \label{compar_Olsen}}
\end{figure}

We do not observe a strong variation of the mean  L$_{\mathrm{[CII]}}$--SFR relation with redshift. The average trend can be represented by the following law: 
\begin{equation}
\label{SFR_CII_eq}
\mathrm{log}\left(\frac{\mathrm{L_{\mathrm{[CII]}}}}{L_{\odot}}\right) = (1.4 - 0.07z)\times \mathrm{log}\left(\frac{\mathrm{SFR}}{M_{\odot} yr^{-1}}\right) + 7.1- 0.07z
\end{equation}
valid for all redshifts explored here. According to our ranges of available values of SFR and L$_{\mathrm{[CII]}}$ this average relation is limited to L$_{\mathrm{[CII]}}>10^7$\,L$_{\odot}$ and SFR$<$1000\,M$_{\odot}$yr$^{-1}$.\\

We see from Fig.\,\ref{SFR_CII} that the \cite{delooze14} L$_{\mathrm{[CII]}}$--SFR  relation for the local dwarf galaxies does not really apply to our simulated galaxies but at low [CII] luminosities (L$_{CII}\sim$10$^7$\,L$_{\odot}$) and z=4. This mismatch cannot be accounted for only by CMB effects. We show in Fig.\,\ref{SFR_LCII_wocmb}, the L$_{\mathrm{[CII]}}$--SFR relation when both heating and attenuation by the CMB are ignored. While the L$_{\mathrm{[CII]}}$--SFR relation becomes more compatible with the local dwarf galaxy sample relation up to z$\simeq$6, we still predict a shallower slope at higher redshift.\\

We show in Fig. \ref{compar_Olsen} the L$_{\mathrm{[CII]}}$--SFR relation associated to a sub-sample of our simulated galaxies. Following \cite{olsen17}, this sub-sample of 575 objects has been extracted at $z=5.9$ based on three criteria: i) stellar mass between 0.7 and 8$\times$10$^9$\,M$_{\odot}$, ii) SFR between 3 et 23\,M$_{\odot}$yr$^{-1}$ and iii) average gas metallicities between 0.15 and 0.45.
Compared to the mean trend of the whole galaxy sample at this redshift (Eq. \ref{SFR_CII_eq}), this sub-sample is biased toward lower $L_{\mathrm{[CII]}}$ luminosities (by 0.1\,dex to 0.5\,dex as L$_{\rm CII}$ increases). The L$_{\mathrm{[CII]}}$--SFR relation based on our simulated galaxy sub-sample and \cite{olsen17} sample are in very good agreement, even if our galaxy sample shows a higher dispersion. This difference could be explained by the different number of objects in the two samples (575 versus 30). Even if \cite{olsen17} galaxies lie mostly in the middle our selection, the mean trend of our sub-sample is shifted to higher $L_{\mathrm{[CII]}}$ by a factor of 0.15\,dex. In the light of the very different approaches used between the two models (\cite{olsen17} modeled the multi-phased interstellar medium using numerical simulations), the agreement between the two samples is noteworthy. A still better agreement would be obtained by reducing in \citet{olsen17} simulations the dust-to-metals ratio by a factor of 2 (consistent with the fact that dust production is less efficient at low metallicities), or by decreasing the slope of the giant molecular cloud mass spectrum from 1.8 (which corresponds to the Galactic mass spectrum) to 1.5 (see the discussion in Sect.\,5.1 of \citet{olsen17}).\\

\begin{figure}
  \centering
  \includegraphics[scale=0.53]{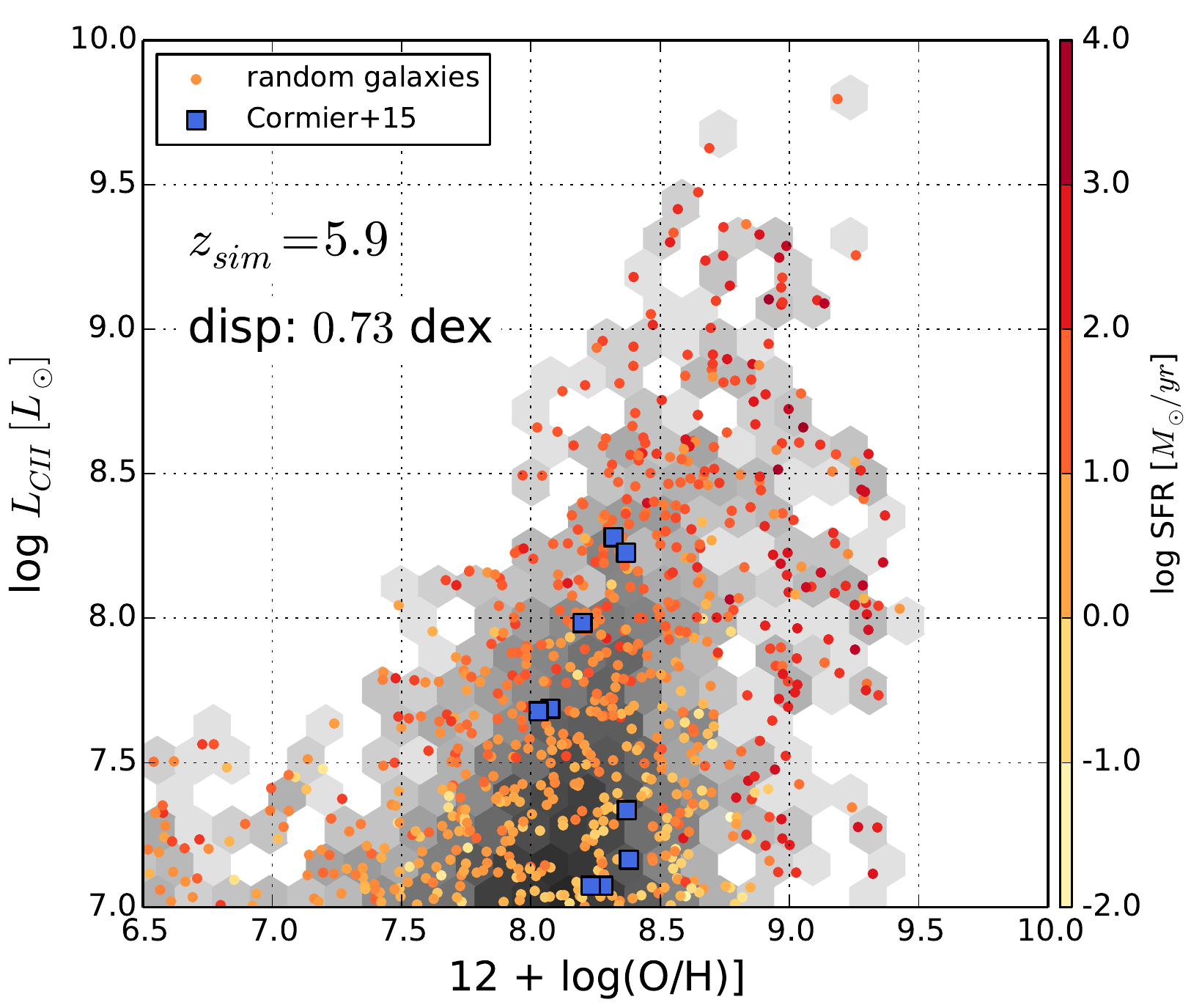}
  \caption{Distribution of the simulated galaxies in [CII] luminosities and metallicities at z=5.9. Coloured points mark the SFR of a randomly selected sample. Blue squares are from the local dwarf galaxy sample \citep{cormier15}.}
  \label{LCII_Z}
\end{figure}

To study the variation of the [CII] luminosities with gas metallicities, we plot on Fig.\,\ref{LCII_Z} the distribution of our simulated galaxies in the L$_{\mathrm{[CII]}}$ -- $Z_g$ plane. There is a broad correlation; galaxies with higher [CII] luminosities tend to have higher metallicities. However, the scatter is quite large, with 0.84 and 0.72\,dex  and z=5 and 7, respectively. We also show the data points from the local dwarf galaxy sample \citep{cormier15}. They are well sampled by the distribution of our simulated galaxies.\\

\begin{figure}
  \centering
  \includegraphics[scale=0.5]{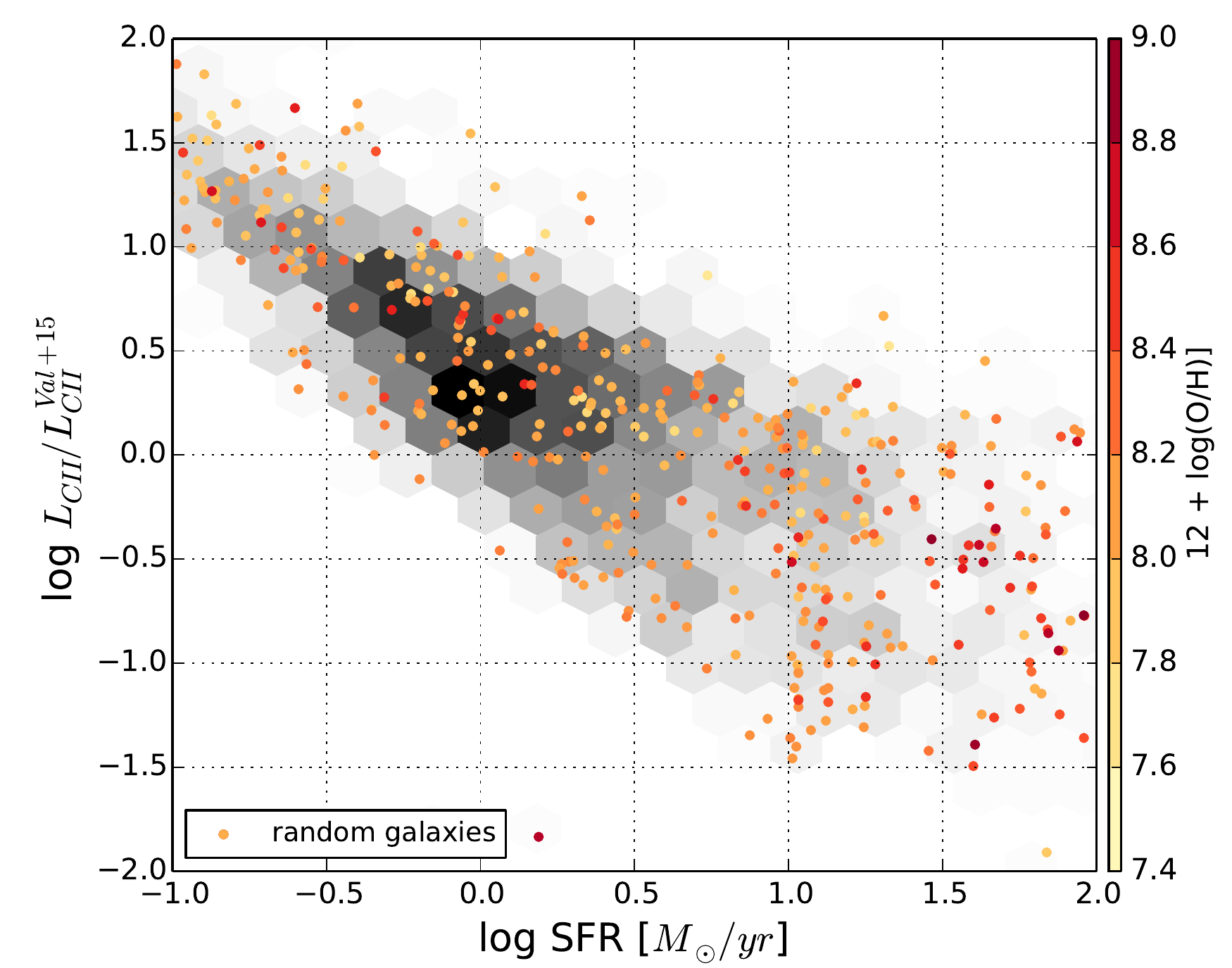}
  \includegraphics[scale=0.5]{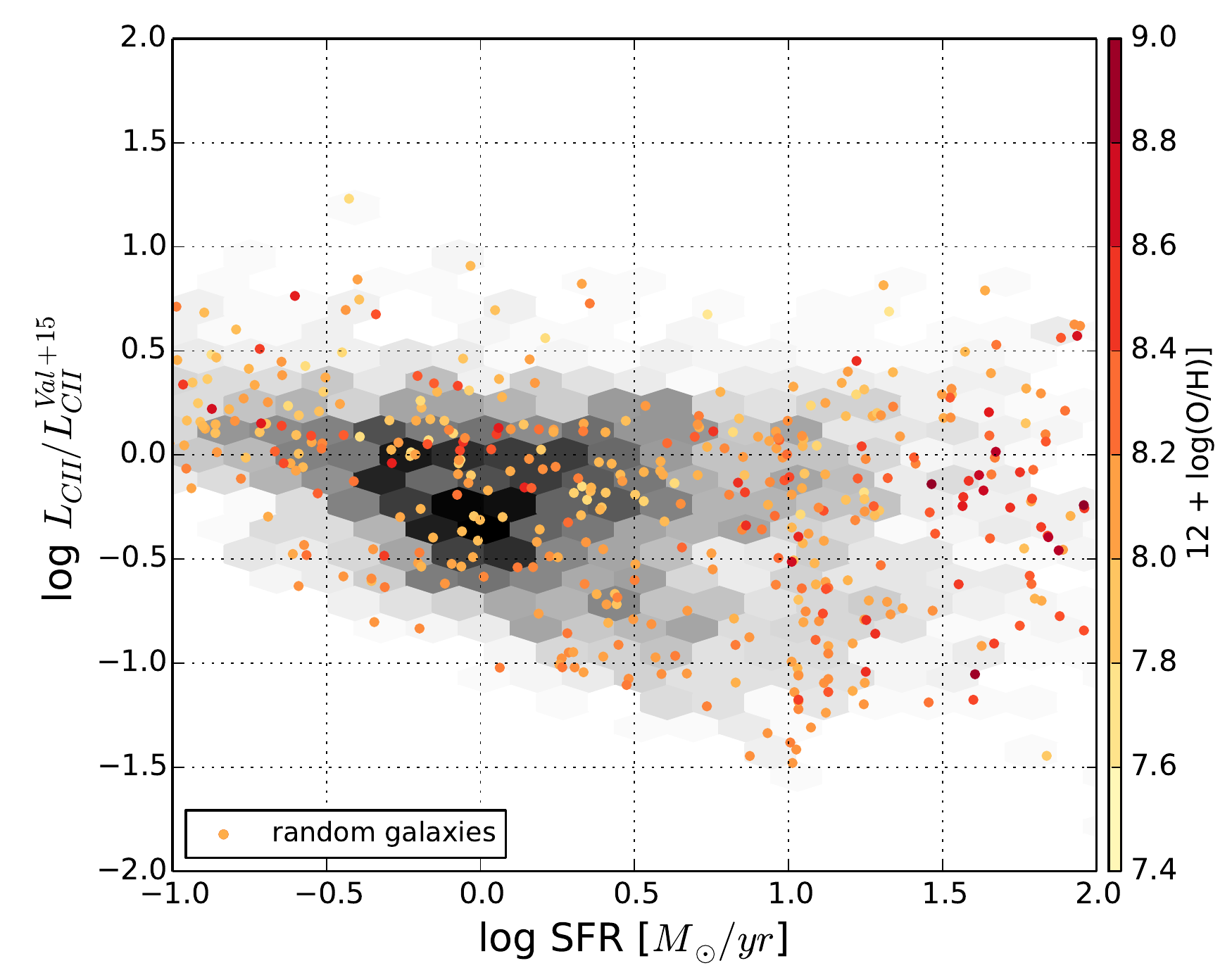}
  \caption{Ratio of our predicted [CII] luminosity at z$\sim$6 (L$_{\mathrm{[CII]}}$) and that predicted by \cite{vallini15} model (L$_{\mathrm{[CII]}}^{\rm Val+15}$). The top panel considers L$_{\mathrm{[CII]}}^{\rm Val+15}$ from Eq.\,\ref{eq_vallini}, and thus N=1, while the bottom panel considers L$_{\mathrm{[CII]}}^{\rm Val+15}$ from Eq.\,\ref{LCII_vallini_N2}, and thus N=2. N is the power law index of the Kennicutt-Schmidt relation, $\Sigma_{\rm SFR}\propto \Sigma_{\rm H_2}^N$.
  \label{LCII_vallini}}
\end{figure}

Finally, we compare our prediction with \cite{vallini15} model (Eq.\,\ref{eq_vallini}) in Fig.\,\ref{LCII_vallini}. For this comparison and to be consistent with \cite{vallini15}, we consider our model at z=5.9 and with CMB effects. We observe a systematic trend with a decreasing L$_{\mathrm{[CII]}}$/L$_{\mathrm{[CII]}}^{\rm Val+15}$ ratio with both SFR and $Z_g$. The bulk of our galaxies has higher L$_{\mathrm{[CII]}}$ than that predicted by Eq.\,\ref{eq_vallini} (by factors of about 1.5-5).  This equation holds when the molecular mass is scaled with the SFR (the Kennicutt-Schmidt relation),  $\Sigma_{\rm SFR}\propto \Sigma_{\rm H_2}^N$ adopting N=1. As discussed in \cite{vallini15}, the range in power law index (N) depends on a variety of factors, among which the most important ones are the observed angular scales, and the calibration of star formation rates. As shown in Fig.\,\ref{LCII_vallini}, we have a better agreement considering a fit of \cite{vallini15} model with N=2:
\vspace{-0.1cm}
\begin{eqnarray}
{\rm log}{L_{\mathrm{[CII]}}}=7.5+0.67~{\log} (\mathrm{SFR}) -0.13~{\log} (Z_g) +\nonumber\\
0.063~{\log (\mathrm{SFR})\log (Z_g)}-0.79~{\log ^2(Z_g)}\,.
  \label{LCII_vallini_N2}
\end{eqnarray}
This is expected as the Kennicutt-Schmidt relation predicted by {\tt G.A.S.} has 1.4$\le$N$\le$2.

\begin{figure}
  \centering
  \includegraphics[scale=0.5]{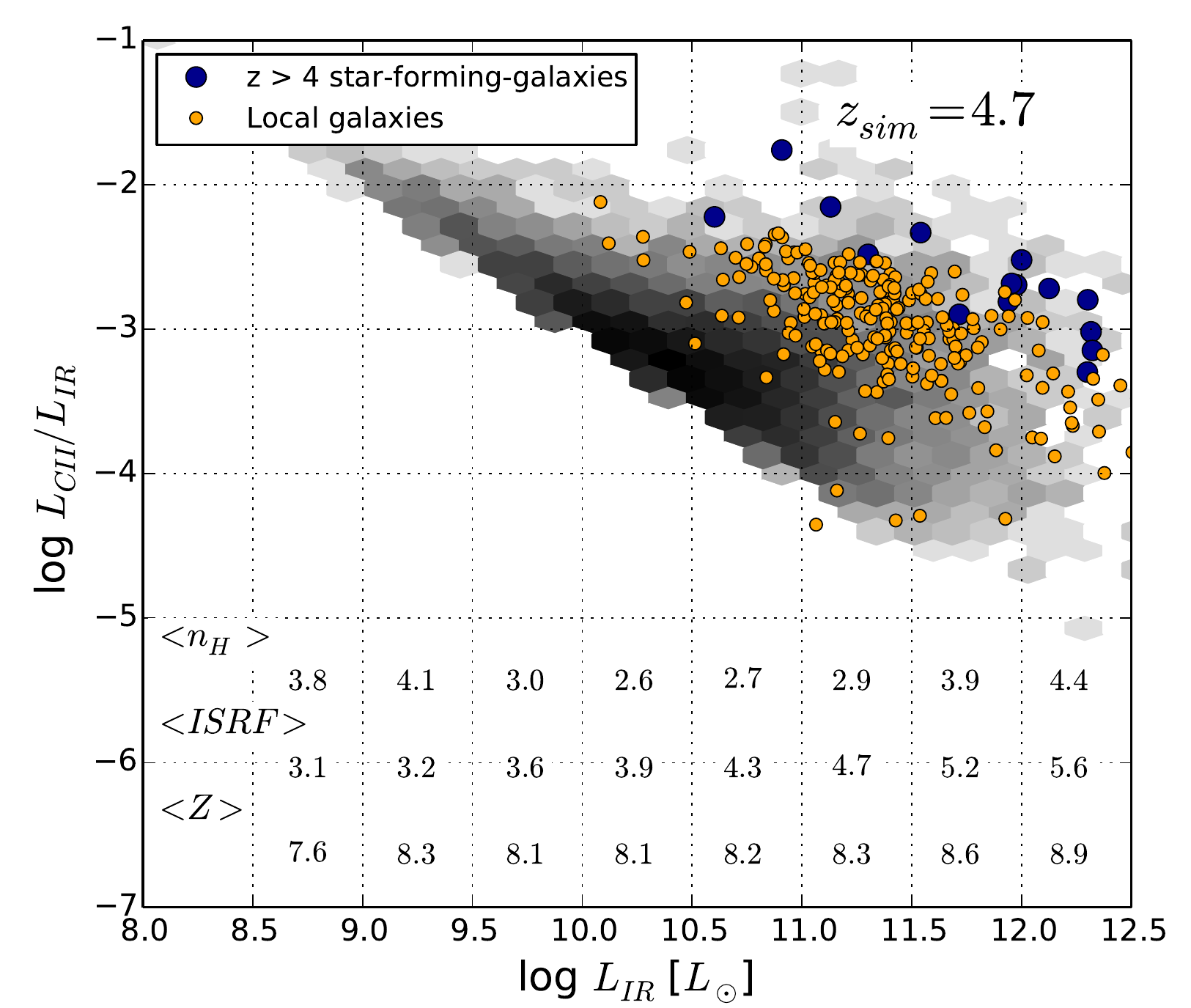}
  \caption{L$_{\mathrm [CII]}$/L$_{IR}$ versus L$_{IR}$ for our sample of simulated galaxies (grey shaded areas) at z=4.7. Local galaxies (orange circles) are the GOALS luminous infrared galaxy sample \citep{diaz17}; a mean correction has been applied to their L$_{\mathrm [CII]}$ to mimic the effects of CMB heating and attenuation. High-resdhift galaxies (blue circles) are extracted from Table\,\ref{CII_compil}. We give on the bottom of the figure the mean densities, ISRF and metallicities (in log), for bins of 0.5\,dex in $\log$\,L$_{IR}$. The [CII] deficit naturally arises in our model. It is well correlated with the intensity of the ISRF.
  \label{CII_deficit}}
\end{figure}

\section{[CII] deficit \label{sect_CII_def}}
In the early days of [CII] observations of low-redshift galaxies with the Infrared Space Observatory, it was observed that very luminous infrared galaxies such as ULIRGs appear to have a deficit in [CII] emission compared to their FIR luminosities \citep{luhman98, malhotra01}. This deficit has been later confirmed with Herschel and extended to lower infrared luminosities, LIRGs. For example, \cite{diaz13} find that LIRGs show a tight correlation of [CII]/FIR with infrared luminosity, with a strong negative trend spanning from $\sim$10$^{-2}$ to 10$^{-4}$, as the infrared luminosity increases. The result from high-redshift objects is more mixed, with a large scatter (2 orders of magnitude) at high luminosities.  Different explanations for this measured decline have been proposed \citep[e.g.,][]{casey14}, including the compactness of the starburst, the AGN activity, optically thick [CII] emission, varying IMF or [CII] saturation at high temperature. \\ 
As shown on Fig.\,\ref{CII_deficit}, the [CII] deficit naturally arises in our model, with a decrease of L$_{\mathrm [CII]}$/L$_{IR}$ by about 2 orders of magnitudes from L$_{IR}$=10$^9$ to 10$^{12}$\,L$_{\odot}$. The dispersion in the L$_{\mathrm [CII]}$/L$_{IR}$ increases with L$_{IR}$. 
The large dispersion at high L$_{IR}$ is also observed for high-redshift objects (e.g. Fig.\,4 of \cite{gullberg15}). Compared to the GOALS sample of local galaxies, the L$_{\mathrm [CII]}$/L$_{IR}$ decrease is stronger in the model and simulated galaxies have on average a lower L$_{\mathrm [CII]}$/L$_{IR}$. This might be due to selection effects (often linked to a limited sensitivity in the observations).\\

We investigate the origin of the deficit using our model parameters. First, we compute the [CII] transition upper level loading to test the hypothesis of \cite{munoz16}, in which the [CII] deficit observed in the highest IR surface-brightness systems is a natural consequence of saturating the upper fine-structure transition state at gas temperatures above 91\,K. We find that from z=4 to 7, the transition upper level loading is not saturated in the region where the bulk of the [CII] intensity is emitted (see Fig.\,\ref{Profile_pdr}), and that 0.01$<n_u/n_l<$0.05. 
The crucial difference between our model and the analytical work of \cite{munoz16} is that we have strong ISRF for our galaxies. \cite{munoz16} ignore the effects of the local (isotropic) radiation field, under the assumption of densities in excess of the critical density for that transition. Therefore, this saturation effect cannot be responsible for the [CII] deficit observed in our model. We searched for correlations of the deficit with the different parameters of our model and found that it is strongly correlated with the intensity of the ISRF (see Fig.\,\ref{CII_deficit}). This is consistent with the analysis of \cite{luhman03} which suggests that a high ISRF incident on a moderate density PDR could explain the deficit in their observed ULIRGs. We extend this analysis to lower luminosities, and show that the deficit still holds at very high redshift. For 10$^{10}<$L$_{IR}<3\times10^{11}$\,L$_{\odot}$, we have a weak correlation of the deficit with the metallicity, with slightly increased metallicity associated with deeper deficits, as observed in \cite{smith17} but for higher metallicities ($Z_g$ between 8.54 and 8.86).

\section{[CII] luminosity function \label{sect_LF}}

\begin{figure*}
  \centering
  \includegraphics[scale=0.46]{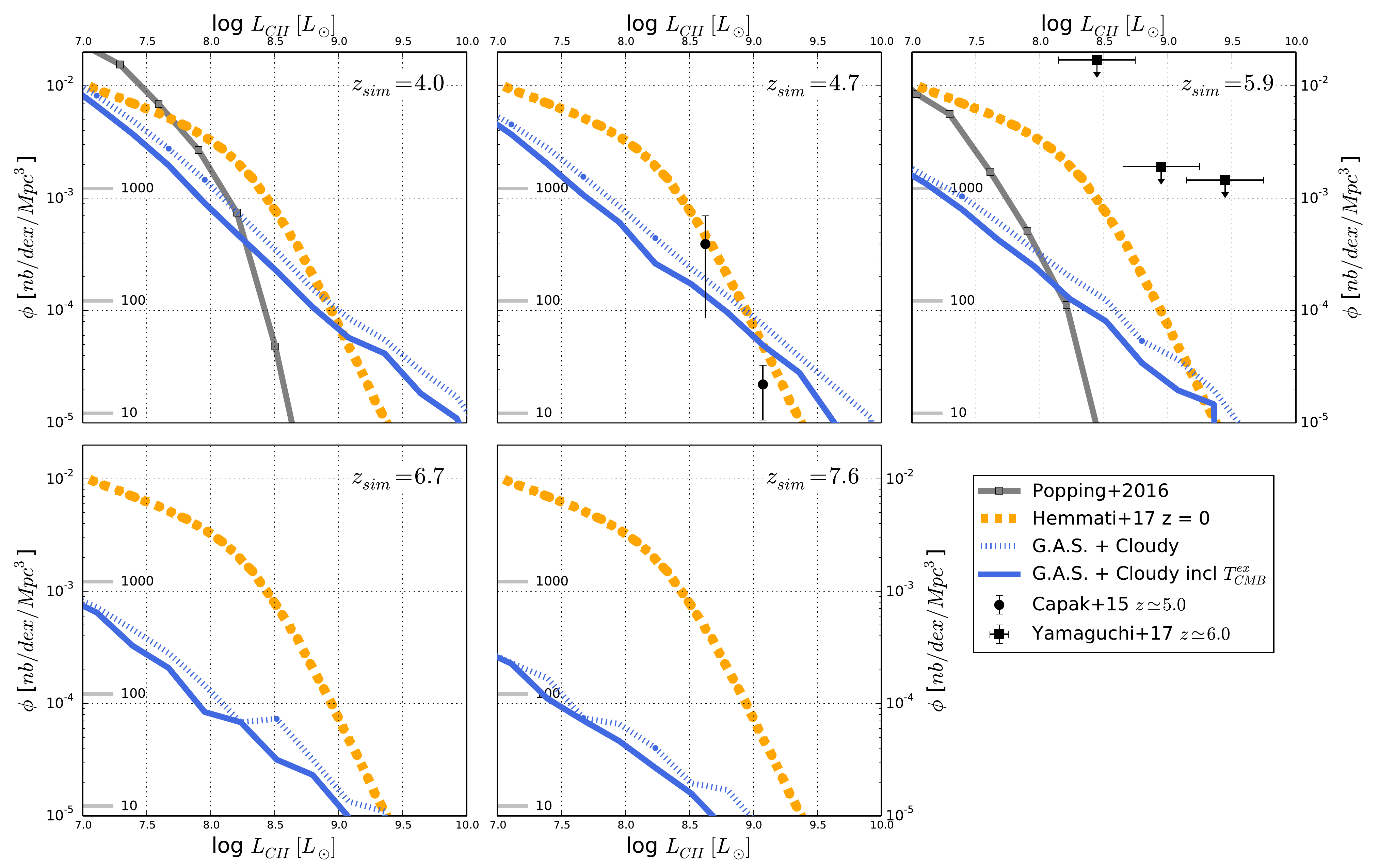}
  \caption{[CII] luminosity function predicted by the {\tt G.A.S.}+{\tt CLOUDY} model from z = 4.0 to z = 7.6. The blue solid curve shows the prediction that accounts for the attenuation of [CII] emission due to the CMB. The blue dotted line would be the luminosity function ignoring the attenuation. 
At z$\simeq$5, we show the observational constraints from \cite{capak15}. At z$\simeq$6, the black squares indicate the observational results of \cite{yamaguchi17}.  We also add the local [CII] luminosity function published by \cite{hemmati17} (orange dotted line)  and model predictions of \cite{popping16} (grey solid line). \label{LFCII}}
\end{figure*}

Fig.\,\ref{LFCII} shows our [CII] luminosity function predicted at $4.0\lesssim z \lesssim8$. We present two different predictions, with and without CMB attenuation. We see a systematic deviation between the two, which is almost constant with redshift (as expected, see Sect\,\ref{CMB_eff}). The attenuation induced by the CMB increases slowly with the [CII] luminosity, from 25\% at L$_{\mathrm{[CII]}}$$\sim$10$^{7}$\,L$_{\sun}$ to 35\% at L$_{\mathrm{[CII]}}$$\sim$10$^{10}$\,L$_{\sun}$. This trend is similar at all redshifts.\\
We also show on Fig. \ref{LFCII} \cite{popping16} luminosity function predictions, that are also based on a semi-analytical model. Compared to \cite{popping16}, we predict a smaller (larger) number of [CII]-emitting galaxies in the faint (bright)-end part, with a crossing point at  L$_{\mathrm{[CII]}} \simeq 2 \times$10$^{8}$\,L$_{\odot}$.\\
We can also compare our [CII] luminosity function with that obtained by using the SFR function from \cite{smit12} and our mean L$_{\mathrm{[CII]}}$--SFR relation. We found that such a combination overestimates the [CII] luminosity function, by e.g., factor $\sim$6 at z=4 for L$_{\mathrm{[CII]}}$= 10$^8$\,L$_{\odot}$.

\subsection{The functional form of the luminosity function}
Our predicted [CII] luminosity function has a power law shape for the whole range of L$_{\mathrm{[CII]}}$ probed in our simulation. This shape is quite different from the [CII] luminosity function measured at z=0 \citep{hemmati17}, which agrees well with the form of the IR luminosity function. This IR luminosity function is better fitted either by a double power law  \citep{magnelli11}, following, \\
\begin{eqnarray}
\Phi(L)=\Phi^{\star}\left(\frac{L}{L^{\star}}\right)^{\alpha_1} \, \mathrm{, for\,} L<L^\star\, \nonumber
\\
\Phi(L)=\Phi^{\star}\left(\frac{L}{L^{\star}}\right)^{\alpha_2}\, \mathrm{, for\,} L>L^\star\,.
\label{eq_DPL}
\end{eqnarray}
or alternatively by a double-exponential function \citep{saunders90, caputi07, gruppioni13}, which is a modified-Schechter function behaving as a power law for $L \ll L^{\star}$ and as a Gaussian in $\log L$ for $L\gg L^{\star}$:
\begin{equation}
\Phi(L)=\Phi^{\star}\left(\frac{L}{L^{\star}}\right)^{\alpha'} \exp\left[-\frac{1}{2\sigma^2}\log_{10}^2\left(1+\frac{L}{L^{\star}}\right)\right] \,.
\label{eq_DEF}
\end{equation}
In these equations, L$^\star$ is the characteristic luminosity where the transition between the faint and bright regimes occurs, and $\Phi^\star$ is the normalization factor\footnote{$\alpha'$ stands for $1-\alpha$ in \cite{saunders90, caputi07} and \cite{gruppioni13}.}. 
Unfortunately, the IR luminosity function has not been measured at z$\ge$4. At lower redshift, it is found that log\,L$_{\mathrm IR}^{\star}$ increases with redshift, from 10.48 (z=0) to 12.35 (z$\sim$2) in \cite{magnelli13}, assuming a double power law, and from 10.12 (z=0) to 11.9 (z$\sim$4) in \cite{gruppioni13} and 11.40 (z$\sim$1) and 11.80 (z$\sim$2) in \cite{caputi07}, assuming a double-exponential function.

Thus, a typical L$_{\mathrm IR}^{\star}$$\sim$10$^{12}$\,${\mathrm L}_{\odot}$ is expected at high redshift. This IR luminosity converts to SFR=100\,M$_{\odot}$\,yr$^{-1}$ (using \cite{kennicutt98} and assuming a \cite{chabrier2003} IMF). Using our L$_{\mathrm{[CII]}}$--SFR relation (Eq.\,\ref{SFR_CII_eq}), we obtain log\,L$^\star_{[CII]}$=8.4 to 9.1, from z=7 to 4, respectively. These characteristic luminosities are quite high and difficult to probe with our model. They fall in a regime where we have less than 50 objects in our simulation.\\

A power law shape is not completely unexpected at these very high redshifts. A single power law provides an equally good fit to the UV luminosity function at z=8, while at z = 6 and 7, an exponential cutoff at the bright end is moderately preferred \citep{finkelstein15}.
For the stellar mass function (SMF), the knee is sharpened as time goes by, and a progressive flattening of the low-mass end is observed from z$\sim$6 to zero \citep{davidzon17}. At z$>$4.5 the SMF is best fit by a single power law function with a cut-off at 3$\times$10$^{11}$\,M$_\odot$. Measurements of the SMF extend down to $\sim$10$^{10}$M$_\odot$. In this range of masses, the [CII] luminosities range from $\sim$2$\times$10$^8$ to 5$\times$10$^9$\,L$_\odot$. Accordingly, a cut-off may be expected in the [CII] luminosity function but outside of the range of luminosities probed by our simulation. This may explain why we do not see a break in our [CII] luminosity function. \\

\begin{figure}
  \centering
  \includegraphics[scale=0.55]{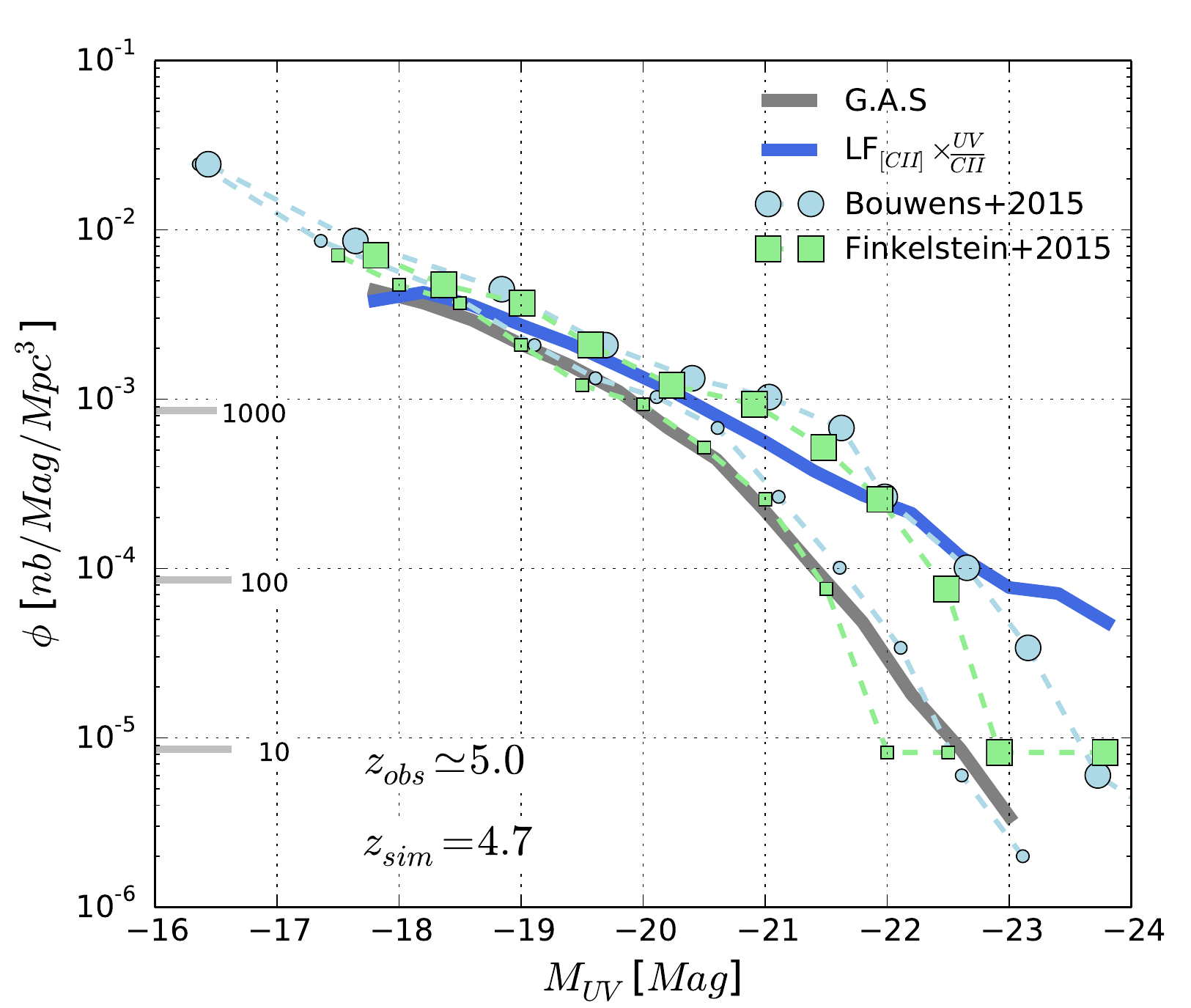}
  \caption{UV luminosity function derived from the [CII] luminosity function at z$\sim$5. 
   Small and large symbols are observational measurements without and with extinction correction, respectively. The correction of extinction of observed measurements has been done using {\tt G.A.S.} (using the two grey cuves of Fig.\,\ref{fig_UVLF}). The solid grey line shows our predicted UV luminosity function from  {\tt G.A.S.}, which is in good agreement with the observed data points coming from \cite{bouwens15} and \cite{finkelstein15}. The UV luminosity function predicted from our [CII] luminosity function assuming a fixed [CII] to UV luminosity ratio is shown in blue. \label{UFLF_from_LCII}}
   \end{figure}

We compare on Fig.\,\ref{UFLF_from_LCII} the observed UV luminosity function at $z=5$ with that predicted from the [CII] luminosity function, applying a UV to [CII] luminosity ratio. For that comparison, it is primordial to correct the observed UV luminosity function for attenuation as dust strongly affects the shape of the observed UV luminosity function. 
We use our model to derive a mean attenuation per UV magnitude bin, and correct the observed UV data points. 
We have a quite good agreement between the corrected UV luminosity function and that predicted from [CII] using a constant luminosity ratio, L$_{UV}^{corr}$/L$_{\mathrm{[CII]}}$ = 1.6$\times$10$^3$. This constant ratio is a crude approximation but it is not worth searching for any variation with luminosity given the large and uncertain corrections for dust attenuation. This ratio is much larger than the L$_{UV}$/L$_{\mathrm{[CII]}}$ ratios of $\sim$100 to 650 obtained for UV-selected galaxies at $z=5$ \citep{barisic17}. Part of the discrepancy may be attributed to dust attenuation of the observed UV light.

\subsection{Redshift evolution}
The slope of the power law that fits the luminosity function is not evolving strongly with redshift for 4$\lesssim$z$\lesssim$8, and is $\simeq$-1. At such high redshift, the slope of the IR luminosity function for L$_{IR}<$L$^{\star}_{IR}$ is not known; at lower redshift, it is usually fixed to -0.6 \citep[e.g.][]{magnelli11}, but a shallower faint-end slope ($\alpha_1$= -0.4) has been recently measured at $1.5<z<2.5$ \citep{koprowski17}. At $z=0$, the slope of the [CII] luminosity function is equal to -0.42 for  ${\mathrm L}_{\mathrm{[CII]}}<{\mathrm L}_{\mathrm{[CII]}}^{\star}$, where L$_{\mathrm{[CII]}}^{\star}$=2.17$\times$10$^8$\,L$_\odot$. The steepening of the faint-end slope of the [CII]  luminosity function between z=0 and $z>4$ may reflect the fact that the galaxy population is richer in faint [CII] emitting galaxies with increasing redshift, which is the natural consequence of the hierarchical formation of galaxies.
In the UV, the faint-end slopes varies from -1.5 at z=4 to -2 at z=7 \citep{bowler15, finkelstein15} but part of the steepening in that case may be explained by a changing impact of dust attenuation with redshift. For [CII], only the CMB is attenuating the luminosity and this does not cause any change in the slope of the power law (see Fig.\,\ref{LFCII}).\\

At a given [CII] luminosity the density of object decreases with redshift following,
\begin{equation}
\label{LF_CII_evol_eq}
\mathrm{log}\left(\frac{\mathrm{\phi}}{\mathrm{dex}^{-1}\,\mathrm{Mpc}^{-3}}\right) = -1.0\times\mathrm{log}\left(\frac{\mathrm{L_{\mathrm{[CII]}}}}{\mathrm{L}_{\odot}}\right) -0.4\times z + 6.7
\end{equation}
valid for redshifts 4.7$\le$z$\le$8.  The slope of the decrease in density is equal to -0.4 ;  it is close to the slope of -0.31 seen in UV \citep{finkelstein15}. The density evolves to higher values by a factor of 20$\times$ from z=7.6 to z=4.7. At the characteristic luminosity of the local [CII] luminosity function (L$_{\mathrm{[CII]}}^{\star}$=2.17$\times$10$^8$\,L$_\odot$), the number density is 3.8 times lower at z=4 than at z=0.

\subsection{Comparison with observational constraints}
\begin{figure}
  \centering
  \includegraphics[scale=0.55]{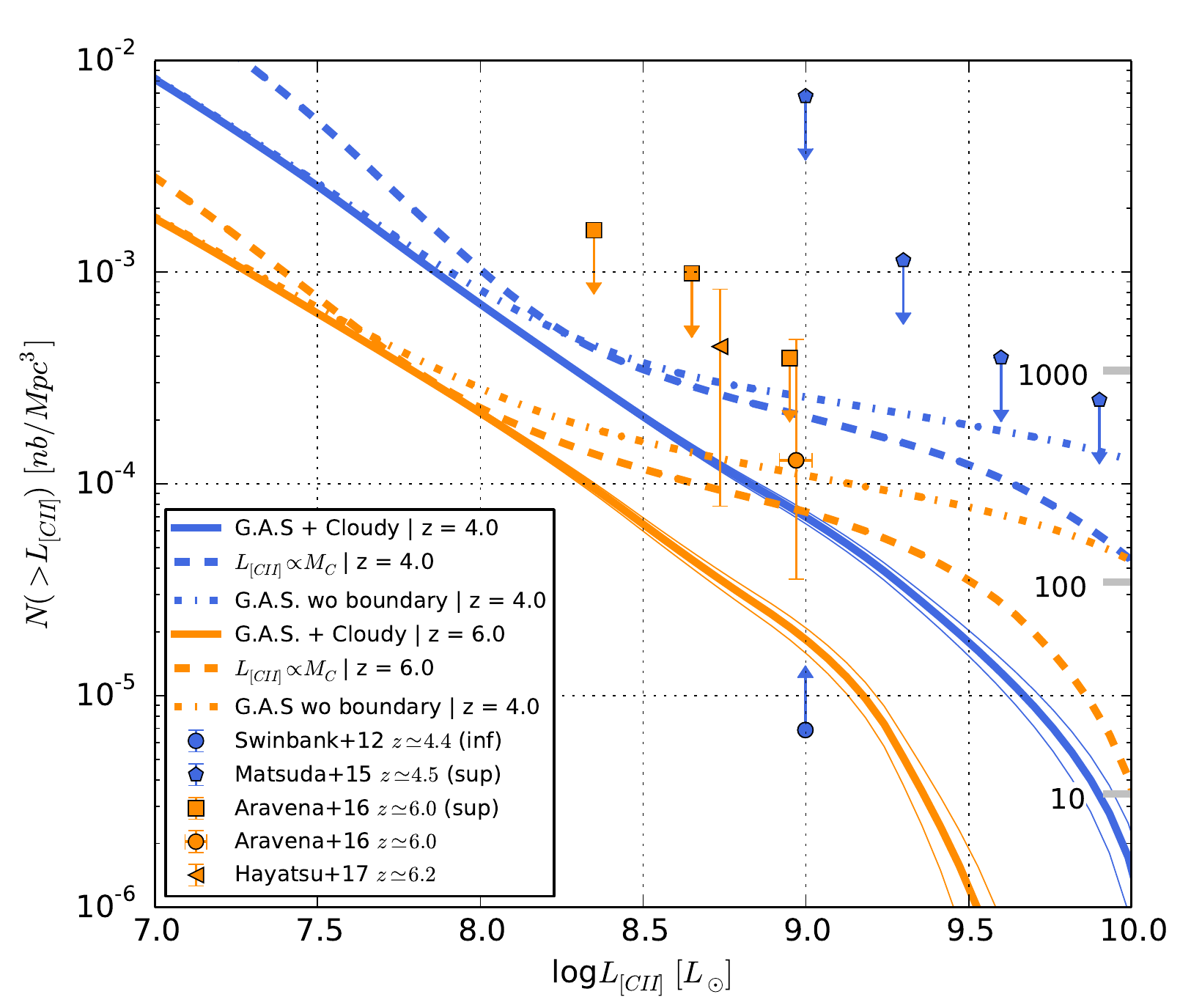}
  \caption{Cumulative [CII] luminosity functions predicted at z$\simeq$4 and z$\simeq$6 (thick continuous blue and orange lines, respectively, with the thin lines representing $\sqrt N_{obj}$), compared to the current observational constraints (data points with error bars).  The dot-dashed lines are derived by relaxing the criterium $r_{eq}<r_g$ which concerns only $<0.8\%$ of the galaxies. The dashed lines show the [CII] luminosity functions derived from the carbon mass function of our model. \label{NsupL}}
\end{figure}

To date, observational constraints are very sparse, with upper limits from \cite{yamaguchi17} at z$\simeq$6 and estimates at z$\simeq$5 derived from \cite{capak15} in \cite{hemmati17}. As explained in \cite{hemmati17}, the z$\simeq$5 estimates are very rough. They are based on observations of nine Lyman-break galaxies in [CII] using ALMA. To see where these measurements sit compared to the luminosity function, \cite{hemmati17}
measure the volume for each observation using the area and the redshift width of each ALMA pointing and correct the volume using the number density of Lyman-break galaxies. These factors and the low number statistics  makes these estimates very sensitive on choice of bins and therefore uncertain. Moreover, there might exist classes of galaxies that are not selected as Lyman-break galaxies at high redshifts but that are bright in [CII] and can contribute to the luminosity function.\\
As we can see from Fig.\,\ref{LFCII}, upper limits are not giving very stringent constraints and our predicted luminosity function is well below. Our predicted [CII] luminosity function at z=4.7 is compatible with the two points estimated from \cite{capak15} (at 1 and 3$\sigma$, respectively).\\

More observational constraints are available on the cumulative [CII] luminosity functions. At z$\simeq$4, there exists both lower \citep{swinbank12} and upper \citep{matsuda15} limits. At 6$<$z$<$8, the ASPECS blind survey gives only upper limits to the bright end of the [CII] luminosity function, as their detections are candidate [CII]-line emitters \citep{aravena16}. We also consider their number density assuming that only the brightest candidate is real.
\cite{hayatsu17} found two [C II] emitter candidates at z=6.0 and 6.5 in the ALMA 1.1-mm survey of the SSA22 field. They estimate the luminosity function at z= 6.2 from blind detection on the assumption that one of the two unconfirmed lines is [CII] at z$\sim$6.
We show on Fig.\,\ref{NsupL} the comparison between our predictions and those constraints. 
We do not consider the measurements from \cite{miller17} as their ALMA sample is biased to fields of extreme objects at z$>$6 and cannot be used to directly constrain the field luminosity function.
At z$\simeq$4, our model falls between the observational upper and lower limits. At z$\simeq$6, it is well below the upper limits. It is also 1.3$\sigma$ below the two estimates from \cite{hayatsu17} and \cite{aravena16}. These estimates are given for L$_{\mathrm{[CII]}}$$\gtrsim 5.4\times$10$^8$\,L$_{\odot}$ and $\gtrsim$9.1$\times$10$^8$\,L$_{\odot}$, where we have few objects in our simulated volume (110 and 60, respectively -- i.e. $<$1.5\%). 
These galaxies are characterized by a high metallicity, $Z_g \ge 8.0$,  a star formation rate $\ge$100\,M$_{\odot}/yr$ and therefore a strong ISRF.  A 1$\sigma$ agreement with the measurements would need an increase of the number density of such [CII]-emitting galaxies by a factor of 2.5.

To investigate this small discrepancy, we have built a [CII] luminosity function derived from the carbon mass function. We assume a constant carbon mass to [CII] luminosity ratio ($R$). By using $R$$\simeq$25L$_{\odot}$/M$_{\odot}$, the mass-derived [CII] luminosity function lies very close to that build with our {\tt G.A.S+CLOUDY} model, for L$_{\mathrm{[CII]}}$$\le$10$^8$\,L$_{\odot}$, as shown on Fig.\,\ref{NsupL}. But we clearly see that this simple model predicts an excess of bright objects (for L$_{\mathrm{[CII]}} \gtrsim 1.5\times 10^8$\,L$_{\odot}$). Some galaxies can reach [CII] luminosities ten times higher than in our fiducial model. The predictions of this simple model is then in agreement with the current observational constraints. This result indicates that the carbon content in our simulated galaxies is sufficient to produce an excess on the bright-end of the luminosity function.

As explained in Sect.\,\ref{PDR_eq_model_SECT} our fiducial model is based on an equivalent PDR structure. For each galaxy, we assumed a metallicity, hydrogen density and ISRF. These three parameters allowed us to compute, for each PDR, a [CII] luminosity per unit of surface area and an effective surface area of emission. In some rare cases ($<$0.8\%) the equivalent radius of the PDR is larger than the radius encompassing the whole mass of the galaxies ($r_g = 11r_d$, containing 99.9\% of the mass). In these cases, we artificially limited the PDR equivalent radius $r_{eq}$ to $r_g$ (while keeping ISRF, n$_H$ and Z$_g$ to their original values). This led to a reduction of the surface area of the emission, $S_{PDR}$ (Eq.\,\ref{eq_S_PDR}), and therefore of the [CII] luminosity. If this size criterion is relaxed, the [CII] luminosity function becomes very close to that obtained with our simple model.  Galaxies that violate the size criterion are very small (disc size $<$1\,kpc), extremely dense (n$_{H}>$10$^5$\,cm$^{-3}$) and have high metallicities ($Z_g>$8.5). These objects (which represent a very small fraction of the sample, $<0.8\%$) are probably not a realistic population of galaxies and cannot account for the difference seen between the predicted and observed cumulative luminosity functions. With our model, we cannot produce much more high  L$_{\mathrm{[CII]}}$ objects by simply changing the parameters (i.e. ISRF, n$_H$ and Z$_g$) in reasonable proportions. Higher [CII] luminosities could be obtained by considering an additional excitation of the [CII] line by other processes, i.e. AGN emission.

\section{Conclusions \label{sect_cl}}

We have used our semi-analytical model of galaxy formation (\verb?G.A.S?)  combined with the photoionisation code {\tt CLOUDY} to compute the [CII] luminosity for a large number of galaxies at $z\ge4$ ($\sim$28,000 at z=5). With such a large statistical sample, we can investigate the dispersion in the L$_{\mathrm{[CII]}}$--SFR relation as well as derive the [CII] luminosity function. Our model takes into account the effects of CMB heating and attenuation that are important at such high redshifts. \\

We showed that our model is able to reproduce the L$_{\mathrm{[CII]}}$--SFR relation observed for $\sim$50 star-forming galaxies at $z\ge4$. However, our model doesn't contain any galaxy with a very strong  [CII]-excess emission as compared to their SFR, as found in the blind ASPECS survey \citep{aravena16}. More generally, we found that the L$_{\mathrm{[CII]}}$--SFR relation is very dispersed (0.51 to 0.62\,dex from z=7.6 to z=4), the large dispersion being due to combined effects of different metallicities, ISRF and gas contents in the simulated high-redshift galaxies. The high dispersion provides an explanation to the upper limits obtained on a number galaxies at $z\ge6$ \citep[e.g.][]{gonzalez14, ota14}.  We found that the dispersion and the fraction of outliers are reduced when the SFR of galaxies is derived from the UV and IR luminosities, following what is being done from the observations (SFR=SFR$_{UV-obs}$ + SFR$_{IR}$). This demonstrates the importance of timescales when using instantaneous quantities as SFR (the timescales being shorter in the model than those assumed in the luminosity-SFR conversions), and the effect of using average conversions. CMB attenuation and heating (which becomes important in the cold gas) also contribute to the dispersion, because its effects depend on the properties of each galaxies (e.g., kinetic temperature and density of the gas). It will be very difficult to correct individual [CII] observations from CMB effects because this would require to know the physical properties of the [CII]-emitting gas.\\

We observed a small evolution of the L$_{\mathrm{[CII]}}$--SFR relation with redshift, with a decrease of the [CII] luminosity of only $\sim$30\% from z=4 to z=7.6 at a given SFR. Our L$_{\mathrm{[CII]}}$--SFR relation at $z\ge5$ is not compatible with the relation for the local dwarf galaxy sample. 
Finally, we also showed that there is a broad correlation, with a scatter $\sim$0.8\,dex, between the [CII] luminosity and gas metallicity. \\

We found that our model naturally predicts the [CII] deficit, with a decrease of L$_{\mathrm [CII]}$/L$_{IR}$ by about 2 orders of magnitudes from L$_{IR}$=10$^9$ to 10$^{12}$\,L$_{\odot}$. We investigated the origin of the deficit and found that it is strongly correlated with the intensity of the ISRF.\\

We then presented the predictions for the [CII] luminosity function for $4\le z \le8$ and ${\rm log}\,{L_{\mathrm{[CII]}}}$$\ge$7, which is our completeness limit. On the bright end, our simulations contain less than 10 objects with ${\rm log}\,{L_{\rm CII}}$ higher than 9.9, 9.9, 9.4, 9.2, and 8.8 at z$\simeq$4.0, 4.7, 5.9, 6.7 and 7.6, and these values are thus the upper bounds of our predictions.
In these ranges of L$_{\rm CII}$, the luminosity function has a power law shape with $\alpha$=-1. This may be explained by a redshift evolution characterized by continued positive luminosity evolution, as seen for the IR luminosity function \citep{koprowski17}. The characteristic luminosity L$_{\mathrm{[CII]}}^{\star}$=2.17$\times$10$^8$\,L$_\odot$, measured at z=0, should then increase by a factor of about 5 at z=4 and recover the local value at z=7.6. In the mean time, the number density decreases by a factor of 20$\times$ from z=4.7 to z=7.6. At those redshifts, we have a reasonable agreement between the UV and [CII] luminosity functions considering a constant luminosity ratio, L$_{UV}^{corr}$/L$_{\mathrm{[CII]}}$ = 1.6$\times$10$^3$, and assuming a correction for attenuation of UV luminosities derived from our model. Finally we compared our predictions with the few observational constraints. We found that our differential luminosity function is in reasonable agreement with the observational estimates, but our cumulative luminosity function is 1.3$\sigma$ below the estimates at z=6 and L$_{\mathrm{[CII]}}$$\simeq$5-9$\times$10$^8$\,L$_{\odot}$. By relaxing a parameter in the model that constrains the size of the effective PDR and that affects only $<0.8\%$ of simulated galaxies, we can increase the number density of bright [CII]-emitters and better match the estimates on the cumulative luminosity function at z=6.\\
 
Our model relies on the assumption that the [CII] line is originating exclusively from PDRs, with one effective PDR defined for each galaxy. It does not take into account the whole complexity of the ISM in galaxies, as the structure of giant clouds or inhomogeneous ISRF, that can affect the [CII] luminosities. So, it is remarquable how this simplified model can reproduce the observations at high redshift. Our {\tt G.A.S.}+{\tt CLOUDY} predictions are also in good agreement with those obtained from cosmological zoom simulations of galaxies combined with a multiphased interstellar medium modeling \citep{olsen17}. However, the main limitations of all current models is that they miss the contribution from [CII] that can be excited (i) on a large scale by the dissipation of mechanical energy (turbulence and shocks) in the early stages of the building of galaxy disks \citep{appleton13}, and (ii) by the AGN.\\

The data from the model presented here are distributed as FITS-formatted files at the CDS (http://cdsweb.u-strasbg.fr). The files (one per redshift) contain for each galaxy all the data used in this paper (e.g., M$_{\star}$, SFR, ISRF, Z$_g$, L$_{\mathrm[CII]}$, L$_{IR}$).

\begin{acknowledgements}
We acknowledge financial support from the "Programme National de Cosmologie and Galaxies" (PNCG) funded by CNRS/INSU-IN2P3-INP, CEA and CNES, France, from the ANR under the contract ANR-15-CE31-0017 and from the OCEVU Labex (ANR-11-LABX-0060) and the A*MIDEX project (ANR-11-IDEX-0001-02) funded by the "Investissements d'Avenir" French government program managed by the ANR.
MC acknowledges the support from the CNES. We warmly thank Maryvonne G\'erin for enlightening discussions, and Maxime Ruaud for providing us his code of [CII] excitation. MC thank Jacques Le Bourlot, Franck Le Petit and Benjamin Godard for their help in using PDR Meudon code (which has been used for crosschecks). We thank Gary Ferland for pointing us the new C17 version of {\tt CLOUDY}, which incorporates the treatment of isotropic backgrounds. We thank Lin Yan and Shoubaneh Hemmati for providing us the data from their paper on the local [CII] luminosity function, Tanio D\'iaz-Santos for providing us the data for the [CII] deficit of local galaxies, and  Steven Finkelstein and Rychard Bouwens for providing us the data points of their high-redshift UV luminosity functions. Finally, we thank the anonymous referee for his/her very useful comments and responsiveness.

\end{acknowledgements}
\appendix

\section{[CII] excitation temperature \label{CII_ex}}

The excitation temperature of the [CII] transition is defined by the relative populations of the upper and lower levels, $n_u$ and $n_l$, respectively, through the standard equation
\be
\frac{n_u}{n_l}  = \frac{g_u }{g_l} e^{-T^* / T^{ex}} \,,
\label{tex}
\ee
where $T^*$ is the equivalent temperature (= $h\nu/k$), and $g_u$ ($g_l$) is the statistical weight of upper (lower) level.\
The upwards and downwards rate coefficients are related by detailed balance
\be
R_{lu}/R_{ul} = (g_u / g_l) e^{-T^* / T^{kin}} \,,
\ee
where $T^{kin}$ is the kinetic temperature.
Due to the wide range of conditions under which it is the dominant form of carbon, collisional excitation of [CII]  by electrons, H, and H$_2$ can all be important. For a single collision partner, the collision rates are equal to the rate coefficients times the density $n$ of that collision partner, thus  
\be
C_{ul} = R_{ul}n {\rm ~and ~} C_{lu} = R_{lu}n \,.
\ee
For a region with multiple collision partners, the upwards and downwards rates are the sum of the rates produced by each.

The energy density in the cloud at  the frequency of the [CII] transition is given by
\be
U = (1 - \beta)U(T^{ex}) + \beta U(T^{bg}) \,,
\ee
where $\beta$\ is the photon escape probability and $T^{bg}$ is the temperature of the isotropic CMB radiation field.

For a spherical cloud with a large velocity gradient of the form v $\propto$ r, the escape probability is given by 
\be
\beta = \frac{1 - e^{-\tau}}{\tau} \,,
\ee
where $\tau$ is the peak optical depth of the transition.

Radiative processes include spontaneous emission (rate $A_{ul}$ $s^{-1}$), stimulated emission (rate $B_{ul}U$), and stimulated absorption (rate $B_{lu}U$). The stimulated rate coefficients are again related by detailed balance through
\be
B_{lu} = (g_u / g_l)B_{ul} \,.
\ee
From the relationship between the stimulated and spontaneous downwards rates, 
\be
B_{ul}U = \frac{(1 - \beta) A_{ul}}{e^{T^*/T^{ex}} - 1} + \frac{\beta A_{ul}}{e^{T^*/T^{bg}} - 1} \,.
\ee
Following \cite{goldsmith12}, for convenience in dealing with the background, we define 
\be
G = \frac{1}{e^{T^*/T^{bg}} - 1} \,.
\label{eq:photon-occ}
\ee

The rate equation that determines the level populations includes collisional and radiative processes, and is
\be
n_u(A_{ul} + B_{ul}U + C_{ul}) = n_l(B_{lu}U + C_{lu}) \,.
\label{rate}
\ee
The expression for the excitation temperature finally becomes
\be
e^{T^* / T^{ex}} = \frac{C_{ul} + \beta(1 + G)A_{ul}}{G\beta A_{ul} + C_{ul}e^{-T^* /T^{kin}}} \,.
\ee
The optical depth can be written 
\be
\label{eq_tau}
\tau = \tau_0 \frac{1 - e^{-T^*/T^{ex}}}{1 + (g_u/g_l)e^{-T^*/T^{ex}}} \,,
\ee
with $\tau_0$ being the optical depth which would occur if there were no excitation, i.e. $T^{ex}$ = 0 and
\be
\tau_0 = \frac{hB_{lu}N(C^+)}{\delta v} \,.
\label{tau_0}
\ee
In this equation, the line profile function at line center is approximated by $\delta v^{-1}$, and $N(C^+)$ is the total column density [CII]. 

\section{Measured L$_{\mathrm{[CII]}}$ and SFR for $z>4$ star-forming galaxies}
We give in Table\,\ref{CII_compil} a compilation of measured L$_{\mathrm{[CII]}}$ and SFR for high-redshift star-forming galaxies.

  \begin{table*}
   \tiny
      \caption[]{\label{cii_lfir} {\small Compilation of z$>$4 star-forming galaxies (i.e. excluding known QSO or AGN) with both [CII] and SFR (or IR luminosity) measurements  (so excluding upper limits). IR luminosities are computed by integrated the SED over rest-frame 8 to 1000\,$\mu$m. The seventh column gives the multiplicative factors that were used to convert the quoted FIR luminosities (defined for a wavelength range given in brackets) to 8-1000 \,$\mu$m luminosities (based on \cite{bethermin15} effective SEDs). When applicable, the lensing magnifications are also given (all luminosities are uncorrected for lensing amplification). }}
         \label{CII_compil}
         \begin{tabular}{lllllllcl}
            \hline
            \noalign{\smallskip}
	Source Name & Redshift & L$_{\mathrm{[CII]}}$  &  L$_{IR}$ & SFR & Ref & From L$_{FIR}$ to L$_{IR}$ & Lensing & Selection \\   
      	& & [10$^{9}$\,L$_{\odot}$] &  [10$^{12}$\,L$_{\odot}$] & M$_{\odot}$ yr$^{-1}$ & & & Magnification & \\ \hline   
	\multicolumn{8}{l}{\underline{$4<z<5$}}\\
	SPT0418-47 & 4.224 & 65$\pm$5 & 67.7$\pm$4.32 &  -- & 1 & 1.08$\times$L[42.5-500]  & 32.7$\pm$2.7  & DSFG\\
	SPT0113-46  & 4.232 & 46$\pm$10 & 22.68$\pm$1.08  &  -- & 1 & 1.08$\times$L[42.5-500] & 23.9$\pm$0.5  & DSFG\\
	ID141 & 4.243 & 61.6$\pm$9.8 & 85$\pm$3 & -- & 2 & -- & 10-30 & DSFG \\ 
	ALMAJ081740.86+135138.2 & 4.260 & 3.02$^{+0.37}_{-0.33}$ & 1.00$^{+2.16}_{-0.68}$ & & 3 & -- & -- & DLA \\ 
	SPT2311-54 & 4.281 & 24$\pm$3 & 35.97$\pm$3.27 & -- & 1 & 1.09$\times$L[42.5-500] &  1.9$\pm$0.1  & DSFG\\
	SPT0345-47 & 4.296 & 33$\pm$4 & 100.28$\pm$8.72 & -- & 1 & 1.09$\times$L[42.5-500] & 7.9$\pm$0.5  & DSFG\\
	SPT2103-60 & 4.435 & 71$\pm$10 & 37.06$\pm$2.18 & -- & 1 & 1.09$\times$L[42.5-500] & 27.8$\pm$1.8  & DSFG\\
	ALESS61.1 & 4.419 & 1.5$\pm$0.3 & 2.1$\pm$0.4 &  -- & 4 & -- & -- &  DSFG \\ 
	SMG1 & 4.424  & 8.3$\pm$0.2 &  16$\pm$3 & & 5 & -- & -- & DSFG\\
	SMG2  & 4.429 & 2.9$\pm$0.2 & 7.9$\pm$0.3 & &  5 & -- & -- & DSFG\\
	ALESS65.1 & 4.445 & 3.2$\pm$0.4  & 2.0$\pm$0.4  & -- & 4 & -- & -- &  DSFG \\ 
	SPT0441-46 & 4.477 & 24$\pm$6 & 40.33$\pm$2.18 & -- & 1 & 1.09$\times$L[42.5-500] & 12.7$\pm$1  & DSFG\\
	SPT2146-55 & 4.567 & 22$\pm$5 & 29.46$\pm$3.27 & -- & 1 & 1.09$\times$L[42.5-500] & 6.6$\pm$0.4$^{1}$ & DSFG\\ 
	BR1202-0725N & 4.691 & 10.0$\pm$1.5 & 12.86$\pm$2.14 & -- & 6 & 1.07$\times$L[40-500] & --  & DSFG \\ 
	SPT2132-58 & 4.768 & 21$\pm$4 & 33.96$\pm$3.29 & -- & 1 & 1.10$\times$L[42.5-500] & 5.7$\pm$0.5 & DSFG\\ 	
	\multicolumn{8}{l}{\underline{$5<z<6$}}\\	
	HZ8 & 5.148 & 0.26$^{+0.13}_{-0.09}$ & -- & 18$^{+5}_{-2}$ &  7 & -- & -- & UV \\
	HDF850.1 & 5.185 & 11$\pm$2.2 & 8.7$\pm$1.0 & -- & 8 & -- & 1.6$\pm$0.1 & DSFG\\ 
	HLSJ091828.6+514223 & 5.243 & 85$\pm$2 & 160$\pm$10 & -- & 9 & -- & 8.9$\pm$1.9 & DSFG\\
	HZ7 & 5.250 & 0.32$^{+0.41}_{-0.23}$ & -- & 21$^{+5}_{-2}$ &  7 & --  & -- & UV \\
	HZ6 & 5.290 & 1.41$^{+0.68}_{-0.46}$ & 0.081$^{+0.019}_{-0.063}$ & 49$^{+44}_{-12}$ & 7 & -- & -- & UV\\
	SPT2319-55 & 5.293 & 14$\pm$2 & 27.69$\pm$2.22 & -- & 1 & 1.11$\times$L[42.5-500] & 13.9$\pm$1.8$^{2}$ & DSFG\\  
	AzTEC-3 & 5.299 & 6.69$\pm$0.23 & 17.34$^{+3.47}_{-3.31}$ & -- & 10 & 1.67$\times$L[42.5-122.5]& -- & DSFG\\
	HZ4 & 5.540 & 0.95$^{+0.63}_{-0.38}$ & 0.135$^{+0.333}_{-0.096}$ & 51$^{+54}_{-18}$ & 7 & -- & -- & UV\\
	HZ3 & 5.546 & 0.47$^{+0.42}_{-0.21}$ & -- & 18$^{+8}_{-3}$ &  7 & --  & -- & UV\\
	HZ9 & 5.548 & 1.62$^{+0.37}_{-0.30}$ &  0.347$^{+0.190}_{-0.123}$ & 67$^{+30}_{-20}$ & 7 & --  & -- & UV\\
	SPT0346-52 & 5.656 & 50$\pm$7 & 	137.5$\pm$5.6 & -- & 1 & 1.12$\times$L[42.5-500] & 5.6$\pm$0.1  & DSFG\\ 
	HZ10 & 5.659 & 1.35$^{+0.47}_{-0.35}$ &  0.871$^{+0.176}_{-0.147}$ & 169$^{+32}_{-27}$ &  7 & --  & -- & UV\\
	HZ2 & 5.670 & 0.36$^{+0.57}_{-0.22}$ & -- & 25$^{+5}_{-2}$ &  7 & -- & -- & UV\\
	HZ1 & 5.690 & 0.25$^{+0.27}_{-0.13}$ & -- & 24$^{+6}_{-3}$ &  7 & --  & -- & UV\\
	\multicolumn{8}{l}{\underline{$6<z<7$}}\\
	ID52 & 6.018 & 0.40$\pm$0.06 & -- & 0.1  & 11 & -- & -- & Blind\\ 
	ID09 & 6.024 & 0.30$\pm$0.07 &  -- & 0.3 & 11 & -- & -- & Blind\\ 
	A383-5.1 & 6.028 & 0.0083 & --  & 3.2 & 12 & -- & 11.4 & UV \\
	ID49 & 6.051 & 0.24$\pm$0.06 & -- & 0.1  & 11 & -- & -- & Blind\\ 
	SDSS J0842+1218 Comp & 6.066  & 1.87$\pm$0.24 & 0.9$\pm$0.3 & 140$\pm$50 & 14 & -- & -- & Blind\\
	WMH5 & 6.070 & 0.66$\pm$0.07 & 0.200 $\pm$0.038 & 43$\pm$5 & 15 & 1.60$\times$L[42.5-122.5]& -- & UV \\
	CFHQ J2100-1715 Comp & 6.080 & 2.45$\pm$0.42 & 5.4$\pm$0.7 & 360$\pm$70 & 14 & -- & -- & Blind\\
	CLM1 & 6.166 & 0.24$\pm$0.03 & 0.040$\pm$0.024 & 37$\pm$4 & 15 & 1.59$\times$L[42.5-122.5] & -- & UV \\
	PSO J308-21 Comp EC & 6.249 & 0.66$\pm$0.13 & 0.52$\pm$0.17 & 77$\pm$26 &  14 & -- & -- & Blind\\
	HFLS3 &  6.337 & 15.5$\pm$3.2 & 45.1$\pm$5.0 & -- & 16 & 1.58$\times$L[42.5-122.5]  & 1.24$^{+0.14}_{-0.11}$ & DSFG \\
	ID41 & 6.346 & 0.39$\pm$0.09 & -- & 0.4  & 11 & -- & -- & Blind\\ 
	PSO J231-20 Comp & 6.590 & 4.47$\pm$0.53 & 5.1$\pm$0.5 &730$\pm$100 & 14 & -- & -- & Blind\\
	ID38 & 6.593 & 0.33$\pm$0.08 & -- & 0.2 & 11 & -- & -- & Blind\\ 
	CR7 & 6.600 & 0.20$\pm$0.043 & -- & 45$\pm$2 & 17 & -- & -- & UV\\
	COSMOS24108 & 6.629 & 0.101 & -- & 29 & 18 & --  & -- & UV\\
	UDS16291 & 6.638 & 0.069 & --  & 15.8 & 18 & -- & -- & UV\\
	NTTDF6345 & 6.701 & 0.178 & -- & 25 & 18 & -- & -- & UV\\
	ID14 & 6.751 & 0.31$\pm$0.07 & -- & 0.7 & 11 & -- & -- & Blind\\ 
	RX J1347-1145 & 6.766 & 0.015$^{+0.002}_{-0.004}$ & -- & 8.5$^{+5.8}_{-1.0}$ & 19 & -- & 5$\pm$0.3 & UV\\
	COS-2987030247 & 6.808 & 0.36$\pm$0.05 & -- & 22.7$\pm$2 & 13 & -- & -- & Optical \\
	COS-3018555981 & 6.854 & 0.47$\pm$0.05 & -- & 19.2$\pm$1.6 & 13 & -- & -- & Optical \\	
	ID30 &  6.854 & 0.70$\pm$0.11 &  -- & 4.0 & 11  & -- & -- & Blind\\ 
	ID04 & 6.867 & 0.92$\pm$ 0.11 &  -- & 0.4 & 11 & -- & -- & Blind\\ 
	SPT0311-58 & 6.900 & 29.81$\pm$0.75 & -- & 4100$\pm$700 & 20 & -- & 1.9 & DSFG\\ 
	\multicolumn{8}{l}{\underline{$z>7$}}\\
	COSMOS13679 & 7.145 & 0.076 & -- & 23.9 & 17 & -- & -- & UV\\
	ID44 & 7.360 & 0.44$\pm$0.11 & -- & 1.2  & 11 & -- & -- & Blind\\ 
	ID31 & 7.494 & 0.33$\pm$0.08 & -- & 12.4  & 11 & -- & -- & Blind\\ 
	ID27 &7.575 & 0.29$\pm$0.07 & -- & 10.5 & 11 & -- & -- & Blind\\ 
	A1689-zD1 & 7.603 & 0.17 & -- & 12$^{+4}_{-3}$ & 21 & -- & 9.5 & UV \\
	ID02 & 7.914 & 0.92$\pm$0.18 & -- & 0.6 & 11 & -- & -- & Blind \\
    \end{tabular}
         
   \tablebib{(1)~\citet{gullberg15}; (2) \citet{cox11}; (3) \citet{neeleman17};  (4) \citet{swinbank12}; (5) \citet{oteo16}; 
   (6) \citet{wagg12}; (7) \citet{capak15};  (8) \citet{walter12}; (9) \citet{rawle14};(10) \citet{riechers14}; (11) \citet{aravena16}; 
   (12) \citet{knudsen16};   (13) \citet{smit17}; (14) \citet{decarli17}; (15)  \citet{willott15}; (16) \citet{riechers13} ;
   (17) \cite{matthee17} ; (18) \cite{pentericci16} ; (19) \cite{bradac17};  (20) \citet{strandet17}; (21) \cite{knudsen17}.}
   \tablefoot{\tablefoottext{1}{Lensing for SPT sources are from \cite{spilker16}.}
   \tablefoottext{2}{We consider the integrated [CII] luminosity to be dominated by comp. B.}} 

   \end{table*}

\section{CMB effect on the L$_{\mathrm{[CII]}}$--SFR relation}

\begin{figure*}[h!]
  \centering
  \includegraphics[scale=0.45]{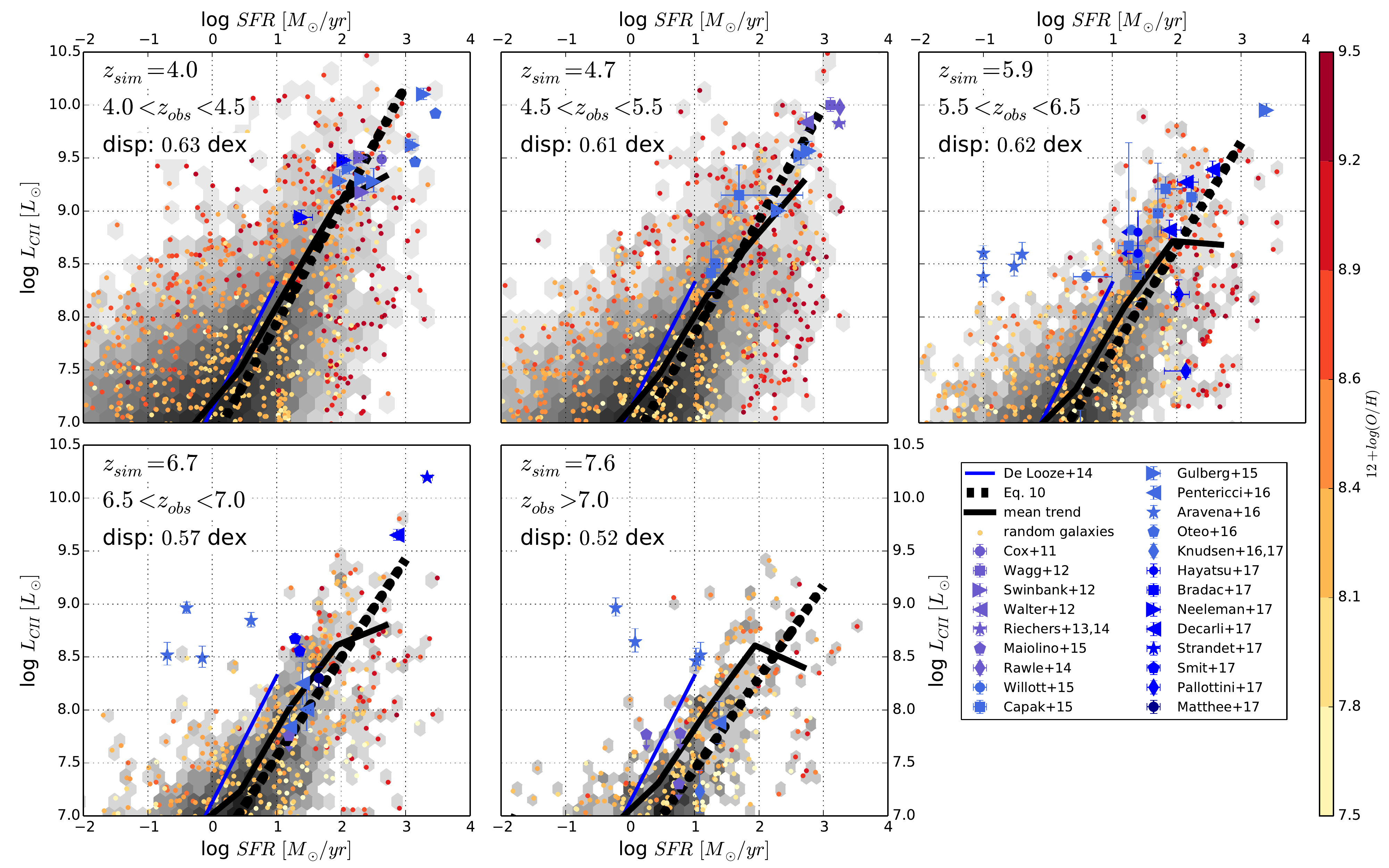}
  \caption{Same as Fig.\,\ref{LCII_grid_Cloudy} but without taking into account CMB effects (both heating and attenuation). \label{SFR_LCII_wocmb}}
\end{figure*}

We show on Fig.\,\ref{SFR_LCII_wocmb} the L$_{\mathrm{[CII]}}$--SFR  relation obtained when ignoring the heating and attenuation of the [CII] line emission by the CMB. 

\bibliographystyle{aa} 
\bibliography{biblio.bib} 

\begin{thebibliography}{125}
\expandafter\ifx\csname natexlab\endcsname\relax\def\natexlab#1{#1}\fi

\bibitem[{{Appleton} {et~al.}(2013){Appleton}, {Guillard}, {Boulanger},
  {Cluver}, {Ogle}, {Falgarone}, {Pineau des For{\^e}ts}, {O'Sullivan}, {Duc},
  {Gallagher}, {Gao}, {Jarrett}, {Konstantopoulos}, {Lisenfeld}, {Lord}, {Lu},
  {Peterson}, {Struck}, {Sturm}, {Tuffs}, {Valchanov}, {van der Werf}, \&
  {Xu}}]{appleton13}
{Appleton}, P.~N., {Guillard}, P., {Boulanger}, F., {et~al.} 2013, \apj, 777,
  66

\bibitem[{{Aravena} {et~al.}(2016){Aravena}, {Decarli}, {Walter}, {Bouwens},
  {Oesch}, {Carilli}, {Bauer}, {Da Cunha}, {Daddi}, {G{\'o}nzalez-L{\'o}pez},
  {Ivison}, {Riechers}, {Smail}, {Swinbank}, {Weiss}, {Anguita}, {Bacon},
  {Bell}, {Bertoldi}, {Cortes}, {Cox}, {Hodge}, {Ibar}, {Inami}, {Infante},
  {Karim}, {Magnelli}, {Ota}, {Popping}, {van der Werf}, {Wagg}, \&
  {Fudamoto}}]{aravena16}
{Aravena}, M., {Decarli}, R., {Walter}, F., {et~al.} 2016, \apj, 833, 71

\bibitem[{{Barinovs} {et~al.}(2005){Barinovs}, {van Hemert}, {Krems}, \&
  {Dalgarno}}]{barinovs05}
{Barinovs}, {\u G}., {van Hemert}, M.~C., {Krems}, R., \& {Dalgarno}, A. 2005,
  \apj, 620, 537

\bibitem[{{Barisic} {et~al.}(2017){Barisic}, {Faisst}, {Capak}, {Pavesi},
  {Riechers}, {Scoville}, {Cooke}, {Kartaltepe}, {Casey}, \&
  {Smolcic}}]{barisic17}
{Barisic}, I., {Faisst}, A.~L., {Capak}, P.~L., {et~al.} 2017, \apj, 845, 41

\bibitem[{{Benson} \& {Bower}(2011)}]{benson2011}
{Benson}, A.~J. \& {Bower}, R. 2011, \mnras, 410, 2653

\bibitem[{{B{\'e}thermin} {et~al.}(2015){B{\'e}thermin}, {Daddi}, {Magdis},
  {Lagos}, {Sargent}, {Albrecht}, {Aussel}, {Bertoldi}, {Buat}, {Galametz},
  {Heinis}, {Ilbert}, {Karim}, {Koekemoer}, {Lacey}, {Le Floc'h}, {Navarrete},
  {Pannella}, {Schreiber}, {Smol{\v c}i{\'c}}, {Symeonidis}, \&
  {Viero}}]{bethermin15}
{B{\'e}thermin}, M., {Daddi}, E., {Magdis}, G., {et~al.} 2015, \aap, 573, A113

\bibitem[{{Black}(1987)}]{black87}
{Black}, J.~H. 1987, in Astrophysics and Space Science Library, Vol. 134,
  Interstellar Processes, ed. D.~J. {Hollenbach} \& H.~A. {Thronson}, Jr.,
  731--744

\bibitem[{{Black} \& {Dalgarno}(1977)}]{black77}
{Black}, J.~H. \& {Dalgarno}, A. 1977, \apjs, 34, 405

\bibitem[{{Bouwens} {et~al.}(2015){Bouwens}, {Illingworth}, {Oesch}, {Trenti},
  {Labb{\'e}}, {Bradley}, {Carollo}, {van Dokkum}, {Gonzalez}, {Holwerda},
  {Franx}, {Spitler}, {Smit}, \& {Magee}}]{bouwens15}
{Bouwens}, R.~J., {Illingworth}, G.~D., {Oesch}, P.~A., {et~al.} 2015, \apj,
  803, 34

\bibitem[{{Bowler} {et~al.}(2015){Bowler}, {Dunlop}, {McLure}, {McCracken},
  {Milvang-Jensen}, {Furusawa}, {Taniguchi}, {Le F{\`e}vre}, {Fynbo}, {Jarvis},
  \& {H{\"a}u{\ss}ler}}]{bowler15}
{Bowler}, R.~A.~A., {Dunlop}, J.~S., {McLure}, R.~J., {et~al.} 2015, \mnras,
  452, 1817

\bibitem[{{Brada{\v c}} {et~al.}(2017){Brada{\v c}}, {Garcia-Appadoo}, {Huang},
  {Vallini}, {Quinn Finney}, {Hoag}, {Lemaux}, {Borello Schmidt}, {Treu},
  {Carilli}, {Dijkstra}, {Ferrara}, {Fontana}, {Jones}, {Ryan}, {Wagg}, \&
  {Gonzalez}}]{bradac17}
{Brada{\v c}}, M., {Garcia-Appadoo}, D., {Huang}, K.-H., {et~al.} 2017, \apjl,
  836, L2

\bibitem[{{Bruzual} \& {Charlot}(2003)}]{Bruzual_2003}
{Bruzual}, G. \& {Charlot}, S. 2003, \mnras, 344, 1000

\bibitem[{{Calzetti} {et~al.}(2000){Calzetti}, {Armus}, {Bohlin}, {Kinney},
  {Koornneef}, \& {Storchi-Bergmann}}]{calzetti_2000}
{Calzetti}, D., {Armus}, L., {Bohlin}, R.~C., {et~al.} 2000, \apj, 533, 682

\bibitem[{{Capak} {et~al.}(2015){Capak}, {Carilli}, {Jones}, {Casey},
  {Riechers}, {Sheth}, {Carollo}, {Ilbert}, {Karim}, {Lefevre}, {Lilly},
  {Scoville}, {Smolcic}, \& {Yan}}]{capak15}
{Capak}, P.~L., {Carilli}, C., {Jones}, G., {et~al.} 2015, \nat, 522, 455

\bibitem[{{Caputi} {et~al.}(2015){Caputi}, {Ilbert}, {Laigle}, {McCracken}, {Le
  F{\`e}vre}, {Fynbo}, {Milvang-Jensen}, {Capak}, {Salvato}, \&
  {Taniguchi}}]{caputi15}
{Caputi}, K.~I., {Ilbert}, O., {Laigle}, C., {et~al.} 2015, \apj, 810, 73

\bibitem[{{Caputi} {et~al.}(2007){Caputi}, {Lagache}, {Yan}, {Dole},
  {Bavouzet}, {Le Floc'h}, {Choi}, {Helou}, \& {Reddy}}]{caputi07}
{Caputi}, K.~I., {Lagache}, G., {Yan}, L., {et~al.} 2007, \apj, 660, 97

\bibitem[{{Carlstrom} \& {Kronberg}(1991)}]{carlstrom91}
{Carlstrom}, J.~E. \& {Kronberg}, P.~P. 1991, \apj, 366, 422

\bibitem[{{Carniani} {et~al.}(2017){Carniani}, {Maiolino}, {Pallottini},
  {Vallini}, {Pentericci}, {Ferrara}, {Castellano}, {Vanzella}, {Grazian},
  {Gallerani}, {Santini}, {Wagg}, \& {Fontana}}]{carniani17}
{Carniani}, S., {Maiolino}, R., {Pallottini}, A., {et~al.} 2017, ArXiv e-prints

\bibitem[{{Casey} {et~al.}(2014){Casey}, {Narayanan}, \& {Cooray}}]{casey14}
{Casey}, C.~M., {Narayanan}, D., \& {Cooray}, A. 2014, \physrep, 541, 45

\bibitem[{{Chabrier}(2003)}]{chabrier2003}
{Chabrier}, G. 2003, \apjl, 586, L133

\bibitem[{{Charlot} \& {Fall}(2000)}]{charlot00}
{Charlot}, S. \& {Fall}, S.~M. 2000, \apj, 539, 718

\bibitem[{{Chatzikos} {et~al.}(2014){Chatzikos}, {Ferland}, {Williams}, \&
  {Fabian}}]{chatzikos14}
{Chatzikos}, M., {Ferland}, G.~J., {Williams}, R.~J.~R., \& {Fabian}, A.~C.
  2014, \apj, 787, 96

\bibitem[{{Chatzikos} {et~al.}(2013){Chatzikos}, {Ferland}, {Williams},
  {Porter}, \& {van Hoof}}]{chatzikos13}
{Chatzikos}, M., {Ferland}, G.~J., {Williams}, R.~J.~R., {Porter}, R., \& {van
  Hoof}, P.~A.~M. 2013, \apj, 779, 122

\bibitem[{{Compi{\`e}gne} {et~al.}(2011){Compi{\`e}gne}, {Verstraete}, {Jones},
  {Bernard}, {Boulanger}, {Flagey}, {Le Bourlot}, {Paradis}, \&
  {Ysard}}]{compiegne_2011}
{Compi{\`e}gne}, M., {Verstraete}, L., {Jones}, A., {et~al.} 2011, \aap, 525,
  A103

\bibitem[{{Cooksy} {et~al.}(1986){Cooksy}, {Saykally}, {Brown}, \&
  {Evenson}}]{cooksy86}
{Cooksy}, A.~L., {Saykally}, R.~J., {Brown}, J.~M., \& {Evenson}, K.~M. 1986,
  \apj, 309, 828

\bibitem[{{Cormier} {et~al.}(2015){Cormier}, {Madden}, {Lebouteiller}, {Abel},
  {Hony}, {Galliano}, {R{\'e}my-Ruyer}, {Bigiel}, {Baes}, {Boselli},
  {Chevance}, {Cooray}, {De Looze}, {Doublier}, {Galametz}, {Hughes},
  {Karczewski}, {Lee}, {Lu}, \& {Spinoglio}}]{cormier15}
{Cormier}, D., {Madden}, S.~C., {Lebouteiller}, V., {et~al.} 2015, \aap, 578,
  A53

\bibitem[{{Cousin} {et~al.}(2016){Cousin}, {Buat}, {Boissier}, {Bethermin},
  {Roehlly}, \& {G{\'e}nois}}]{cousin16}
{Cousin}, M., {Buat}, V., {Boissier}, S., {et~al.} 2016, \aap, 589, A109

\bibitem[{{Cousin} {et~al.}(to be submitted){Cousin}, {Guillard}, {Lagache},
  {toto}, {titi}, \& {tata}}]{cousin17a}
{Cousin}, M., {Guillard}, P., {Lagache}, G., {et~al.} to be submitted, \aap

\bibitem[{{Cousin} {et~al.}(2015{\natexlab{a}}){Cousin}, {Lagache},
  {Bethermin}, {Blaizot}, \& {Guiderdoni}}]{cousin15b}
{Cousin}, M., {Lagache}, G., {Bethermin}, M., {Blaizot}, J., \& {Guiderdoni},
  B. 2015{\natexlab{a}}, \aap, 575, A32

\bibitem[{{Cousin} {et~al.}(2015{\natexlab{b}}){Cousin}, {Lagache},
  {Bethermin}, \& {Guiderdoni}}]{cousin15a}
{Cousin}, M., {Lagache}, G., {Bethermin}, M., \& {Guiderdoni}, B.
  2015{\natexlab{b}}, \aap, 575, A33

\bibitem[{{Cox} {et~al.}(2011){Cox}, {Krips}, {Neri}, {Omont}, {G{\"u}sten},
  {Menten}, {Wyrowski}, {Wei{\ss}}, {Beelen}, {Gurwell}, {Dannerbauer},
  {Ivison}, {Negrello}, {Aretxaga}, {Hughes}, {Auld}, {Baes}, {Blundell},
  {Buttiglione}, {Cava}, {Cooray}, {Dariush}, {Dunne}, {Dye}, {Eales},
  {Frayer}, {Fritz}, {Gavazzi}, {Hopwood}, {Ibar}, {Jarvis}, {Maddox},
  {Micha{\l}owski}, {Pascale}, {Pohlen}, {Rigby}, {Smith}, {Swinbank}, {Temi},
  {Valtchanov}, {van der Werf}, \& {de Zotti}}]{cox11}
{Cox}, P., {Krips}, M., {Neri}, R., {et~al.} 2011, \apj, 740, 63

\bibitem[{{Croxall} {et~al.}(2017){Croxall}, {Smith}, {Pellegrini}, {Groves},
  {Bolatto}, {Herrera-Camus}, {Sandstrom}, {Draine}, {Wolfire}, {Armus},
  {Boquien}, {Brandl}, {Dale}, {Galametz}, {Hunt}, {Kennicutt}, {Kreckel},
  {Rigopoulou}, {van der Werf}, \& {Wilson}}]{croxall17}
{Croxall}, K., {Smith}, J.~D.~T., {Pellegrini}, E., {et~al.} 2017, ArXiv
  e-prints

\bibitem[{{Davidzon} {et~al.}(2017){Davidzon}, {Ilbert}, {Laigle}, {Coupon},
  {McCracken}, {Delvecchio}, {Masters}, {Capak}, {Hsieh}, {Le F{\`e}vre},
  {Tresse}, {Bethermin}, {Chang}, {Faisst}, {Le Floc'h}, {Steinhardt}, {Toft},
  {Aussel}, {Dubois}, {Hasinger}, {Salvato}, {Sanders}, {Scoville}, \&
  {Silverman}}]{davidzon17}
{Davidzon}, I., {Ilbert}, O., {Laigle}, C., {et~al.} 2017, \aap, 605, A70

\bibitem[{{De Looze} {et~al.}(2014){De Looze}, {Cormier}, {Lebouteiller},
  {Madden}, {Baes}, {Bendo}, {Boquien}, {Boselli}, {Clements}, {Cortese},
  {Cooray}, {Galametz}, {Galliano}, {Graci{\'a}-Carpio}, {Isaak}, {Karczewski},
  {Parkin}, {Pellegrini}, {R{\'e}my-Ruyer}, {Spinoglio}, {Smith}, \&
  {Sturm}}]{delooze14}
{De Looze}, I., {Cormier}, D., {Lebouteiller}, V., {et~al.} 2014, \aap, 568,
  A62

\bibitem[{{Decarli} {et~al.}(2014){Decarli}, {Walter}, {Carilli}, {Bertoldi},
  {Cox}, {Ferkinhoff}, {Groves}, {Maiolino}, {Neri}, {Riechers}, \&
  {Weiss}}]{decarli14}
{Decarli}, R., {Walter}, F., {Carilli}, C., {et~al.} 2014, \apjl, 782, L17

\bibitem[{{Decarli} {et~al.}(2017){Decarli}, {Walter}, {Venemans},
  {Ba{\~n}ados}, {Bertoldi}, {Carilli}, {Fan}, {Farina}, {Mazzucchelli},
  {Riechers}, {Rix}, {Strauss}, {Wang}, \& {Yang}}]{decarli17}
{Decarli}, R., {Walter}, F., {Venemans}, B.~P., {et~al.} 2017, \nat, 545, 457

\bibitem[{{Devriendt} {et~al.}(1999){Devriendt}, {Guiderdoni}, \&
  {Sadat}}]{devriendt99}
{Devriendt}, J.~E.~G., {Guiderdoni}, B., \& {Sadat}, R. 1999, \aap, 350, 381

\bibitem[{{D{\'{\i}}az-Santos} {et~al.}(2017){D{\'{\i}}az-Santos}, {Armus},
  {Charmandaris}, {Lu}, {Stierwalt}, {Stacey}, {Malhotra}, {van der Werf},
  {Howell}, {Privon}, {Mazzarella}, {Goldsmith}, {Murphy}, {Barcos-Mu{\~n}oz},
  {Linden}, {Inami}, {Larson}, {Evans}, {Appleton}, {Iwasawa}, {Lord},
  {Sanders}, \& {Surace}}]{diaz17}
{D{\'{\i}}az-Santos}, T., {Armus}, L., {Charmandaris}, V., {et~al.} 2017, \apj,
  846, 32

\bibitem[{{D{\'{\i}}az-Santos} {et~al.}(2013){D{\'{\i}}az-Santos}, {Armus},
  {Charmandaris}, {Stierwalt}, {Murphy}, {Haan}, {Inami}, {Malhotra},
  {Meijerink}, {Stacey}, {Petric}, {Evans}, {Veilleux}, {van der Werf}, {Lord},
  {Lu}, {Howell}, {Appleton}, {Mazzarella}, {Surace}, {Xu}, {Schulz},
  {Sanders}, {Bridge}, {Chan}, {Frayer}, {Iwasawa}, {Melbourne}, \&
  {Sturm}}]{diaz13}
{D{\'{\i}}az-Santos}, T., {Armus}, L., {Charmandaris}, V., {et~al.} 2013, \apj,
  774, 68

\bibitem[{{Dole} {et~al.}(2006){Dole}, {Lagache}, {Puget}, {Caputi},
  {Fern{\'a}ndez-Conde}, {Le Floc'h}, {Papovich}, {P{\'e}rez-Gonz{\'a}lez},
  {Rieke}, \& {Blaylock}}]{dole06}
{Dole}, H., {Lagache}, G., {Puget}, J.-L., {et~al.} 2006, \aap, 451, 417

\bibitem[{{Duncan} {et~al.}(2014){Duncan}, {Conselice}, {Mortlock}, {Hartley},
  {Guo}, {Ferguson}, {Dav{\'e}}, {Lu}, {Ownsworth}, {Ashby}, {Dekel},
  {Dickinson}, {Faber}, {Giavalisco}, {Grogin}, {Kocevski}, {Koekemoer},
  {Somerville}, \& {White}}]{duncan14}
{Duncan}, K., {Conselice}, C.~J., {Mortlock}, A., {et~al.} 2014, \mnras, 444,
  2960

\bibitem[{{Ferland} {et~al.}(2017){Ferland}, {Chatzikos}, {Guzm{\'a}n},
  {Lykins}, {van Hoof}, {Williams}, {Abel}, {Badnell}, {Keenan}, {Porter}, \&
  {Stancil}}]{C17}
{Ferland}, G.~J., {Chatzikos}, M., {Guzm{\'a}n}, F., {et~al.} 2017, ArXiv
  e-prints

\bibitem[{{Ferland} {et~al.}(2013){Ferland}, {Porter}, {van Hoof}, {Williams},
  {Abel}, {Lykins}, {Shaw}, {Henney}, \& {Stancil}}]{ferland13}
{Ferland}, G.~J., {Porter}, R.~L., {van Hoof}, P.~A.~M., {et~al.} 2013, \rmxaa,
  49, 137

\bibitem[{{Finkelstein} {et~al.}(2015){Finkelstein}, {Ryan}, {Papovich},
  {Dickinson}, {Song}, {Somerville}, {Ferguson}, {Salmon}, {Giavalisco},
  {Koekemoer}, {Ashby}, {Behroozi}, {Castellano}, {Dunlop}, {Faber}, {Fazio},
  {Fontana}, {Grogin}, {Hathi}, {Jaacks}, {Kocevski}, {Livermore}, {McLure},
  {Merlin}, {Mobasher}, {Newman}, {Rafelski}, {Tilvi}, \&
  {Willner}}]{finkelstein15}
{Finkelstein}, S.~L., {Ryan}, Jr., R.~E., {Papovich}, C., {et~al.} 2015, \apj,
  810, 71

\bibitem[{{Giallongo} {et~al.}(2015){Giallongo}, {Grazian}, {Fiore}, {Fontana},
  {Pentericci}, {Vanzella}, {Dickinson}, {Kocevski}, {Castellano}, {Cristiani},
  {Ferguson}, {Finkelstein}, {Grogin}, {Hathi}, {Koekemoer}, {Newman}, \&
  {Salvato}}]{giallongo15}
{Giallongo}, E., {Grazian}, A., {Fiore}, F., {et~al.} 2015, \aap, 578, A83

\bibitem[{{Glassgold} \& {Langer}(1975)}]{glassgold75}
{Glassgold}, A.~E. \& {Langer}, W.~D. 1975, \apj, 197, 347

\bibitem[{{Gnedin}(2000)}]{gnedin_2000}
{Gnedin}, N.~Y. 2000, \apj, 542, 535

\bibitem[{{Goldsmith} {et~al.}(2012){Goldsmith}, {Langer}, {Pineda}, \&
  {Velusamy}}]{goldsmith12}
{Goldsmith}, P.~F., {Langer}, W.~D., {Pineda}, J.~L., \& {Velusamy}, T. 2012,
  \apjs, 203, 13

\bibitem[{{Gong} {et~al.}(2012){Gong}, {Cooray}, {Silva}, {Santos}, {Bock},
  {Bradford}, \& {Zemcov}}]{gong12}
{Gong}, Y., {Cooray}, A., {Silva}, M., {et~al.} 2012, \apj, 745, 49

\bibitem[{{Gonz{\'a}lez-L{\'o}pez} {et~al.}(2014){Gonz{\'a}lez-L{\'o}pez},
  {Riechers}, {Decarli}, {Walter}, {Vallini}, {Neri}, {Bertoldi}, {Bolatto},
  {Carilli}, {Cox}, {da Cunha}, {Ferrara}, {Gallerani}, \&
  {Infante}}]{gonzalez14}
{Gonz{\'a}lez-L{\'o}pez}, J., {Riechers}, D.~A., {Decarli}, R., {et~al.} 2014,
  \apj, 784, 99

\bibitem[{{Grazian} {et~al.}(2015){Grazian}, {Fontana}, {Santini}, {Dunlop},
  {Ferguson}, {Castellano}, {Amorin}, {Ashby}, {Barro}, {Behroozi}, {Boutsia},
  {Caputi}, {Chary}, {Dekel}, {Dickinson}, {Faber}, {Fazio}, {Finkelstein},
  {Galametz}, {Giallongo}, {Giavalisco}, {Grogin}, {Guo}, {Kocevski},
  {Koekemoer}, {Koo}, {Lee}, {Lu}, {Merlin}, {Mobasher}, {Nonino}, {Papovich},
  {Paris}, {Pentericci}, {Reddy}, {Renzini}, {Salmon}, {Salvato}, {Sommariva},
  {Song}, \& {Vanzella}}]{grazian15}
{Grazian}, A., {Fontana}, A., {Santini}, P., {et~al.} 2015, \aap, 575, A96

\bibitem[{{Gruppioni} {et~al.}(2013){Gruppioni}, {Pozzi}, {Rodighiero},
  {Delvecchio}, {Berta}, {Pozzetti}, {Zamorani}, {Andreani}, {Cimatti},
  {Ilbert}, {Le Floc'h}, {Lutz}, {Magnelli}, {Marchetti}, {Monaco}, {Nordon},
  {Oliver}, {Popesso}, {Riguccini}, {Roseboom}, {Rosario}, {Sargent},
  {Vaccari}, {Altieri}, {Aussel}, {Bongiovanni}, {Cepa}, {Daddi},
  {Dom{\'{\i}}nguez-S{\'a}nchez}, {Elbaz}, {F{\"o}rster Schreiber}, {Genzel},
  {Iribarrem}, {Magliocchetti}, {Maiolino}, {Poglitsch}, {P{\'e}rez
  Garc{\'{\i}}a}, {Sanchez-Portal}, {Sturm}, {Tacconi}, {Valtchanov},
  {Amblard}, {Arumugam}, {Bethermin}, {Bock}, {Boselli}, {Buat}, {Burgarella},
  {Castro-Rodr{\'{\i}}guez}, {Cava}, {Chanial}, {Clements}, {Conley}, {Cooray},
  {Dowell}, {Dwek}, {Eales}, {Franceschini}, {Glenn}, {Griffin},
  {Hatziminaoglou}, {Ibar}, {Isaak}, {Ivison}, {Lagache}, {Levenson}, {Lu},
  {Madden}, {Maffei}, {Mainetti}, {Nguyen}, {O'Halloran}, {Page}, {Panuzzo},
  {Papageorgiou}, {Pearson}, {P{\'e}rez-Fournon}, {Pohlen}, {Rigopoulou},
  {Rowan-Robinson}, {Schulz}, {Scott}, {Seymour}, {Shupe}, {Smith}, {Stevens},
  {Symeonidis}, {Trichas}, {Tugwell}, {Vigroux}, {Wang}, {Wright}, {Xu},
  {Zemcov}, {Bardelli}, {Carollo}, {Contini}, {Le F{\'e}vre}, {Lilly},
  {Mainieri}, {Renzini}, {Scodeggio}, \& {Zucca}}]{gruppioni13}
{Gruppioni}, C., {Pozzi}, F., {Rodighiero}, G., {et~al.} 2013, \mnras, 432, 23

\bibitem[{{Guiderdoni} \& {Rocca-Volmerange}(1987)}]{guiderdoni87}
{Guiderdoni}, B. \& {Rocca-Volmerange}, B. 1987, \aap, 186, 1

\bibitem[{{Gullberg} {et~al.}(2015){Gullberg}, {De Breuck}, {Vieira},
  {Wei{\ss}}, {Aguirre}, {Aravena}, {B{\'e}thermin}, {Bradford}, {Bothwell},
  {Carlstrom}, {Chapman}, {Fassnacht}, {Gonzalez}, {Greve}, {Hezaveh},
  {Holzapfel}, {Husband}, {Ma}, {Malkan}, {Marrone}, {Menten}, {Murphy},
  {Reichardt}, {Spilker}, {Stark}, {Strandet}, \& {Welikala}}]{gullberg15}
{Gullberg}, B., {De Breuck}, C., {Vieira}, J.~D., {et~al.} 2015, \mnras, 449,
  2883

\bibitem[{{Habing}(1968)}]{habing68}
{Habing}, H.~J. 1968, \bain, 19, 421

\bibitem[{{Hayatsu} {et~al.}(2017){Hayatsu}, {Matsuda}, {Umehata}, {Yoshida},
  {Smail}, {Swinbank}, {Ivison}, {Kohno}, {Tamura}, {Kubo}, {Iono},
  {Hatsukade}, {Nakanishi}, {Kawabe}, {Nagao}, {Inoue}, {Takeuchi}, {Lee},
  {Ao}, {Fujimoto}, {Izumi}, {Yamaguchi}, {Ikarashi}, \& {Yamada}}]{hayatsu17}
{Hayatsu}, N.~H., {Matsuda}, Y., {Umehata}, H., {et~al.} 2017, \pasj, 69, 45

\bibitem[{{Hemmati} {et~al.}(2017){Hemmati}, {Yan}, {Diaz-Santos}, {Armus},
  {Capak}, {Faisst}, \& {Masters}}]{hemmati17}
{Hemmati}, S., {Yan}, L., {Diaz-Santos}, T., {et~al.} 2017, \apj, 834, 36

\bibitem[{{Herrera-Camus} {et~al.}(2015){Herrera-Camus}, {Bolatto}, {Wolfire},
  {Smith}, {Croxall}, {Kennicutt}, {Calzetti}, {Helou}, {Walter}, {Leroy},
  {Draine}, {Brandl}, {Armus}, {Sandstrom}, {Dale}, {Aniano}, {Meidt},
  {Boquien}, {Hunt}, {Galametz}, {Tabatabaei}, {Murphy}, {Appleton}, {Roussel},
  {Engelbracht}, \& {Beirao}}]{herrera15}
{Herrera-Camus}, R., {Bolatto}, A.~D., {Wolfire}, M.~G., {et~al.} 2015, \apj,
  800, 1

\bibitem[{{Hollenbach} \& {Tielens}(1999)}]{HollenbachTielens1999}
{Hollenbach}, D.~J. \& {Tielens}, A.~G.~G.~M. 1999, Reviews of Modern Physics,
  71, 173

\bibitem[{{Hollenbach} {et~al.}(1971){Hollenbach}, {Werner}, \&
  {Salpeter}}]{hollenbach71}
{Hollenbach}, D.~J., {Werner}, M.~W., \& {Salpeter}, E.~E. 1971, \apj, 163, 165

\bibitem[{{Inoue} {et~al.}(2016){Inoue}, {Tamura}, {Matsuo}, {Mawatari},
  {Shimizu}, {Shibuya}, {Ota}, {Yoshida}, {Zackrisson}, {Kashikawa}, {Kohno},
  {Umehata}, {Hatsukade}, {Iye}, {Matsuda}, {Okamoto}, \&
  {Yamaguchi}}]{inoue16}
{Inoue}, A.~K., {Tamura}, Y., {Matsuo}, H., {et~al.} 2016, Science, 352, 1559

\bibitem[{{Joy} {et~al.}(1987){Joy}, {Lester}, \& {Harvey}}]{joy87}
{Joy}, M., {Lester}, D.~F., \& {Harvey}, P.~M. 1987, \apj, 319, 314

\bibitem[{{Jura}(1974)}]{jura74}
{Jura}, M. 1974, \apj, 191, 375

\bibitem[{{Karakas}(2010)}]{karakas2010}
{Karakas}, A.~I. 2010, \mnras, 403, 1413

\bibitem[{{Katz} {et~al.}(2016){Katz}, {Kimm}, {Sijacki}, \&
  {Haehnelt}}]{katz17}
{Katz}, H., {Kimm}, T., {Sijacki}, D., \& {Haehnelt}, M. 2016, ArXiv e-prints

\bibitem[{{Kaufman} {et~al.}(1999){Kaufman}, {Wolfire}, {Hollenbach}, \&
  {Luhman}}]{kaufman99}
{Kaufman}, M.~J., {Wolfire}, M.~G., {Hollenbach}, D.~J., \& {Luhman}, M.~L.
  1999, \apj, 527, 795

\bibitem[{{Kennicutt}(1998)}]{kennicutt98}
{Kennicutt}, Jr., R.~C. 1998, \araa, 36, 189

\bibitem[{{Khochfar} \& {Silk}(2009)}]{khochfar2009}
{Khochfar}, S. \& {Silk}, J. 2009, \apjl, 700, L21

\bibitem[{{Knudsen} {et~al.}(2016){Knudsen}, {Richard}, {Kneib}, {Jauzac},
  {Cl{\'e}ment}, {Drouart}, {Egami}, \& {Lindroos}}]{knudsen16}
{Knudsen}, K.~K., {Richard}, J., {Kneib}, J.-P., {et~al.} 2016, \mnras, 462, L6

\bibitem[{{Knudsen} {et~al.}(2017){Knudsen}, {Watson}, {Frayer}, {Christensen},
  {Gallazzi}, {Micha{\l}owski}, {Richard}, \& {Zavala}}]{knudsen17}
{Knudsen}, K.~K., {Watson}, D., {Frayer}, D., {et~al.} 2017, \mnras, 466, 138

\bibitem[{{Koprowski} {et~al.}(2017){Koprowski}, {Dunlop}, {Micha{\l}owski},
  {Coppin}, {Geach}, {McLure}, {Scott}, \& {van der Werf}}]{koprowski17}
{Koprowski}, M.~P., {Dunlop}, J.~S., {Micha{\l}owski}, M.~J., {et~al.} 2017,
  \mnras, 471, 4155

\bibitem[{{Kravtsov} {et~al.}(2004){Kravtsov}, {Gnedin}, \&
  {Klypin}}]{kravtsov_2004}
{Kravtsov}, A.~V., {Gnedin}, O.~Y., \& {Klypin}, A.~A. 2004, \apj, 609, 482

\bibitem[{{Lagache} {et~al.}(2005){Lagache}, {Puget}, \& {Dole}}]{lagache05}
{Lagache}, G., {Puget}, J.-L., \& {Dole}, H. 2005, \araa, 43, 727

\bibitem[{{Luhman} {et~al.}(1998){Luhman}, {Satyapal}, {Fischer}, {Wolfire},
  {Cox}, {Lord}, {Smith}, {Stacey}, \& {Unger}}]{luhman98}
{Luhman}, M.~L., {Satyapal}, S., {Fischer}, J., {et~al.} 1998, \apjl, 504, L11

\bibitem[{{Luhman} {et~al.}(2003){Luhman}, {Satyapal}, {Fischer}, {Wolfire},
  {Sturm}, {Dudley}, {Lutz}, \& {Genzel}}]{luhman03}
{Luhman}, M.~L., {Satyapal}, S., {Fischer}, J., {et~al.} 2003, \apj, 594, 758

\bibitem[{{Madau} \& {Dickinson}(2014)}]{madau14}
{Madau}, P. \& {Dickinson}, M. 2014, \araa, 52, 415

\bibitem[{{Magnelli} {et~al.}(2011){Magnelli}, {Elbaz}, {Chary}, {Dickinson},
  {Le Borgne}, {Frayer}, \& {Willmer}}]{magnelli11}
{Magnelli}, B., {Elbaz}, D., {Chary}, R.~R., {et~al.} 2011, \aap, 528, A35

\bibitem[{{Magnelli} {et~al.}(2013){Magnelli}, {Popesso}, {Berta}, {Pozzi},
  {Elbaz}, {Lutz}, {Dickinson}, {Altieri}, {Andreani}, {Aussel},
  {B{\'e}thermin}, {Bongiovanni}, {Cepa}, {Charmandaris}, {Chary}, {Cimatti},
  {Daddi}, {F{\"o}rster Schreiber}, {Genzel}, {Gruppioni}, {Harwit}, {Hwang},
  {Ivison}, {Magdis}, {Maiolino}, {Murphy}, {Nordon}, {Pannella}, {P{\'e}rez
  Garc{\'{\i}}a}, {Poglitsch}, {Rosario}, {Sanchez-Portal}, {Santini}, {Scott},
  {Sturm}, {Tacconi}, \& {Valtchanov}}]{magnelli13}
{Magnelli}, B., {Popesso}, P., {Berta}, S., {et~al.} 2013, \aap, 553, A132

\bibitem[{{Maiolino} {et~al.}(2015){Maiolino}, {Carniani}, {Fontana},
  {Vallini}, {Pentericci}, {Ferrara}, {Vanzella}, {Grazian}, {Gallerani},
  {Castellano}, {Cristiani}, {Brammer}, {Santini}, {Wagg}, \&
  {Williams}}]{maiolino15}
{Maiolino}, R., {Carniani}, S., {Fontana}, A., {et~al.} 2015, \mnras, 452, 54

\bibitem[{{Malhotra} {et~al.}(2001){Malhotra}, {Kaufman}, {Hollenbach},
  {Helou}, {Rubin}, {Brauher}, {Dale}, {Lu}, {Lord}, {Stacey}, {Contursi},
  {Hunter}, \& {Dinerstein}}]{malhotra01}
{Malhotra}, S., {Kaufman}, M.~J., {Hollenbach}, D., {et~al.} 2001, \apj, 561,
  766

\bibitem[{{Matsuda} {et~al.}(2015){Matsuda}, {Nagao}, {Iono}, {Hatsukade},
  {Kohno}, {Tamura}, {Yamaguchi}, \& {Shimizu}}]{matsuda15}
{Matsuda}, Y., {Nagao}, T., {Iono}, D., {et~al.} 2015, \mnras, 451, 1141

\bibitem[{{Matthee} {et~al.}(2017){Matthee}, {Sobral}, {Boone},
  {R{\"o}ttgering}, {Schaerer}, {Girard}, {Pallottini}, {Vallini}, {Ferrara},
  {Darvish}, \& {Mobasher}}]{matthee17}
{Matthee}, J., {Sobral}, D., {Boone}, F., {et~al.} 2017, ArXiv e-prints

\bibitem[{{Miller} {et~al.}(2016){Miller}, {Chapman}, {Hayward}, {Behroozi},
  {Bradford}, {Willott}, \& {Wagg}}]{miller17}
{Miller}, T.~B., {Chapman}, S.~C., {Hayward}, C.~C., {et~al.} 2016, ArXiv
  e-prints

\bibitem[{{Mu{\~n}oz} \& {Furlanetto}(2014)}]{munoz14}
{Mu{\~n}oz}, J.~A. \& {Furlanetto}, S.~R. 2014, \mnras, 438, 2483

\bibitem[{{Mu{\~n}oz} \& {Oh}(2016)}]{munoz16}
{Mu{\~n}oz}, J.~A. \& {Oh}, S.~P. 2016, \mnras, 463, 2085

\bibitem[{{Nagamine} {et~al.}(2006){Nagamine}, {Wolfe}, \&
  {Hernquist}}]{nagamine06}
{Nagamine}, K., {Wolfe}, A.~M., \& {Hernquist}, L. 2006, \apj, 647, 60

\bibitem[{{Neeleman} {et~al.}(2017){Neeleman}, {Kanekar}, {Prochaska},
  {Rafelski}, {Carilli}, \& {Wolfe}}]{neeleman17}
{Neeleman}, M., {Kanekar}, N., {Prochaska}, J.~X., {et~al.} 2017, Science, 355,
  1285

\bibitem[{{Neri} {et~al.}(2014){Neri}, {Downes}, {Cox}, \& {Walter}}]{neri14}
{Neri}, R., {Downes}, D., {Cox}, P., \& {Walter}, F. 2014, \aap, 562, A35

\bibitem[{{Okamoto} {et~al.}(2008){Okamoto}, {Gao}, \& {Theuns}}]{okamoto_2008}
{Okamoto}, T., {Gao}, L., \& {Theuns}, T. 2008, \mnras, 390, 920

\bibitem[{{Olsen} {et~al.}(2017){Olsen}, {Greve}, {Narayanan}, {Thompson},
  {Dav{\'e}}, {Rios}, \& {Stawinski}}]{olsen17}
{Olsen}, K.~P., {Greve}, T.~R., {Narayanan}, D., {et~al.} 2017, ArXiv e-prints

\bibitem[{{Olsen} {et~al.}(2015){Olsen}, {Greve}, {Narayanan}, {Thompson},
  {Toft}, \& {Brinch}}]{olsen15}
{Olsen}, K.~P., {Greve}, T.~R., {Narayanan}, D., {et~al.} 2015, ArXiv e-prints

\bibitem[{{Ota} {et~al.}(2014){Ota}, {Walter}, {Ohta}, {Hatsukade}, {Carilli},
  {da Cunha}, {Gonz{\'a}lez-L{\'o}pez}, {Decarli}, {Hodge}, {Nagai}, {Egami},
  {Jiang}, {Iye}, {Kashikawa}, {Riechers}, {Bertoldi}, {Cox}, {Neri}, \&
  {Weiss}}]{ota14}
{Ota}, K., {Walter}, F., {Ohta}, K., {et~al.} 2014, \apj, 792, 34

\bibitem[{{Oteo} {et~al.}(2016){Oteo}, {Ivison}, {Dunne}, {Smail}, {Swinbank},
  {Zhang}, {Lewis}, {Maddox}, {Riechers}, {Serjeant}, {Van der Werf}, {Biggs},
  {Bremer}, {Cigan}, {Clements}, {Cooray}, {Dannerbauer}, {Eales}, {Ibar},
  {Messias}, {Micha{\l}owski}, {P{\'e}rez-Fournon}, \& {van Kampen}}]{oteo16}
{Oteo}, I., {Ivison}, R.~J., {Dunne}, L., {et~al.} 2016, \apj, 827, 34

\bibitem[{{Pallottini} {et~al.}(2017{\natexlab{a}}){Pallottini}, {Ferrara},
  {Bovino}, {Vallini}, {Gallerani}, {Maiolino}, \& {Salvadori}}]{pallottini17b}
{Pallottini}, A., {Ferrara}, A., {Bovino}, S., {et~al.} 2017{\natexlab{a}},
  ArXiv e-prints

\bibitem[{{Pallottini} {et~al.}(2017{\natexlab{b}}){Pallottini}, {Ferrara},
  {Gallerani}, {Vallini}, {Maiolino}, \& {Salvadori}}]{pallottini17}
{Pallottini}, A., {Ferrara}, A., {Gallerani}, S., {et~al.} 2017{\natexlab{b}},
  \mnras, 465, 2540

\bibitem[{{Pentericci} {et~al.}(2016){Pentericci}, {Carniani}, {Castellano},
  {Fontana}, {Maiolino}, {Guaita}, {Vanzella}, {Grazian}, {Santini}, {Yan},
  {Cristiani}, {Conselice}, {Giavalisco}, {Hathi}, \&
  {Koekemoer}}]{pentericci16}
{Pentericci}, L., {Carniani}, S., {Castellano}, M., {et~al.} 2016, \apjl, 829,
  L11

\bibitem[{{Pineda} {et~al.}(2014){Pineda}, {Langer}, \& {Goldsmith}}]{pineda14}
{Pineda}, J.~L., {Langer}, W.~D., \& {Goldsmith}, P.~F. 2014, \aap, 570, A121

\bibitem[{{Pineda} {et~al.}(2013){Pineda}, {Langer}, {Velusamy}, \&
  {Goldsmith}}]{pineda13}
{Pineda}, J.~L., {Langer}, W.~D., {Velusamy}, T., \& {Goldsmith}, P.~F. 2013,
  \aap, 554, A103

\bibitem[{{Planck Collaboration} {et~al.}(2016){Planck Collaboration}, {Adam},
  {Aghanim}, {Ashdown}, {Aumont}, {Baccigalupi}, {Ballardini}, {Banday},
  {Barreiro}, {Bartolo}, {Basak}, {Battye}, {Benabed}, {Bernard}, {Bersanelli},
  {Bielewicz}, {Bock}, {Bonaldi}, {Bonavera}, {Bond}, {Borrill}, {Bouchet},
  {Boulanger}, {Bucher}, {Burigana}, {Calabrese}, {Cardoso}, {Carron},
  {Chiang}, {Colombo}, {Combet}, {Comis}, {Couchot}, {Coulais}, {Crill},
  {Curto}, {Cuttaia}, {Davis}, {de Bernardis}, {de Rosa}, {de Zotti},
  {Delabrouille}, {Di Valentino}, {Dickinson}, {Diego}, {Dor{\'e}}, {Douspis},
  {Ducout}, {Dupac}, {Elsner}, {En{\ss}lin}, {Eriksen}, {Falgarone}, {Fantaye},
  {Finelli}, {Forastieri}, {Frailis}, {Fraisse}, {Franceschi}, {Frolov},
  {Galeotta}, {Galli}, {Ganga}, {G{\'e}nova-Santos}, {Gerbino}, {Ghosh},
  {Gonz{\'a}lez-Nuevo}, {G{\'o}rski}, {Gruppuso}, {Gudmundsson}, {Hansen},
  {Helou}, {Henrot-Versill{\'e}}, {Herranz}, {Hivon}, {Huang}, {Ili{\'c}},
  {Jaffe}, {Jones}, {Keih{\"a}nen}, {Keskitalo}, {Kisner}, {Knox},
  {Krachmalnicoff}, {Kunz}, {Kurki-Suonio}, {Lagache}, {L{\"a}hteenm{\"a}ki},
  {Lamarre}, {Langer}, {Lasenby}, {Lattanzi}, {Lawrence}, {Le Jeune},
  {Levrier}, {Lewis}, {Liguori}, {Lilje}, {L{\'o}pez-Caniego}, {Ma},
  {Mac{\'{\i}}as-P{\'e}rez}, {Maggio}, {Mangilli}, {Maris}, {Martin},
  {Mart{\'{\i}}nez-Gonz{\'a}lez}, {Matarrese}, {Mauri}, {McEwen}, {Meinhold},
  {Melchiorri}, {Mennella}, {Migliaccio}, {Miville-Desch{\^e}nes}, {Molinari},
  {Moneti}, {Montier}, {Morgante}, {Moss}, {Naselsky}, {Natoli}, {Oxborrow},
  {Pagano}, {Paoletti}, {Partridge}, {Patanchon}, {Patrizii}, {Perdereau},
  {Perotto}, {Pettorino}, {Piacentini}, {Plaszczynski}, {Polastri}, {Polenta},
  {Puget}, {Rachen}, {Racine}, {Reinecke}, {Remazeilles}, {Renzi}, {Rocha},
  {Rossetti}, {Roudier}, {Rubi{\~n}o-Mart{\'{\i}}n}, {Ruiz-Granados},
  {Salvati}, {Sandri}, {Savelainen}, {Scott}, {Sirri}, {Sunyaev}, {Suur-Uski},
  {Tauber}, {Tenti}, {Toffolatti}, {Tomasi}, {Tristram}, {Trombetti},
  {Valiviita}, {Van Tent}, {Vielva}, {Villa}, {Vittorio}, {Wandelt}, {Wehus},
  {White}, {Zacchei}, \& {Zonca}}]{planck16}
{Planck Collaboration}, {Adam}, R., {Aghanim}, N., {et~al.} 2016, \aap, 596,
  A108

\bibitem[{{Popping} {et~al.}(2016){Popping}, {van Kampen}, {Decarli}, {Spaans},
  {Somerville}, \& {Trager}}]{popping16}
{Popping}, G., {van Kampen}, E., {Decarli}, R., {et~al.} 2016, ArXiv e-prints

\bibitem[{{Rawle} {et~al.}(2014){Rawle}, {Egami}, {Bussmann}, {Gurwell},
  {Ivison}, {Boone}, {Combes}, {Danielson}, {Rex}, {Richard}, {Smail},
  {Swinbank}, {Altieri}, {Blain}, {Clement}, {Dessauges-Zavadsky}, {Edge},
  {Fazio}, {Jones}, {Kneib}, {Omont}, {P{\'e}rez-Gonz{\'a}lez}, {Schaerer},
  {Valtchanov}, {van der Werf}, {Walth}, {Zamojski}, \& {Zemcov}}]{rawle14}
{Rawle}, T.~D., {Egami}, E., {Bussmann}, R.~S., {et~al.} 2014, \apj, 783, 59

\bibitem[{{Riechers} {et~al.}(2013){Riechers}, {Bradford}, {Clements},
  {Dowell}, {P{\'e}rez-Fournon}, {Ivison}, {Bridge}, {Conley}, {Fu}, {Vieira},
  {Wardlow}, {Calanog}, {Cooray}, {Hurley}, {Neri}, {Kamenetzky}, {Aguirre},
  {Altieri}, {Arumugam}, {Benford}, {B{\'e}thermin}, {Bock}, {Burgarella},
  {Cabrera-Lavers}, {Chapman}, {Cox}, {Dunlop}, {Earle}, {Farrah}, {Ferrero},
  {Franceschini}, {Gavazzi}, {Glenn}, {Solares}, {Gurwell}, {Halpern},
  {Hatziminaoglou}, {Hyde}, {Ibar}, {Kov{\'a}cs}, {Krips}, {Lupu}, {Maloney},
  {Martinez-Navajas}, {Matsuhara}, {Murphy}, {Naylor}, {Nguyen}, {Oliver},
  {Omont}, {Page}, {Petitpas}, {Rangwala}, {Roseboom}, {Scott}, {Smith},
  {Staguhn}, {Streblyanska}, {Thomson}, {Valtchanov}, {Viero}, {Wang},
  {Zemcov}, \& {Zmuidzinas}}]{riechers13}
{Riechers}, D.~A., {Bradford}, C.~M., {Clements}, D.~L., {et~al.} 2013, \nat,
  496, 329

\bibitem[{{Riechers} {et~al.}(2014){Riechers}, {Carilli}, {Capak}, {Scoville},
  {Smol{\v c}i{\'c}}, {Schinnerer}, {Yun}, {Cox}, {Bertoldi}, {Karim}, \&
  {Yan}}]{riechers14}
{Riechers}, D.~A., {Carilli}, C.~L., {Capak}, P.~L., {et~al.} 2014, \apj, 796,
  84

\bibitem[{{Robertson} {et~al.}(2015){Robertson}, {Ellis}, {Furlanetto}, \&
  {Dunlop}}]{robertson15}
{Robertson}, B.~E., {Ellis}, R.~S., {Furlanetto}, S.~R., \& {Dunlop}, J.~S.
  2015, \apjl, 802, L19

\bibitem[{{Sargsyan} {et~al.}(2014){Sargsyan}, {Samsonyan}, {Lebouteiller},
  {Weedman}, {Barry}, {Bernard-Salas}, {Houck}, \& {Spoon}}]{Sargasyan14}
{Sargsyan}, L., {Samsonyan}, A., {Lebouteiller}, V., {et~al.} 2014, \apj, 790,
  15

\bibitem[{{Saunders} {et~al.}(1990){Saunders}, {Rowan-Robinson}, {Lawrence},
  {Efstathiou}, {Kaiser}, {Ellis}, \& {Frenk}}]{saunders90}
{Saunders}, W., {Rowan-Robinson}, M., {Lawrence}, A., {et~al.} 1990, \mnras,
  242, 318

\bibitem[{{Smit} {et~al.}(2017){Smit}, {Bouwens}, {Carniani}, {Oesch},
  {Labb{\'e}}, {Illingworth}, {van der Werf}, {Bradley}, {Gonzalez}, {Hodge},
  {Holwerda}, \& {Maiolino}}]{smit17}
{Smit}, R., {Bouwens}, R.~J., {Carniani}, S., {et~al.} 2017, ArXiv e-prints

\bibitem[{{Smit} {et~al.}(2012){Smit}, {Bouwens}, {Franx}, {Illingworth},
  {Labb{\'e}}, {Oesch}, \& {van Dokkum}}]{smit12}
{Smit}, R., {Bouwens}, R.~J., {Franx}, M., {et~al.} 2012, \apj, 756, 14

\bibitem[{{Smith} {et~al.}(2017){Smith}, {Croxall}, {Draine}, {De Looze},
  {Sandstrom}, {Armus}, {Beir{\~a}o}, {Bolatto}, {Boquien}, {Brandl},
  {Crocker}, {Dale}, {Galametz}, {Groves}, {Helou}, {Herrera-Camus}, {Hunt},
  {Kennicutt}, {Walter}, \& {Wolfire}}]{smith17}
{Smith}, J.~D.~T., {Croxall}, K., {Draine}, B., {et~al.} 2017, \apj, 834, 5

\bibitem[{{Song} {et~al.}(2016){Song}, {Finkelstein}, {Ashby}, {Grazian}, {Lu},
  {Papovich}, {Salmon}, {Somerville}, {Dickinson}, {Duncan}, {Faber}, {Fazio},
  {Ferguson}, {Fontana}, {Guo}, {Hathi}, {Lee}, {Merlin}, \&
  {Willner}}]{song16}
{Song}, M., {Finkelstein}, S.~L., {Ashby}, M.~L.~N., {et~al.} 2016, \apj, 825,
  5

\bibitem[{{Spilker} {et~al.}(2016){Spilker}, {Marrone}, {Aravena},
  {B{\'e}thermin}, {Bothwell}, {Carlstrom}, {Chapman}, {Crawford}, {de Breuck},
  {Fassnacht}, {Gonzalez}, {Greve}, {Hezaveh}, {Litke}, {Ma}, {Malkan},
  {Rotermund}, {Strandet}, {Vieira}, {Weiss}, \& {Welikala}}]{spilker16}
{Spilker}, J.~S., {Marrone}, D.~P., {Aravena}, M., {et~al.} 2016, \apj, 826,
  112

\bibitem[{{Stacey} {et~al.}(1991){Stacey}, {Geis}, {Genzel}, {Lugten},
  {Poglitsch}, {Sternberg}, \& {Townes}}]{stacey91}
{Stacey}, G.~J., {Geis}, N., {Genzel}, R., {et~al.} 1991, \apj, 373, 423

\bibitem[{{Stacey} {et~al.}(2010){Stacey}, {Hailey-Dunsheath}, {Ferkinhoff},
  {Nikola}, {Parshley}, {Benford}, {Staguhn}, \& {Fiolet}}]{stacey10}
{Stacey}, G.~J., {Hailey-Dunsheath}, S., {Ferkinhoff}, C., {et~al.} 2010, \apj,
  724, 957

\bibitem[{{Strandet} {et~al.}(2017){Strandet}, {Wei{\ss}}, {De Breuck},
  {Marrone}, {Vieira}, {Aravena}, {Ashby}, {B{\'e}thermin}, {Bothwell},
  {Bradford}, {Carlstrom}, {Chapman}, {Cunningham}, {Chen}, {Fassnacht},
  {Gonzalez}, {Greve}, {Gullberg}, {Hayward}, {Hezaveh}, {Litke}, {Ma},
  {Malkan}, {Menten}, {Miller}, {Murphy}, {Narayanan}, {Phadke}, {Rotermund},
  {Spilker}, \& {Sreevani}}]{strandet17}
{Strandet}, M.~L., {Wei{\ss}}, A., {De Breuck}, C., {et~al.} 2017, ArXiv
  e-prints

\bibitem[{{Swinbank} {et~al.}(2012){Swinbank}, {Karim}, {Smail}, {Hodge},
  {Walter}, {Bertoldi}, {Biggs}, {de Breuck}, {Chapman}, {Coppin}, {Cox},
  {Danielson}, {Dannerbauer}, {Ivison}, {Greve}, {Knudsen}, {Menten},
  {Simpson}, {Schinnerer}, {Wardlow}, {Wei{\ss}}, \& {van der
  Werf}}]{swinbank12}
{Swinbank}, A.~M., {Karim}, A., {Smail}, I., {et~al.} 2012, \mnras, 427, 1066

\bibitem[{{Tweed} {et~al.}(2009){Tweed}, {Devriendt}, {Blaizot}, {Colombi}, \&
  {Slyz}}]{Tweed_2009}
{Tweed}, D., {Devriendt}, J., {Blaizot}, J., {Colombi}, S., \& {Slyz}, A. 2009,
  \aap, 506, 647

\bibitem[{{Vallini} {et~al.}(2013){Vallini}, {Gallerani}, {Ferrara}, \&
  {Baek}}]{vallini13}
{Vallini}, L., {Gallerani}, S., {Ferrara}, A., \& {Baek}, S. 2013, \mnras, 433,
  1567

\bibitem[{{Vallini} {et~al.}(2015){Vallini}, {Gallerani}, {Ferrara},
  {Pallottini}, \& {Yue}}]{vallini15}
{Vallini}, L., {Gallerani}, S., {Ferrara}, A., {Pallottini}, A., \& {Yue}, B.
  2015, \apj, 813, 36

\bibitem[{{Wagg} {et~al.}(2012){Wagg}, {Wiklind}, {Carilli}, {Espada}, {Peck},
  {Riechers}, {Walter}, {Wootten}, {Aravena}, {Barkats}, {Cortes}, {Hills},
  {Hodge}, {Impellizzeri}, {Iono}, {Leroy}, {Mart{\'{\i}}n}, {Rawlings},
  {Maiolino}, {McMahon}, {Scott}, {Villard}, \& {Vlahakis}}]{wagg12}
{Wagg}, J., {Wiklind}, T., {Carilli}, C.~L., {et~al.} 2012, \apjl, 752, L30

\bibitem[{{Walter} {et~al.}(2012){Walter}, {Decarli}, {Carilli}, {Bertoldi},
  {Cox}, {da Cunha}, {Daddi}, {Dickinson}, {Downes}, {Elbaz}, {Ellis}, {Hodge},
  {Neri}, {Riechers}, {Weiss}, {Bell}, {Dannerbauer}, {Krips}, {Krumholz},
  {Lentati}, {Maiolino}, {Menten}, {Rix}, {Robertson}, {Spinrad}, {Stark}, \&
  {Stern}}]{walter12}
{Walter}, F., {Decarli}, R., {Carilli}, C., {et~al.} 2012, \nat, 486, 233

\bibitem[{{Watson} {et~al.}(2015){Watson}, {Christensen}, {Knudsen}, {Richard},
  {Gallazzi}, \& {Micha{\l}owski}}]{watson15}
{Watson}, D., {Christensen}, L., {Knudsen}, K.~K., {et~al.} 2015, \nat, 519,
  327

\bibitem[{{Wiesenfeld} \& {Goldsmith}(2014)}]{wiesenfeld14}
{Wiesenfeld}, L. \& {Goldsmith}, P.~F. 2014, \apj, 780, 183

\bibitem[{{Willott} {et~al.}(2015){Willott}, {Carilli}, {Wagg}, \&
  {Wang}}]{willott15}
{Willott}, C.~J., {Carilli}, C.~L., {Wagg}, J., \& {Wang}, R. 2015, \apj, 807,
  180

\bibitem[{{Wolfire} {et~al.}(2003){Wolfire}, {McKee}, {Hollenbach}, \&
  {Tielens}}]{wolfire03}
{Wolfire}, M.~G., {McKee}, C.~F., {Hollenbach}, D., \& {Tielens}, A.~G.~G.~M.
  2003, \apj, 587, 278

\bibitem[{{Yamaguchi} {et~al.}(2017){Yamaguchi}, {Kohno}, {Tamura}, {Oguri},
  {Ezawa}, {Hayatsu}, {Kitayama}, {Matsuda}, {Matsuo}, {Oshima}, {Ota},
  {Izumi}, \& {Umehata}}]{yamaguchi17}
{Yamaguchi}, Y., {Kohno}, K., {Tamura}, Y., {et~al.} 2017, ArXiv e-prints

\end{thebibliography}

\end{document}